\documentclass[dutch]{book}
\usepackage{babel,palatino,a4wide}
\usepackage[pdftex]{color,graphicx,hyperref}
\usepackage{style,links}

\advance\topmargin by -1cm

\begin{document}
\begin{titlepage}

\href{http://www.niii.kun.nl}{
{\ExtraHuge \NIIITekst} 
\begin{tabular}[b]{l}
 \Large \sf nijmeegs instituut\\[0.2mm]
 \Large \sf voor informatica en informatiekunde
\end{tabular}}

\begin{center}
  \vfill

  \vfill

  \textcolor{blue}{\ExtraHuge Informatiekunde}
 
  \vspace{0.5cm}

  \textcolor{red}{\Huge Curriculum 2003}

  \vfill

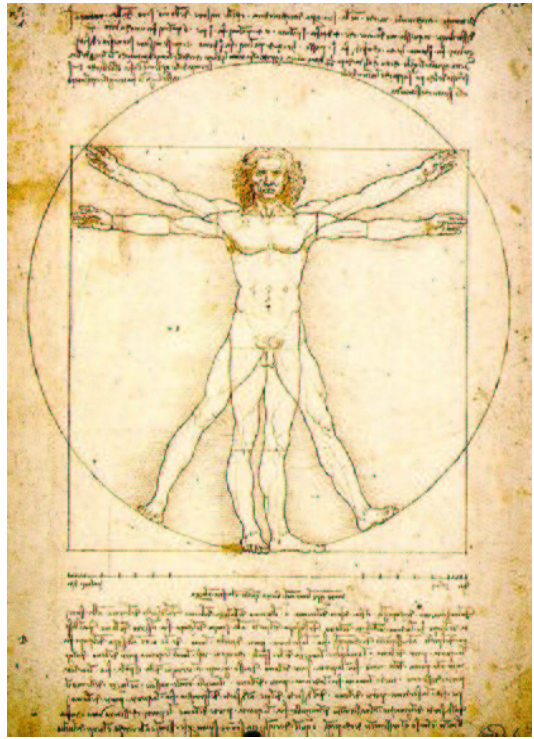
  \Image{width=7cm}{DaVinci}

  \vfill
\end{center}

\textcolor{blue}{\begin{tabular}{ll}
  {\large Vormgevers van de digitale samenleving:}\\[0.1cm]
  \mbox{--~}
  Informatiekundigen met een $\beta$-opleiding, een $\gamma$-feeling
  en een gezonde dosis creativiteit
\end{tabular}
}
\vfill

\begin{center}
  \textcolor{red}{\large Versie van: \today}
\end{center}

\end{titlepage}

\newpage
\tableofcontents

\chapter{Introductie}
\label{h:introductie}

\section{Doel van dit document}
Dit document heeft betrekking op het curriculum van de opleiding
informatiekunde van het Nijmeegs Instituut voor Informatica en Informatiekunde
(\NIII). Het doel is het bieden van een `repository' met betrekking tot de
inrichting van het curriculum wat vanaf 2003 zal gaan gelden.
De visie op het vakgebied informatiekunde, die ten grondslag ligt aan dit
curriculum document is te vinden in~\cite{Visie2003}.

In de afgelopen drie jaar is zowel op landelijk als op Nijmeegs niveau het
beeld van informatiekunde als vakgebied nader geconcretiseerd. Het Curriculum
2003 is enerzijds het resultaat van deze concretisering en anderzijds van de
drie jaar ervaring die inmiddels binnen het \NIII\ is opgebouwd met de
informatiekunde opleiding.  In dit document zal daarom ook expliciet aandacht
besteed worden aan de `migratie' vanuit de bestaande `opstart' curricula: 2000,
2001 en 2002.  Hierbij moet opgemerkt worden dat de studenten van cohort 2000
in principe dit jaar (2003) de bachelorfase van de Informatiekunde opleiding
afronden. Voor deze lichting studenten is dus geen specifieke `migratie' nodig.

In overeenstemming met het doel van dit document om als `repository' dienst te
doen ligt het in de verwachting dat dit document jaarlijks een update zal
ondergaan.  Dit betekent echter niet dat het de bedoeling is jaarlijks het
curriculum te wijzigen.  De ambitie met betrekking tot de stabiliteit van
hetgeen in dit document worden besproken is als volgt (de tijdsspanne geeft aan
na hoeveel tijd het betreffende onderdeel naar verwachting herzien zal gaan
worden):
\begin{itemize}
   \item Visie op het vakgebied, zoals te vinden in~\cite{Visie2003}:
   5 \`a 6 jaar
   \item Doelstellingen en eindtermen: 5 \`a 6 jaar
   \item Hoofdlijn van het programma: 4 \`a 5 jaar
   \item Programma: 3 \`a 4 jaar
   \item Migratie programma's: 2 \`a 3 jaar
\end{itemize}

\section{Doelgroep}
De doelgroep van dit document bestaat in eerste instantie uit de bestuurders
van de Faculteit NWI, de collega's binnen het \NIII\, en de collega's van
externe opleidingen die bij het informatiekunde curriculum dan wel het
vakgebied betrokken zijn (de interne klankbordgroep, inclusief 4 studentleden).
In tweede instantie behoren daartoe ook vertegenwoordigers van
zusterfaculteiten en -opleidingen elders, en vertegenwoordigers uit het
werkveld (de externe klankbordgroep). Het stuk kan in een later stadium als
basis dienen voor het genereren van verschillende teksten voor andere
doelgroepen, zoals (potenti\"ele) studenten, de visitatiecommissie e.d.  Op dit
moment wordt echter nog geen volledigheid/leesbaarheid voor die doelgroepen
nagestreefd.

\section{Ontstaan van dit document}
Het startpunt van dit document bestond uit:
\begin{itemize}
  \item Diverse presentaties en documenten met betrekking tot de
        informatiekunde opleiding, van de hand van diverse auteurs.
  \item Ervaringen met Curricula 2000, 2001 en 2002.
  \item Standaard curriculum Informatiemanagement van de ACM en de IEEE.
\end{itemize}
Het huidige document voor Curriculum 2003 werd geproduceerd door een kernteam
bestaande uit (in alfabetische volgorde):
\begin{itemize}
  \item \ErikB\ (vertegenwoordiger van de afdeling \AfdGS\ en liaison informatica).
  \item \StijnH\ (editor).
  \item \VeraK\ (editor).
  \item \ErikP\ (editor en eindverantwoordelijke).
  \item \JanT\ (vertegenwoordiger van de afdeling \AfdST).
  \item \TvdW\ (vertegenwoordiger van de afdeling \AfdIRIS).
  \item \Hanno\ (vertegenwoordiger van de afdeling \AfdITT).
\end{itemize}
Het daadwerkelijke schrijfwerk is met name door \VeraK, \ErikP, en \StijnH\
gedaan. Tevens is gebruik gemaakt van twee klankbordgroepen.  Een (KUN) interne
klankbordgroep bestaande uit:
\begin{itemize}
  \item Informatiekundestudenten:
            Jeroen Groenen (1e jaars),
            Mark Jenniskens (2e jaars),
            Arnoud Vermeij (3e jaars),
            Wout Lemmens (HBO-instromer).
  \item \OCie\ (Opleidingscommissie) liaison: Mark Jenniskens.
  \item \NSM: Bart Prakken.
  \item \MIK: Hans ten Hoopen.
  \item \IOWO: Bea Edlinger, G\'e Ophelders.
  \item \NIII: Jeroen Bruijning, Bas van Gils,
        Franc Grootjen, Stijn Hoppenbrouwers, Pieter Koopman, Martijn Oostdijk,
        Ger Paulussen, Rinus Plasmeijer, Eric Schabell, Frits Vaandrager,
        Gert Veldhuijzen van Zanten, Paul de Vrieze.
\end{itemize}
Tevens werd er een externe klankbordgroep ingeschakeld bestaande uit vertegenwoordigers
van zusterfaculteiten en vertegenwoordigers uit het werkveld:
\begin{itemize}
  \item Hans Bossenbroek, \Luminis.
  \item Jeroen Top, \Belastingdienst.
  \item Victor van Reijswoud, \UMU.
  \item Roel Wieringa, \UT.
\end{itemize}

In eerste instantie was dit curriculumdocument volledig ge\"{\i}ntegreerd met
het document~\cite{Visie2003} wat de visie op het vakgebied informatiekunde,
de opleiding en het onderzoek beschrijft. Wegens de uitgebreidheid van de visie
op het vakgebied, en de zelfstandige rol die deze visie kan vervullen richting
opleiding \emph{en} onderzoek, is uiteindelijk besloten deze documenten te
splitsen.

\section{Gevolgde redeneerlijn}
Bij het structureren van dit document is er voor gekozen om een structuur
te gebruiken die het mogelijk maakt om:
\begin{itemize}
   \item in een aantal controleerbare stappen vanuit een visie op het vakgebied
         te komen tot de uiteindelijke inrichting van de opleiding,
   \item een duidelijke relatie te leggen tussen de structuur van de opleiding
         en de rol (en bijbehorende vaardigheden) die studenten na afloop van
         hun studie kunnen vervullen.
\end{itemize}
Het laatste punt komt voort uit de hypothese dat de motivatie van een student
tijdens de studie hoger zal zijn als steeds duidelijk is hoe een vak bijdraagt
aan het doel waarmee de student gekozen heeft om de opleiding te volgen.

De redeneerlijn die werd aangehouden bij het opstellen van het document is daarom
als volgt.
\begin{enumerate}
    \item Er werd een aantal visies geformuleerd: op het vakgebied van de
    informatiekunde, op de opleiding informatiekunde, het onderzoek, op
    onderwijs en leren, op beoordeling en op kwaliteitsbeheersing.
    \item Er werden randvoorwaarden beschreven: vaststaande factoren waar
    rekening gehouden zal moeten worden.
    \item Samen waren alle visies plus de randvoorwaarden input voor de
    inrichting van het curriculum.
    \item Samen waren alle visies plus de randvoorwaarden input voor de
    inrichting van het curriculum.
    \item Vanuit de visies en de randvoorwaarden werden expliciet verbanden
    gelegd naar de inrichting. Dit gebeurde met behulp van:
    \begin{itemize}
       \item formulering van \emph{vereiste vaardigheden};
       \item formulering van \emph{inrichtingsprincipes}.
    \end{itemize}
\end{enumerate}
De resultaten van stappen 1 en 2 zijn terug te vinden in het
visiedocument~\cite{Visie2003}. In dit document richten we ons
primair op stappen 3 tot en met 5.

We zullen nu eerst nader ingaan op wat we bedoelen met `vaardigheden' en
`inrichtingsprincipes'. In navolging van huidige didactische inzichten wordt
het wenselijk geacht het curriculum te richten op het aanleren van de
vaardigheden die een informatiekundige moet beheersen. Vaardigheden betreffen
`dingen die iemand moet \emph{kunnen}' (niet zo zeer \emph{kennen}). Het
formuleren van vaardigheden legt dus de nadruk op het actief toepassen van
kennis, en minder op het kunnen `beschrijven' van die kennis (dit wordt wel
eens omschreven als `knowledge how' versus `knowledge that'). De standaard
frase die bij het formuleren van vaardigheden steeds wordt gebruikt is `Een
informatiekundige moet in staat zijn om ...'. Zelfs vereisten op het gebied van
theoretische kennis worden daarbij in actieve zin verwoord, bijvoorbeeld `een
informatiekundige moet in staat zijn om theorie $X$ toe te passen op een casus,
en hierover op constructieve wijze te reflecteren'.

Er kunnen verschillen zijn in het \emph{ambitieniveau} waarop een bepaalde
vaardigheid moet worden aangeleerd. Met andere woorden, er moet worden
vastgesteld \emph{hoe goed} een informatiekundige bepaalde vaardigheden dient
te beheersen. Dat kan bijvoorbeeld uiteenlopen van oppervlakkige, passieve beheersing
tot aan diepgaande, actieve beheersing. Pas wanneer een vaardigheid gekoppeld wordt
aan een ambitieniveau kan bepaald worden wat de \emph{eindtermen} van het curriculum
zijn. Een eindterm kan dus beschreven worden als `vaardigheid + ambitieniveau'; het
is een duidelijk omschreven onderwijsdoel.

De vaardigheden mogen dan de basis zijn van de inrichting van het curriculum, ze zijn
niet het enige ingredient. Vanuit de verschillende visies en randvoorwaarden die aan
het curriculum ten grondslag liggen worden ook een aantal \emph{inrichtingsprincipes}
geformuleerd waaraan het curriculum moet voldoen. Die principes bepalen grotendeels
\emph{hoe} de onderwijsdoelen moeten worden bereikt. Inrichtingsprincipes dienen in het
algemeen te voldoen aan twee basisvoorwaarden:
\begin{itemize}
    \item Ze moeten zo SMART mogelijk zijn (Specific, Measurable, Attainable, Realisable,
    Timely).
    \item Ze mogen niet triviaal zijn: ze vertegenwoordigen een standpunt dat
    mogelijk discussie aantrekt.
\end{itemize}

Er is bijvoorbeeld een inrichtingsprincipe geformuleerd met
betrekking tot het doel van de propedeuse: `Het eerste jaar biedt,
naast een methodologische basis, een brede ori\"entatie op de
verschillende verbredingsgebieden, en illustreert de samenhang van
de verschillende kanten van de studie. Er is nog geen sprake van
specialisatie- of keuzevakken.'

Samengevat kunnen alle onderdelen van de visie en alle randvoorwaarden dus op
twee manieren input voor het curriculum vormen: in de vorm van vaardigheden van
de informatiekundige (\emph{wat} het curriculum moet bevatten) en in de vorm
van inrichtingsprincipes voor het curriculum (\emph{hoe} het ingericht moet
worden).  De vaardigheden en inrichtingsprincipes vormen een expliciete
verbinding tussen de visies en randvoorwaarden aan de ene kant, en de
eindtermen aan de andere. Zo kunnen de eindtermen van de inrichting altijd
herleid worden tot een bepaalde visie of een bepaalde randvoorwaarde.

\section{Structuur van dit document}
Dit document begint, in hoofdstuk~\ref{p:samenvatting}, met een samenvatting
van de belangrijkste elementen van de visie op het vakgebied~\cite{Visie2003}.

De doelstellingen en eindtermen van de opleiding informatiekunde (zowel de
bachelorfase als de masterfase) worden besproken in
hoofdstuk~\ref{h:wat-doelstellingen}.  Voordat we de uiteindelijke programma's
van de opleiding kunnen toelichten, is het nodig om eerst stil te staan bij de
inhoudelijke structuur die gekozen is voor de opleiding in termen van thema's
die als een rode draad door de opleiding zullen lopen. Dit is het doel van
hoofdstuk~\ref{p:hoe-structuur}.  Voor het Curriculum 2003 van Informatiekunde
zijn er feitelijk drie opleidingsprogramma's nodig:
\begin{itemize}
  \item Bachelor programma.
  \item Master programma voor doorstromers/instromers met een academische
        Bachelor.
  \item Master programma voor doorstromers/instromers met een HBO Bachelor.
\end{itemize}
Deze programma's worden in hoofdstukken~\ref{h:BachelorProgramma},
\ref{h:VWOMasterProgramma} en~\ref{h:HBOMasterProgramma} besproken.

Daarnaast zijn er vervolgprogramma's nodig voor de nu reeds studerende
cohorten:
\begin{itemize}
   \item Bachelor cohort 2002
   \item Bachelor cohort 2001
\end{itemize}
Deze programma's worden respectievelijk in hoofdstukken~\ref{h:Bachelor2002}
en~\ref{h:Bachelor2001} besproken.
Studenten uit Bachelor cohort 2000 en de HBO Master cohort 2002,
zouden in principe dit jaar hun opleidingen moeten afronden.
Hierbij is het de bedoeling dat de Bachelor's van cohort 2000 instromen
in het Master programma van 2003.

Behalve het cohort ``HBO Master 2002'' zijn er nog geen studenten met hun
masterfase bezig.  De invulling van de HBO Master 2002 was, om diverse redenen,
wat prematuur van aard.  Alle reden om met de invoering van het curriculum 2003
meteen over te gaan op het nieuwe Master programma, en tevens om (met inspraak
van de huidige HBO instromers) het schakelprogramma voor de HBO instromers te
herzien en beter af te stemmen op hun specifieke behoeften.

In~\cite{Visie2003} zijn er diverse inrichtingsprincipes geformuleerd waar
de informatiekunde opleiding aan moet voldoen. Daarom wordt in
hoofdstuk~\ref{h:hoe-principes} de gekozen inrichting getoetst aan deze
principes.

In appendix~\ref{h:Vakken} zijn de beschrijvingen te vinden van de vakken
zoals deze zijn opgenomen in het nieuwe curriculum. Omdat dit document ook de
migratie vanuit de bestaande curricula bespreekt, zijn in
appendix~\ref{h:OudeVakken} voor de volledigheid ook nog de beschrijvingen
opgenomen van vakken uit de bestaande curricula.  Verder wordt er in de
appendix uitgebreid stilgestaan met de relatie tussen het hier gepresenteerde
curriculum 2003, twee standaard curricula en de uitkomsten van de visitatie
Informatiekunde uit 2002:
\begin{itemize}
   \item Model curriculum voor Bacheloropleidingen op het gebied van
   Informatiesystemen (appendix~\ref{IS2002}).
   \item Model curriculum voor Masteropleidingen op het gebied van
   Informatiesystemen (appendix~\ref{MSIS2000}).
   \item Resultaten van de visitatie Informatiekunde 2002
   (appendix~\ref{Visitatie2002}).
\end{itemize}

\chapter{Samenvatting van de visie op het vakgebied}
\label{p:samenvatting}

Vera Kamphuis werkt momenteel aan een samenvatting.

\Ragged{\chapter{Doelstellingen en eindtermen}
\label{h:wat-doelstellingen}

In het visiedocument~\cite{Visie2003} zijn een aantal maatschappelijke
ontwikkelingen besproken die een rol hebben gespeeld bij het ontstaan van het
vakgebied informatiekunde, en die hebben geleid tot de behoefte aan
Informatiekundigen. Tevens is er een inventarisatie gemaakt van de conclusies
die daaraan verbonden moeten worden ten aanzien van de informatiekunde
opleiding. In dit hoofdstuk wordt deze inventarisatie vertaald naar een
concretisering van de doelstellingen en eindtermen van de opleiding
informatiekunde. Achtereenvolgens wordt aandacht besteed aan:
\bi
\item Terminologische definities, zoals leerdoelen, vaardigheden,
doelstellingen en eindtermen (paragraaf~\ref{s:Termen});
\item De leerdoelen \& doelstellingen van de opleiding (paragraaf~\ref{s:Leerdoelen});
\item De vaardigheden \& eindtermen van de opleiding (paragraaf~\ref{s:Vaardigheden}).
\ei

\section{Terminologie}
\label{s:Termen}
Alvorens concreter in te gaan op de doelstellingen en eindtermen van de
opleiding, zullen we deze concepten eerst nader defini\"eren. Hiertoe zullen we
tevens gebruik maken van de notie `ambitieniveau'. We gebruiken deze term om te
verwijzen naar de mate van zelfstandigheid waarmee de latere afgestudeerde het
beoogde beroep of de beoogde vaardigheid kan uitvoeren (bijvoorbeeld,
functioneren in teamverband, zelfstandig of als leidinggevende). Met behulp van
deze notie kunnen we de volgende definities formuleren.

{\bf Vaardigheden} betreffen
`dingen die iemand moet \emph{kunnen}' (niet zo zeer \emph{kennen}). Het
formuleren van vaardigheden legt dus de nadruk op het actief toepassen van
kennis, en minder op het kunnen `beschrijven' van die kennis (dit wordt wel
eens omschreven als `knowledge how' versus `knowledge that'). De standaard
frase die bij het formuleren van vaardigheden steeds wordt gebruikt is `Een
informatiekundige moet in staat zijn om ...'. Zelfs vereisten op het gebied van
theoretische kennis worden daarbij in actieve zin verwoord, bijvoorbeeld `een
informatiekundige moet in staat zijn om theorie $X$ toe te passen op een casus,
en hierover op constructieve wijze te reflecteren'.

{\bf Leerdoelen} beschouwen we als zijnde aggregaties van vaardigheden.
Er kunnen verschillen zijn in het {\bf ambitieniveau} waarop een bepaalde
vaardigheid/leerdoel moet worden aangeleerd. Met andere woorden, er moet worden
vastgesteld \emph{hoe goed} een informatiekundige bepaalde vaardigheden dient
te beheersen. Dat kan bijvoorbeeld uiteenlopen van oppervlakkige, passieve
beheersing tot aan diepgaande, actieve beheersing. Pas wanneer een vaardigheid
gekoppeld wordt aan een ambitieniveau kan bepaald worden wat de
\emph{eindtermen} zijn. Een {\bf eindterm} kan dus beschreven worden als
`vaardigheid + ambitieniveau'. Onder een {\bf doelstelling} verstaan we een
\emph{leerdoel} (aggregatie van vaardigheden) met een daaraan verbonden een
ambitieniveau.

Voor het beschrijven van ambitieniveaus gebruiken we een onderverdeling maken
in 4 niveaus, die we kunnen koppelen aan de verschillende fases van de
opleiding. Merk op dat we in deze schaal tevens iets kunnen terugzien van de
traditionele meester-gezel constructie, waarin de gezel zich onder begeleiding
van een meester gaandeweg zelf tot
meester ontwikkelt.
\begin{description}
   \item [Belangstellend --] bewust zijn van de aard en de positie in de context
   van een vaardigheid. Dit ambitieniveau is typerend voor de Propedeuse-fase;
   \item [Uitvoerend -- ] onder leiding uit kunnen voeren van een vaardigheid,
   binnen een gegeven kader. Dit ambitieniveau is typerend voor de Bachelor-fase.
   \item [Richtinggevend -- ] zelfstandig uit kunnen voeren van een vaardigheid
   binnen een gegeven kader.  Dit ambitieniveau is typerend voor de Master-fase.
   \item [Kaderscheppend --] Zelfstandig uit kunnen voeren van een vaardigheid
   binnen een zelfstandig afgebakend kader.  Dit ambitieniveau is typerend voor
   een Post-Graduate opleiding.
\end{description}

\section{Leerdoelen \& doelstellingen}
\label{s:Leerdoelen}
Informatiekundigen worden in hun vier-jarige opleiding niet exclusief
voorgesorteerd op een rol als onderzoeker, docent of professional.  We hebben al
aangegeven dat informatiekundigen een goede basis mee moeten krijgen in ieder
van deze drie richtingen. Dit leidt tot de onderstaande drie leerdoelen voor de
opleiding informatiekunde. Een informatiekunde kan:
\begin{description}
   \item[Onderzoeken --] ... als \emph{onderzoeker} in een onderzoeksteam een
   theoretische bijdrage leveren aan de ontwikkeling van het eigen vakgebied;
   \item[Doceren \& voorlichten --] ... als \emph{docent} of voorlichter
   anderen onderwijzen in, of voorlichten over, het
   vakgebied der Informatiekunde;
   \item[Veranderen \& bestendigen --] ... als \emph{professional} in diverse rollen
   in een gegeven praktische context een creatieve en praktische bijdrage
   leveren aan het veranderen en bestendigen van informatiesystemen in
   hun menselijke, organisatorische, informationele, technologische en
   systemische context.
\end{description}
De vaardigheden zoals die in de volgende paragraaf worden besproken zijn feitelijk
te zien als nadere concretiseringen van deze leerdoelen.

\def\B{Belangstellend} \def\U{Uitvoerend} \def\R{Richtinggevend} \def\K{Kaderzettend}

In onderstaande tabel zijn de leerdoelen gekoppeld aan ambitieniveau's, leidende
tot doelstellingen. Er is bewust voor gekozen om ook een kolom `Post Graduate' op
te nemen om de ambitie te schetsen van iemand die 4 \'a 5 jaar in het vakgebied
(als docent, onderzoeker of professional) werkzaam is. Daarnaast hebben we deze laatste
kolom opgesplitst op bass van een rol als onderzoeker, docent of professional.
\begin{center} \footnotesize
     \begin{tabular}{l|l|l|l|l|l}
     Leerdoel      & Propedeuse & Bachelor & Master & Post Graduate & Rol\\
     \hline
     \hline
                   &            &          &        & \K            & Onderzoeker\\
     \cline{5-6}
     Onderzoeken   & \B         & \U       & \R     & \R            & Docent\\
     \cline{5-6}
                   &            &          &        & \R            & Professional\\
     \hline
     \hline
                   &            &          &        & \R            & Onderzoeker\\
     \cline{5-6}
     Doceren \& voorlichten
                   & \B         & \U       & \R     & \K            & Docent\\
     \cline{5-6}
                   &            &          &        & \R            & Professional\\
     \hline
     \hline
                   &            &          &        & \R            & Onderzoeker\\
     \cline{5-6}
     Veranderen \& bestendigen
                   & \B         & \U       & \R     & \R            & Docent\\
     \cline{5-6}
                   &            &          &        & \K            & Professional\\
     \hline
   \end{tabular}
\end{center}
Merk op dat de rollen onderzoeker, docent en professional, direct overeenkomen met
de door de faculteit benoemde uitstroomprofielen voor de 5-jarige masteropleidingen:
\begin{description}
   \item[Onderzoeker --] onderzoek profiel.
   \item[Docent --] communicatie \& educatie profiel.
   \item[Professional --] management \& toepassing profiel.
\end{description}

\section{Vaardigheden \& eindtermen}
\label{s:Vaardigheden}
De vaardigheden van een informatiekundige zijn in~\cite{Visie2003} reeds op
hoofdlijnen besproken. Hierbij is gebruik gemaakt van een opdeling van de
vaardigheden naar:
\begin{itemize}
   \item Proces van verandering \& bestendiging
   \item Subject van verandering \& bestendiging
   \item Verbredingsgebieden
   \item Onderzoek \& reflectie
\end{itemize}
Wat hierbij nog ontbreekt zijn de algemene vaardigheden.
De resulterende vijf clusters van vaardigheden, en bijbehorende eindtermen,
worden hieronder per stuk besproken.

\subsection{Algemeen}
Een informatiekundige is na afloop van de studie in staat om:
\begin{description}
   \item[Academisch --] ... op een academisch niveau te werken en te denken;
   \item[Zelf leren --] ... zelfstandig en onder eigen verantwoordelijkheid te leren;
   \item[Kennis ontsluiten --] ... kennis- en ervaringsbronnen te ontsluiten,
   voorzover deze aansluiten bij hun reeds bestaande kennis;
   \item[Kennis inzetten --] ... voor informatiekundige problemen, relevante
   kennisgebieden aan te geven en hun mogelijke bijdrage aan de oplossing van
   het probleem te identificeren, en waar relevant deze kennis inzetten bij het
   oplossen van het probleem.
   \item[Reflectie op leren --] ... te reflecteren op het eigen leerproces (of
   dat van een college) en de daarin gebruikte leerstrategie\"en en -stijlen, en
   indien nodig deze leerprocessen bij te sturen;
   \item[Reflectie op handelen --] ... te reflecteren op hun potenti\"ele rol
   (of die van een collega) en kunnen participeren in een maatschappelijk debat
   over kwesties die samenhangen met het eigen vakgebied.
\end{description}
Wanneer we deze vaardigheden vervolgens koppelen aan ambitieniveau's, dan leidt dit
tot de volgende tabel:
\begin{center} \footnotesize
     \begin{tabular}{l|l|l|l|l|l}
     Vaardigheid              & Propedeuse & Bachelor & Master & Post Graduate & Rol\\
     \hline
     \hline
                              &            &          &        &               & Onderzoeker\\
     Academisch               & \B         & \U       & \R     & \K            & Docent\\
                              &            &          &        &               & Professional\\
     \hline
     \hline
                              &            &          &        &               & Onderzoeker\\
     Zelf leren               & \B         & \U       & \R     & \K            & Docent\\
                              &            &          &        &               & Professional\\
     \hline
     \hline
                              &            &          &        &               & Onderzoeker\\
     Kennis ontsluiten        & \B         & \U       & \R     & \K            & Docent\\
                              &            &          &        &               & Professional\\
     \hline
     \hline
                              &            &          &        &               & Onderzoeker\\
     Kennis inzetten          & \B         & \U       & \R     & \K            & Docent\\
                              &            &          &        &               & Professional\\
     \hline
     \hline
                              &            &          &        &               & Onderzoeker\\
     Reflectie op leren       & \B         & \U       & \R     & \K            & Docent\\
                              &            &          &        &               & Professional\\
     \hline
     \hline
                              &            &          &        &               & Onderzoeker\\
     Reflectie op handelen    & \B         & \U       & \R     & \K            & Docent\\
                              &            &          &        &               & Professional\\
     \hline
  \end{tabular}
\end{center}

\subsection{Proces van verandering \& bestendiging}
Een informatiekundige is na afloop van de studie in staat om:
\begin{description}
   \item[Probleemoplossend vermogen --] ... op verantwoorde wijze
   informatiekundige problemen op te lossen, in het bijzonder:
   \begin{itemize}
      \item ... een accurate diagnose te stellen, die te vertalen naar
      probleemstellingen, en de maatschappelijke relevantie van de onderkende
      problemen vast te stellen;

      \item ... problemen te analyseren, een synthese van oplossingsrichtingen te
      maken, en een solide oplossing te construeren;

      \item ... zich bewust te zijn van en kunnen reflecteren op het
      \emph{proces} van formuleren van probleemstellingen en het ontwikkelen van
      oplossingen, en over de rol die verschillende belanghebbenden hierin spelen;

      \item ... keuzes te maken voor geschikte onderzoeksmethoden, en op basis
      hiervan een onderzoeksplanning kunnen maken en uitvoeren;

      \item ... resultaten te verantwoorden en te presenteren;
   \end{itemize}
   \item[Defini\"eren --] ... een evenwichtig pakket van eisen op te kunnen stellen
   met betrekking tot de relaties van een informatiesysteem met haar omgeving en
   met betrekking tot de relaties tussen de systeemcomponenten onderling;
   \item[Ontwerpen --] ... een ontwerp van de essentie van een informatiesysteem te
   maken dat voldoet aan de gestelde eisen;
   \item[Constru\"eren --] ... de daadwerkelijke constructie van een
   informatiesysteem te begeleiden en te bewaken;
   \item[Invoeren --] ... te kunnen meewerken aan de invoering van een
   informatiesysteem in een gegeven context, en deze te begeleiden en bewaken;
   \item[Bestendigen --] ... mee te kunnen werken aan de bestendiging van een
   bestaand informatiesysteem, en deze te begeleiden.
   \item[Aanbesteding van verandering en bestendiging --] ... te kunnen meewerken
   aan de uitvoering, begeleiding of bewaking van de \emph{aanbesteding} van
   delen van het proces van het defini\"eren, ontwerpen, constru\"eren,
   invoeren of bestendigen van informatiesystemen;
   \item[Besturing van verandering en bestendiging --] ... voor een gegeven
   situatie een adequaat \emph{projectplan} op te stellen voor een project
   waarbinnen een proces van verandering of bestendiging van (dan wel
   aanbesteding hiervan) zal plaatsvinden, en de daadwerkelijke uitvoering van
   een dergelijk project te kunnen \emph{begeleiden};
   \item[Analyseren en modelleren --] ... in een gegeven probleemsituatie een voor
   de informatiekunde relevant domein te:
   \begin{itemize}
       \item ... \emph{analyseren};
       \item ... en de belangrijkste kenmerken van het domein met betrekking tot
       die probleemsituatie in kaart te brengen in termen van een geschikt \emph{model};
       \item ... door te \emph{abstraheren} van irrelevante details/aspecten;
       \item ... tevens dient men het resulterende model te kunnen \emph{valideren}.
   \end{itemize}
   \item[Belangen behartigen --] ... de belangen van de verschillende
   belanghebbenden te behartigen;
   \item[Onderhandelen --] ... de voor het defini\"eren noodzakelijke
   onderhandelingen met de verschillende belanghebbende partijen te voeren, te
   faciliteren en waar nodig bij te sturen;
   \item[Leven met vaagheden --] ... om te gaan met `vaagheden' en al dan niet
   schijnbare tegenstrijdigheden, en hier toch (op het juiste moment) en
   compleet en precies (formeel) pakket van eisen uit af te leiden;
   \item [Communiceren --] ... effectief en op gepaste wijze te communiceren, meer
   concreet:
   \begin{itemize}
      \item ... verschillende communicatie-rollen aan te nemen, zoals leiding
      geven aan een discussie, actief luisteren, open luisteren, van gedachten
      wisselen;
      \item ... vakinhoudelijke informatie op een heldere manier mondeling en
      schriftelijk te presenteren;
   \end{itemize}
   \item[Balans tussen product en proces --] ... een gemotiveerde afweging te
   maken tussen kwaliteit en compleetheid van de bij systeemontwikkeling op te
   leveren producten en van de voortgang en haalbaarheid van het
   daadwerkelijke ontwikkelings- en invoeringsproces.
\end{description}
Wanneer we deze vaardigheden vervolgens koppelen aan ambitieniveau's, dan leidt dit
tot de volgende tabel:
\begin{center} \footnotesize
     \begin{tabular}{l|l|l|l|l|l}
     Vaardigheid                & Propedeuse & Bachelor & Master & Post Graduate & Rol\\
     \hline
     \hline
                                &            &          &        & \K            & Onderzoeker\\
     \cline{5-6}
     Probleem oplossend         & \B         & \U       & \R     & \R            & Docent\\
     \cline{5-6}
                                &            &          &        & \K            & Professional\\
     \hline
     \hline
     Uitvoeren van              &            &          &        & \R            & Onderzoeker\\
     \cline{5-6}
     veranderen \& bestendigen\footnotemark
                                & \B         & \U       & \R     & \U            & Docent\\
     \cline{5-6}
                                &            &          &        & \K            & Professional\\
     \hline
     \hline
     Aanbesteden van            &            &          &        & \U            & Onderzoeker\\
     \cline{5-6}
     veranderen \& bestendigen  & \B         & \U       & \R     & \U            & Docent\\
     \cline{5-6}
                                &            &          &        & \K            & Professional\\
     \hline
     \hline
     Besturen van               &            &          &        & \U            & Onderzoeker\\
     \cline{5-6}
     veranderen \& bestendigen  & \B         & \U       & \R     & \U            & Docent\\
     \cline{5-6}
                                &            &          &        & \K            & Professional\\
     \hline
     \hline
                                &            &          &        & \K            & Onderzoeker\\
     \cline{5-6}
     Analyseren \& modelleren   & \B         & \U       & \R     & \R            & Docent\\
     \cline{5-6}
                                &            &          &        & \K            & Professional\\
     \hline
     \hline
                                &            &          &        & \U            & Onderzoeker\\
     \cline{5-6}
     Belangen behartigen        & \B         & \U       & \R     & \U            & Docent\\
     \cline{5-6}
                                &            &          &        & \K            & Professional\\
     \hline
     \hline
                                &            &          &        & \U            & Onderzoeker\\
     \cline{5-6}
     Onderhandelen              & \B         & \U       & \R     & \U            & Docent\\
     \cline{5-6}
                                &            &          &        & \K            & Professional\\
     \hline
     \hline
                                &            &          &        & \K            & Onderzoeker\\
     \cline{5-6}
     Leven met vaagheden        & \B         & \U       & \R     & \R            & Docent\\
     \cline{5-6}
                                &            &          &        & \K            & Professional\\
     \hline
     \hline
                                &            &          &        & \K            & Onderzoeker\\
     \cline{5-6}
     Communicatief              & \B         & \U       & \R     & \K            & Docent\\
     \cline{5-6}
                                &            &          &        & \K            & Professional\\
     \hline
     \hline
     Balans tussen              &            &          &        & \R            & Onderzoeker\\
     \cline{5-6}
     product en proces          & \B         & \U       & \R     & \U            & Docent\\
     \cline{5-6}
                                &            &          &        & \K            & Professional\\
     \hline
  \end{tabular}
\end{center}
\footnotetext{Uitvoeren van veranderen \& bestendigen is een
   samenvoeging van de voor Informatiekundigen benoemde vaardigheden:
   $\{$ defini\"eren, ontwerpen, constru\"eren, invoeren, bestendigen $\}$
}

\subsection{Subject van verandering \& bestendiging}
Een informatiekundige is na afloop van de studie in staat om:
\begin{description}
   \item[Gezichtspunten --] ... op basis van een gedegen kennis van de
   \emph{organisatorische, menswetenschappelijke, informationele,
   technologische en systemische} gezichtspunten op informatiesystemen, de
   bijbehorende \emph{theorie\"en, methoden, technieken en hulpmiddelen}:
   \begin{itemize}
      \item ... te beoordelen op hun mogelijkheden en gedrag in een concrete
      toepassingssituatie;
      \item ... deze op een adequate wijze in te zetten;
   \end{itemize}
   \item[Integrale visie --] ... vanuit de verschillende gezichtspunten een
   integrale visie op informatiesystemen te hebben, en te redeneren over de
   onderlinge impact tussen en samenhang van de verschillende gezichtspunten.
\end{description}
Wanneer we deze vaardigheden vervolgens koppelen aan ambitieniveau's, dan leidt dit
tot de volgende tabel:
\begin{center} \footnotesize
     \begin{tabular}{l|l|l|l|l|l}
     Vaardigheid                & Propedeuse & Bachelor & Master & Post Graduate & Profiel\\
     \hline
     \hline
                                &            &          &        & \K           & Onderzoeker\\
     \cline{5-6}
     Gezichtspunten             & \B         & \U       & \R     & \K           & Docent\\
     \cline{5-6}
                                &            &          &        & \K           & Professional\\
     \hline
     \hline
                                &            &          &        & \K           & Onderzoeker\\
     \cline{5-6}
     Integrale visie            & \B         & \U       & \R     & \K           & Docent\\
     \cline{5-6}
                                &            &          &        & \K           & Professional\\
     \hline
  \end{tabular}
\end{center}

\subsection{Verbredingsgebieden}
Een informatiekundige is na afloop van de studie in staat om:
\begin{description}
   \item[Inwerken in verbredingsgebieden --] ... zich in een verbredingsgebied in
   te werken teneinde in ieder geval in staat te zijn om:
   \begin{itemize}
      \item ... het gedachtegoed van dat verbredingsgebied te kunnen waarderen, en
      het betreffende domein object van analyse en modellering te maken;
      \item ... met domeinexperts te communiceren over, en zich in te leven in, de
      voor de informatiekundige essenti\"ele eigenschappen van het
      verbredingsgebied;
   \end{itemize}
   \item[Reflecteren over verbredingsgebieden --] ... te reflecteren over de
   \emph{verschillen} en \emph{overeenkomsten} tus\-sen diverse
   verbredingsgebieden.
\end{description}
Wanneer we deze vaardigheden vervolgens koppelen aan ambitieniveau's, dan leidt dit
tot de volgende tabel:
\begin{center} \footnotesize
     \begin{tabular}{l|l|l|l|l|l}
     Vaardigheid                & Propedeuse & Bachelor & Master & Post Graduate & Profiel\\
     \hline
     \hline
     Inwerken in                &            &          &        & \R           & Onderzoeker\\
     \cline{5-6}
     verbredingsgebieden        & \B         & \U       & \R     & \U           & Docent\\
     \cline{5-6}
                                &            &          &        & \K           & Professional\\
     \hline
     \hline
     Reflectie over             &            &          &        & \K           & Onderzoeker\\
     \cline{5-6}
     verbredingsgebieden        & \B         & \U       & \R     & \U           & Docent\\
     \cline{5-6}
                                &            &          &        & \K           & Professional\\
     \hline
  \end{tabular}
\end{center}

\subsection{Onderzoek \& reflectie}
Een informatiekundige is na afloop van de studie in staat om:
\begin{description}
   \item[Onderzoeksvragen --] ... voor de maatschappelijke omgeving relevante
   onderzoeksvragen te kunnen formuleren met betrekking tot het
   informatiekundige vakgebied;
   \item[Besturen van onderzoek --] ... een voor een gegeven onderzoeksvraag
   passende onderzoeksaanpak te formuleren in termen van een projectplan, en de
   uitvoering van dit onderzoek te begeleiden;
   \item[Uitvoeren van onderzoek --] ... conform een opgestelde onderzoeksaanvraag
   onderzoek uit te voeren naar een voor de informatiekunde relevante
   onderzoeksvraag.
\end{description}
Wanneer we deze vaardigheden vervolgens koppelen aan ambitieniveau's, dan leidt dit
tot de volgende tabel:
\begin{center} \footnotesize
     \begin{tabular}{l|l|l|l|l|l}
     Vaardigheid                & Propedeuse & Bachelor & Master & Post Graduate & Profiel\\
     \hline
     \hline
                                &            &          &        & \K            & Onderzoeker\\
     \cline{5-6}
     Onderzoeksvraag            & \B         & \U       & \R     & \U            & Docent\\
     \cline{5-6}
                                &            &          &        & \R            & Professional\\
     \hline
     \hline
                                &            &          &        & \K            & Onderzoeker\\
     \cline{5-6}
     Besturen van onderzoek     & \B         & \U       & \R     & \U            & Docent\\
     \cline{5-6}
                                &            &          &        & \R            & Professional\\
     \hline
     \hline
                                &            &          &        & \K            & Onderzoeker\\
     \cline{5-6}
     Uitvoeren van onderzoek    & \B         & \U       & \R     & \U            & Docent\\
     \cline{5-6}
                                &            &          &        & \R            & Professional\\
     \hline
  \end{tabular}
\end{center}
}
\chapter{Inhoudelijke structuur van de opleiding}
\label{p:hoe-structuur}

Gegeven de doelstellingen en eindtermen uit
hoofdstuk~\ref{h:wat-doelstellingen}, is het doel van deze sectie een
inhoudelijke structuur van de opleiding aan te dragen die een helder verband
heeft met de benoemde doelstellingen en eindtermen.  Op basis van deze
inhoudelijke structuur zal in de rest van dit hoofdstuk de programmatische
structuur van de opleiding worden opgebouwd.

We zullen de inhoudelijke structuur in drie stappen opstellen:
\begin{itemize}
   \item allereerst bespreken we de concepten die we zullen gebruiken bij het opstellen
   van de inhoudelijke structuur (paragraaf~\ref{p:model});
   \item vervolgens benoemen we de belangrijkste inhoudelijke aandachtsgebieden van
   de opleiding (paragraaf~\ref{p:aspecten});
   \item tenslotte zullen we per aandachtsgebied een nadere inhoudelijke invulling
   geven middels het geven van specifieke thema's (paragraaf~\ref{p:invulling});
\end{itemize}

\section{Inrichting van de opleiding}
\label{p:model}
Er is voor gekozen om de opleiding in te delen op basis van vier lagen van
`inhoudelijke componenten'. Het gaat hierbij om een logische indeling en nog
geen fysieke indeling. Met andere woorden, de benoemde inhoudelijke componenten
kunnen in diverse vakken in het curriculum terugkomen, en andersom kan
een vak een bijdrage leveren aan meerdere inhoudelijke componenten. We hebben
hierbij dus te maken met een many-to-many relatie.

De lagen die we zullen onderscheiden zijn:
\begin{enumerate}
   \item \emph{Opleiding:} de opleiding als geheel, i.c.\ de
   Informatiekunde opleiding.
   \item \emph{Aspecten:} de belangrijkste aspecten binnen de opleiding.
   Voor de Informatiekunde opleiding is dit bijvoorbeeld: `proces van
   verandering \& bestendiging'.
   \item \emph{Themas:} de specifieke thema's die binnen een aspect aan bod
   zullen komen. Voor het aspect `proces van verandering \& bestendiging' zou
   dit kunnen zijn: `methoden, technieken \& theorie\"en voor uitvoering'
   \item \emph{Trefwoorden:} een nadere concretisering van de thema's middels
   trefwoorden.
\end{enumerate}
Bij de vertaling naar het daadwerkelijke programma, zoals dit in het volgende
hoofdstuk zal gebeuren, zal blijken dat de thema's op het niveau van vakken zitten,
en de trefwoorden op het niveau van individuele college's. Merk echter wederom
op dat er sprake is van een many-to-many relatie.  De indeling van opleiding,
via aspecten en thema's naar trefwoorden is in principe een boomstructuur. Op
trefwoord niveau kan er echter wel enige overlap ontstaan tussen de
verschillende thema's.

\section{Portfolio van aspecten}
\label{p:aspecten}
Bij het benoemen van de belangrijkste aspecten binnen de opleiding laten we ons
leiden door de clustering van vaardigheden zoals deze uit de visie op het
vakgebied naar voren zijn gekomen~\cite{Visie2003}, en zoals die ook zijn
gebruikt bij het benoemen van de eindtermen van de opleiding als geheel in
hoofdstuk~\ref{h:wat-doelstellingen}:
\begin{itemize}
   \item Algemeen
   \item Proces van verandering \& bestendiging
   \item Subject van verandering \& bestendiging
   \item Verbreding
   \item Onderzoek \& reflectie
\end{itemize}
Het gebruik van deze clustering is een weloverwogen keuze. Het
beoogde effect is hierbij juist ook dat studenten een duidelijk verband gaan
zien tussen de inhoudelijke aspecten van de opleiding, en uiteindelijk de
vakken, en het geschetste beeld van de rol die zij na afloop van de studie
kunnen vervullen. Met andere woorden, we hopen dat het hiermee voor de
studenten ook vooraf duidelijk wordt wat de bijdrage van een bepaald vak aan de
opleiding is.

De algemene eindtermen zullen gedistribueerd ingevuld worden, met name door
onderzoek \& reflectie. Daarnaast ontbreekt in dit lijstje de grondslagen van
het vakgebied. Er zal binnen de opleiding expliciet aandacht besteed worden aan
thema's waarvan we vinden dat deze tot de grondslagen van het vakgebied horen.
Met deze wijzigingen hebben we het volgende portfolio van aspecten:
\begin{itemize}
   \item Proces van verandering \& bestendiging
   \item Subject van verandering \& bestendiging
   \item Verbreding
   \item Grondslagen
   \item Onderzoek \& reflectie
\end{itemize}

\section{Eindtermen per aspect}
Voor de aspecten:
\begin{itemize}
   \item Proces van verandering \& bestendiging
   \item Subject van verandering \& bestendiging
   \item Verbreding
   \item Onderzoek \& reflectie
\end{itemize}
komen de eindtermen overeen met de eindtermen zoals deze zijn aangegeven
in hoofdstuk~\ref{h:wat-doelstellingen}. Het enige aspect waar de eindtermen
nog nader van gedefinieerd moeten worden zijn de grondslagen.
Tot de grondslagen van de informatiekunde rekenen we de volgende gebieden:
\begin{itemize}
  \item Systeemtheorie
  \item Informatietheorie
  \item Communicatietheorie
\end{itemize}
Met betrekking tot de grondslagen van de informatiekunde, is een
informatiekundige na afloop van de studie in staat om:
\begin{description}
   \item[Redeneren over grondslagen --] ... te redeneren over de grondslagen
   van de informatiekunde;
   \item[Grondslagen toepassen --] ... in staat om de bij de grondslagen
   behorende methoden, technieken en theorie\"en toe te passen in concrete
   situaties;
\end{description}
\def\B{Belangstellend} \def\U{Uitvoerend} \def\R{Richtinggevend} \def\K{Kaderzettend}
Wanneer we deze vaardigheden vervolgens koppelen aan ambitieniveau's, dan leidt
dit tot de volgende tabel:
\begin{center} \footnotesize
     \begin{tabular}{l|l|l|l|l|l}
     Vaardigheid                & Propedeuse & Bachelor & Master & Post Graduate & Profiel\\
     \hline
     \hline
                                &            &          &        & \K           & Onderzoeker\\
     \cline{5-6}
     Redeneren over grondslagen & \B         & \U       & \R     & \K           & Docent\\
     \cline{5-6}
                                &            &          &        & \K           & Professional\\
     \hline
     \hline
                                &            &          &        & \K           & Onderzoeker\\
     \cline{5-6}
     Grondslagen toepassen      & \B         & \U       & \R     & \K           & Docent\\
     \cline{5-6}
                                &            &          &        & \K           & Professional\\
     \hline
  \end{tabular}
\end{center}

\section{Inhoudelijke invulling per aspect}
\label{p:invulling} Hieronder is per aspect een nadere invulling
gegegeven. Hierbij is er voor gekozen om de thema's te benoemen
die binnen een aspect aan de orde zullen komen, en tevens per
thema een lijstje trefwoorden om de thema's nader te
concretiseren. De enige uitzondering hierop vormt het aspect
verbreding.
\begin{description}
 \item[Proces van verandering \& bestendiging] ~~
 \begin{itemize}
   \item Methoden, technieken en theori\"en voor uitvoering:

   Definitie, Ontwerp, Constructie, Invoering, Bestendiging

   \item Methoden, technieken en theori\"en voor besturing \& aanbesteding:

   ICT Governance, Projectmanagement, Programmamanagement,
   Kwaliteitsmanagement, Aan\-be\-ste\-ding

   \item Integreren \& ervaren:

   Ge\"{\i}ntegreerd practicum

   \item Algemene vaardigheden

   Communiceren, Presenteren, Onderhandelen, Veranderprocessen
 \end{itemize}

 \item[Subject van verandering \& bestendiging] ~~
 \begin{itemize}
   \item Organisaties

   Organisatieleer, Business modellen, Bedrijfsstrategie

   \item Mens

   Menselijke maat, ethiek, ergonomie

   \item Informatie

   Communicatie \& Informatietheorie,
   Modelleren van informatie, Modelleren van kennis

   \item Technologie

   Software-architectuur, Middleware, Besturingssystemen, Applicatie-integratie,
   Computer architecturen, Security, Telematica

   \item Informatiesysteem

   Informatie-architectuur, Gegevensbeheersystemen, Informatiesystemen, Kennissystemen,
   Communicatiesystemen
 \end{itemize}

 \item[Verbreding] ~~

 Nader te benoemen vakken uit \'e\'en of twee verbredingsgebieden:
 \begin{itemize}
   \item Medische informatiekunde
   \item Taal \& spraaktechnologie
   \item Beslissingsondersteuning
   \item Kennis- \& informatiemanagement
   \item Juridisch kennisbeheer
   \item Recht \& Informatica
   \item Bio-informatica
   \item ...
 \end{itemize}
 Merk op: zoals ook al in~\cite{Visie2003} is aangegeven kunnen we een
 onderscheid maken tussen verbredingssgebieden in de zin van:
 \begin{description}
   \item[Informatiekunde toegepast op ...] waarbij de informatiekunde
   wordt toegepast op een een bepaald toepassingsgebied. Bijvoorbeeld:
   \begin{description}
      \item[Medische informatiekunde] richt zich op het toepassing van ICT in
      een medische setting.
      \item[Beslissingsondersteuning] betreft de inzet van ICT ter ondersteuning van
      besluitvormingsprocessen; meestal in de context van bedrijfsprocessen.
      \item[Rechtsinformatica] betreft het gebruik van ICT methoden, technieken en
      hulpmiddelen, zoals ``Kunstmatige Intelligentie'', ten behoeve van de juridische
      praktijk en theorie.
   \end{description}
   \item[Informatiekunde toepassen onder ....] waarbij de informatiekunde
   bekeken wordt vanuit een bepaald ander domein. Bijvoorbeeld:
   \begin{description}
      \item[Informaticarecht] betreft de juridische aspecten van ICT.
      Informaticarecht betreft de juridische aspecten van informatietechnologie.
      \item[Kennis- \& informatiemanagement] richt zich op de vraag hoe organisaties
      het beste kunnen omgaan met het besturen van hun kennis- en informatiestromen.
      Een belangrijk aspect hierbij is uiteraard de besturing (Governance) van de inrichting
      van de ICT middelen.
   \end{description}
 \end{description}
 \item[Grondslagen] ~~
 \begin{itemize}
   \item Formele methoden

         Formeel denken, Beweren \& bewijzen, Statistiek, Kansberekening,
         Besliskunde

   \item Systeemdenken

         Systeemdynamica, Systeemtheorie, Patterns, Co\"ordinatie
 \end{itemize}

 \item[Onderzoek \& reflectie] ~~
 \begin{itemize}
  \item Onderzoeksvaardigheden
  \item Inhoudelijke integratie \& reflectie
  \item Scripties
 \end{itemize}
\end{description}

\Ragged{\chapter{Bachelor Informatiekunde 2003}
\label{h:BachelorProgramma}
\def\Ja{$\times$}

Hieronder staat het studieprogramma voor de Bachelor Informatiekunde
zoals dat vanaf cohort 2003 zal gaan gelden.
Enkele opmerkingen vooraf:
\begin{itemize}
   \item De {\it cursiefgedrukte} vakken zullen zoveel mogelijk met
         Informatica samen gegeven worden.
   \item De {\bf vetgedrukte} vakken kunnen, wegens verschuivingen bij
         externe faculteiten, van plaats veranderen in het daadwerkelijke
         programma.
   \item De ``Verbredingsvakken'' dienen ingevuld te worden met vakken uit
         \'e\'en of hooguit twee Verbredingsgebieden van de informatiekunde.
         Deze zullen per jaar nader bepaald worden, mede in samenwerking met
         externe faculteiten.
   \item Vakken met een ``\Ja'' in de kolom ``Wisselvak'', zijn door de
         studenten in te wisselen voor andere vakken.
         Dit biedt de studenten de ruimte voor persoonlijke profilering binnen
         de voor Informatica en Informatiekunde relevante vakgebieden.
         Dit inwisselen is aan regels gebonden.
         Deze regels worden hieronder nader besproken.
   \item Voor elke Bachelor opleiding geldt dat er 6 ECTS volledig vrije
         keuze dient te zijn.
         Informatiekunde studenten mogen daarom \'e\'en van de
         ``Verbredingsvakken'' vervangen door een willekeurig ander vak
         van een willekeurige andere Bachelor opleiding van de Universiteit.
\end{itemize}

\begin{center} \small
 \begin{tabular}{|llrr|}
   \hline
   Sem. & Vak & ECTS & Wisselvak\\
   \hline
   \hline
   1.1      & {\it Introductie Informatica \& Informatiekunde}      &  3 & \\
            & {\it Domeinmodellering}                               &  6 & \\
            & {\bf Cognitie}                                        &  6 & \\
            & Formeel Denken                                        &  6 & \\
            & Algoritmiek                                           &  6 & \\
            & Orientatiecollege Toepassingsgebieden                 &  3 & \\
   \hline
            &                                                       & 30 & \\
   \hline
   \hline
   1.2      & {\bf Organisatiekunde}                                &  6 & \\
            & {\it Beweren \& Bewijzen}                             &  6 & \\
            & Opslaan \& Terugvinden                                &  6 & \\
            & Datastructuren                                        &  6 & \\
            & R\&D 1                                                &  6 & \\
   \hline
            &                                                       & 30 & \\
   \hline
   \hline
   2.1      & Verbreding                                            &  6 & \\
            & Kansrekening                                          &  3 & \\
            & {\it Informatiesystemen}                              &  6 & \\
            & Requirements Engineering                              &  6 & \\
            & ICT Infrastructuren                                   &  3 & \\
            & Gedistribueerde software systemen                     &  6 & \\
   \hline
            &                                                       & 30 & \\
   \hline
   \hline
   2.2      & Verbreding                                            &  6 & \\
            & {\bf Soft-Systems Methodology}                        &  6 & \Ja \\
            & Onderhandelen \& veranderen                           &  6 & \\
            & Integratie van Software Systemen                      &  6 & \Ja \\
            & R\&D 2                                                &  6 & \\
   \hline
            &                                                       & 30 & \\
   \hline
   \hline
   3.1      & Verbreding                                            &  6 & \\
            & Statistiek                                            &  3 & \\
            & {\it Software Engineering 1}                          &  3 & \\
            & {\it Intelligente Systemen}                           &  6 & \\
            & Introductie CEM                                       &  6 & \\
            & Systeemtheorie: Ontwerp \& Evolutie                   &  3 & \\
            & {\it Security}                                        &  3 & \Ja \\
   \hline
            &                                                       & 30 & \\
   \hline
   \hline
   3.2      & Verbreding                                            &  6 & \\
            & {\it Software Engineering 2}                          &  6 & \Ja \\
            & {\it Lerende \& Redenerende Systemen}                 &  6 & \Ja \\
            & {\it Mens-Machine Interactie}                         &  3 & \Ja \\
            & {\it ICT \& Samenleving 1}                            &  3 & \\
            & Scriptie                                              &  6 & \\
   \hline
            &                                                       & 30 & \\
   \hline
   \hline
            & TOTAAL Bachelor                                       &180 & \\
   \hline
 \end{tabular}
\end{center}

Diverse vakken uit het bovenstaande programma zijn nieuwe vakken.
In appendix~\ref{h:Vakken} zijn de vakbeschrijvingen te vinden van alle
vakken van het informatiekunde curriculum 2003.
Om inzicht te krijgen in het tempo waarin de eventuele nieuwe vakken ontwikkeld
dienen te worden, staat hieronder een tabel met het programma, gecombineerd
met informatie met betrekking tot de invoering van het vak.
Hierbij is gebruik gemaakt van de de volgende conventie:
\begin{itemize}
   \item De status kolom geeft aan of het een bestaand vak is (B),
         een herschikking van bestaande vakken van voor 2003 is (V),
         of het (vanaf 2003 gemeten) een geheel nieuw vak is (N).
\end{itemize}

\begin{center} \small
 \begin{tabular}{|llrllr|}
   \hline
   Sem. & Vak & ECTS & Status & Afdeling & Invoering\\
   \hline
   \hline
   1.1      & {\it Introductie Informatica \& Informatiekunde}    &  3 & B & \NIII    &     -\\
            & {\bf Organisatiekunde}                              &  6 & B & Extern   &     -\\
            & {\it Domeinmodellering}                             &  6 & V & \AfdIRIS & 03/04\\
            & Formeel Denken                                      &  6 & B & \AfdGS   &     -\\
            & Algoritmiek                                         &  6 & B & \AfdST   &     -\\
            & Orientatiecollege  Toepassingen                     &  3 & B & \NIII    &     -\\
   \hline
            &                                                     & 30 &   &          &      \\
   \hline
   \hline
   1.2      & Cognitie                                            &  6 & B & Extern   &     -\\
            & {\it Beweren \& Bewijzen}                           &  6 & B & \AfdITT  &     -\\
            & Opslaan \& Terugvinden                              &  6 & V & \AfdIRIS & 03/04\\
            & Datastructuren                                      &  6 & B & \AfdST   &     -\\
            & R\&D 1                                              &  6 & N & \NIII    & 03/04\\
   \hline
            &                                                     & 30 &   &          &      \\
   \hline
   \hline
   2.1      & Verbreding                                          &  6 & B & -        &     -\\
            & Kansrekening                                        &  3 & B & Extern   &     -\\
            & {\it Informatiesystemen}                            &  6 & V & \AfdIRIS & 04/05\\
            & Requirements Engineering                            &  6 & N & \AfdIRIS & 04/05\\
            & ICT Infrastructuren                                 &  3 & N & \AfdITT  & 04/05\\
            & Gedistribueerde software systemen                   &  6 & N & \AfdST   & 04/05\\
   \hline
            &                                                     & 30 &   &          &      \\
   \hline
   \hline
   2.2      & Verbreding                                          &  6 & B & -        &     -\\
            & {\bf Soft-Systems Methodology}                      &  6 & B & Extern   &     -\\
            & Onderhandelen \& veranderen                         &  6 & N & \AfdIRIS & 03/04\\
            & Integratie van Software Systemen                    &  6 & N & \AfdST   & 04/05\\
            & R\&D 2                                              &  6 & N & \NIII    & 04/05\\
   \hline
            &                                                     & 30 &   &          &      \\
   \hline
   \hline
   3.1      & Verbreding                                          &  6 & E & -        &     -\\
            & Statistiek                                          &  3 & B & Extern   &     -\\
            & {\it Software Engineering 1}                        &  3 & B & \NIII    &     -\\
            & {\it Intelligente Systemen}                         &  6 & V & \AfdIRIS & 03/04\\
            & Introductie CEM                                     &  6 & E & \FNWI     &    -\\
            & Systeemtheorie: Ontwerp \& Evolutie                 &  3 & V & \AfdIRIS & 05/06\\
            & {\it Security}                                      &  3 & B & \AfdITT  &     -\\
   \hline
            &                                                     & 30 &   &          &      \\
   \hline
   \hline
   3.2      & Verbreding                                          &  6 & E & -        &     -\\
            & {\it Software Engineering 2}                        &  6 & B & \NIII    &     -\\
            & {\it Lerende \& Redenerende Systemen}               &  6 & N & \AfdIRIS & 05/06\\
            & {\it Mens-Machine Interactie}                       &  3 & V & \AfdIRIS & 03/04\\
            & {\it ICT \& Samenleving 1}                          &  3 & B & \FNWI    &     -\\
            & Scriptie                                            &  6 & B & -        &     -\\
   \hline
            &                                                     & 30 &   &          &      \\
   \hline
   \hline
            & TOTAAL Bachelor                                     &180 &   &          &      \\
   \hline
\end{tabular}
\end{center}

Merk op: de vakken ``Intelligente Systemen'', ``Mens-Machine
Interactie'' en ``Onderhandelen \& Veranderen'' eerder gegeven
worden dan voor de cohort 2003 noodzakelijk is. Dit is nodig in
verband met de migratie van de lopende curricula. Waar mogelijk
zullen deze vakken als 3 ECTS ``voorloper'' vak gegeven worden, om
de extra onderwijslast in verband met de nieuwe vakken te
spreiden.

Voor het Informatiekunde programma is het belangrijk om een goede belans te
bewaken tussen de verschillende aspecten \& thema's van het vakgebied zoals die
zijn benoemd in paragraaf~\ref{p:hoe-structuur}.  In de onderstaande tabel, worden
de vakken uit het programma gepositioneerd ten opzichte van deze opdeling.
Merk op dat, zoals reeds eerder gemeld, \'e\'en vak een bijdrage kan leveren
aan meerdere thema's binnen de opleiding.

{\scriptsize \begin{tabular}{|l||r||r|r|r|r||r|r|r|r|r||r||r|r||r|r||l}
   \cline{2-17}
   \multicolumn{1}{l||}{}&\multicolumn{16}{|l|}{Totaal}\\
   \cline{3-17}
   \multicolumn{1}{l||}{}&&\multicolumn{15}{|l|}{MTT voor besturing \& aanbesteding}\\
   \cline{4-17}
   \multicolumn{1}{l||}{}&&&\multicolumn{14}{|l|}{MTT voor uitvoering}\\
   \cline{5-17}
   \multicolumn{1}{l||}{}&&&&\multicolumn{13}{|l|}{Integreren \& ervaren}\\
   \cline{6-17}
   \multicolumn{1}{l||}{}&&&&&\multicolumn{12}{|l|}{Algemene vaardigheden}\\
   \cline{7-17}
   \multicolumn{1}{l||}{}&&&&&&\multicolumn{11}{|l|}{Organisaties}\\
   \cline{8-17}
   \multicolumn{1}{l||}{}&&&&&&&\multicolumn{10}{|l|}{Mens}\\
   \cline{9-17}
   \multicolumn{1}{l||}{}&&&&&&&&\multicolumn{ 9}{|l|}{Informatie }\\
   \cline{10-17}
   \multicolumn{1}{l||}{}&&&&&&&&&\multicolumn{ 8}{|l|}{Technologie}\\
   \cline{11-17}
   \multicolumn{1}{l||}{}&&&&&&&&&&\multicolumn{ 7}{|l|}{Informatiesystemen}\\
   \cline{12-17}
   \multicolumn{1}{l||}{}&&&&&&&&&&&\multicolumn{ 6}{|l|}{Verbreding}\\
   \cline{13-17}
   \multicolumn{1}{l||}{}&&&&&&&&&&&&\multicolumn{ 5}{|l|}{Formele methoden}\\
   \cline{14-17}
   \multicolumn{1}{l||}{}&&&&&&&&&&&&&\multicolumn{ 4}{|l|}{Systeemdenken}\\
   \cline{15-17}
   \multicolumn{1}{l||}{}&&&&&&&&&&&&&&\multicolumn{ 3}{|l|}{R\&D Lab}\\
   \cline{16-17}
   \multicolumn{1}{l||}{}&&&&&&&&&&&&&&&\multicolumn{ 2}{|l|}{Afst.}\\
   \cline{1-17}
   Organisatiekunde                  & 6 &   &   &   &   & 6 &   &   &   &   &   &   &   &   &  \\
   Domeinmodellering                 & 6 &   &   &   &   & 2 &   & 1 & 1 &   &   &   & 2 &   &  \\
   Formeel Denken                    & 6 &   &   &   &   &   &   &   &   &   &   & 6 &   &   &  \\
   Algoritmiek                       & 6 &   & 3 &   &   &   &   &   & 3 &   &   &   &   &   &  \\
   Orientatiecollege Toepassingen    & 3 &   &   &   &   &   &   &   &   &   & 3 &   &   &   &  \\
   Introductie I \& I                & 3 &   &   & 3 &   &   &   &   &   &   &   &   &   &   &  \\
   \cline{1-16}
   Cognitie                          & 6 &   &   &   &   &   & 6 &   &   &   &   &   &   &   &  \\
   Beweren \& Bewijzen               & 6 &   &   &   &   &   &   &   &   &   &   & 6 &   &   &  \\
   Opslaan \& Terugvinden            & 6 &   &   &   &   &   &   & 2 &   & 4 &   &   &   &   &  \\
   Datastructuren                    & 6 &   & 3 &   &   &   &   &   & 3 &   &   &   &   &   &  \\
   R\&D 1                            & 6 &   &   &   &   &   &   &   &   &   &   &   &   & 6 &  \\
   \cline{1-16}
   Verbreding                        & 6 &   &   &   &   &   &   &   &   &   & 6 &   &   &   &  \\
   Kansrekening                      & 3 &   &   &   &   &   &   &   &   &   &   & 3 &   &   &  \\
   Informatiesystemen                & 6 &   &   &   &   &   &   & 3 &   & 3 &   &   &   &   &  \\
   Requirements Engineering          & 6 &   & 6 &   &   &   &   &   &   &   &   &   &   &   &  \\
   ICT Infrastructuren               & 3 &   &   &   &   &   &   &   & 3 &   &   &   &   &   &  \\
   Gedistribueerde software systemen & 6 &   &   &   &   &   &   &   & 6 &   &   &   &   &   &  \\
   \cline{1-16}
   Verbreding                        & 6 &   &   &   &   &   &   &   &   &   & 6 &   &   &   &  \\
   Soft-Systems Methodology          & 6 &   & 2 &   &   & 2 &   &   &   &   &   &   & 2 &   &  \\
   Onderhandelen \& veranderen       & 6 & 2 & 2 &   &   &   &   &   &   &   &   &   & 2 &   &  \\
   Integratie van softwaresystemen   & 6 &   &   &   &   &   &   &   & 6 &   &   &   &   &   &  \\
   R\&D 2                            & 6 &   &   &   &   &   &   &   &   &   &   &   &   & 6 &  \\
   \cline{1-16}
   Verbreding                        & 6 &   &   &   &   &   &   &   &   &   & 6 &   &   &   &  \\
   Statistiek                        & 3 &   &   &   &   &   &   &   &   &   &   & 3 &   &   &  \\
   Software Engineering 1            & 3 &   & 1 & 2 &   &   &   &   &   &   &   &   &   &   &  \\
   Intelligente Systemen             & 6 &   &   &   &   &   &   & 2 &   & 4 &   &   &   &   &  \\
   Introductie CEM                   & 6 &   &   &   & 6 &   &   &   &   &   &   &   &   &   &  \\
   Systeemtheorie: Ontwerp \& Evol.  & 3 &   &   &   &   &   &   &   &   &   &   &   & 3 &   &  \\
   Security                          & 3 &   &   &   &   &   &   &   & 3 &   &   &   &   &   &  \\
   \cline{1-16}
   Verbreding                        & 6 &   &   &   &   &   &   &   &   &   & 6 &   &   &   &  \\
   Software Engineering 2            & 6 &   & 3 & 3 &   &   &   &   &   &   &   &   &   &   &  \\
   Lerende \& Redenerende Systemen   & 6 &   &   &   &   &   &   & 2 &   & 4 &   &   &   &   &  \\
   Mens-Machine Interactie           & 3 &   &   &   &   &   & 1 &   & 2 &   &   &   &   &   &  \\
   ICT \& Samenleving 1              & 3 &   &   &   &   &   & 3 &   &   &   &   &   &   &   &  \\
   Scriptie                          & 6 &   &   &   &   &   &   &   &   &   &   &   &   &   & 6\\
   \cline{1-16}
   Totaal Bachelor                   &180& 2 & 20&  8&  6& 10& 10& 10& 27& 15& 27& 18& 9 & 12& 6\\
   \cline{2-16}
                                     &180&\multicolumn{4}{c||}{36}&\multicolumn{5}{c||}{72}&27&\multicolumn{2}{c||}{27}&\multicolumn{2}{c||}{18}\\
   \cline{1-17}
   \multicolumn{1}{l||}{}&&\multicolumn{4}{c||}{}&\multicolumn{5}{c||}{}&&\multicolumn{2}{c||}{}&\multicolumn{3}{|l|}{Onderzk \& Refl}\\
   \cline{15-17}
   \multicolumn{1}{l||}{}&&\multicolumn{4}{c||}{}&\multicolumn{5}{c||}{}&&\multicolumn{5}{|l|}{Grondslagen}\\
   \cline{13-17}
   \multicolumn{1}{l||}{}&&\multicolumn{4}{c||}{}&\multicolumn{5}{c||}{}&\multicolumn{6}{|l|}{Verbreding}\\
   \cline{12-17}
   \multicolumn{1}{l||}{}&&\multicolumn{4}{c||}{}&\multicolumn{11}{l|}{Subject van verandering \& bestendiging}\\
   \cline{7-17}
   \multicolumn{1}{l||}{}&&\multicolumn{15}{|l|}{Proces van verandering \& bestendiging}\\
   \cline{3-17}
   \multicolumn{1}{l||}{}&\multicolumn{16}{|l|}{Totaal}\\
   \cline{2-17}
\end{tabular}}

De als wisselvak aangemerkte vakken zijn te vervangen door andere
vakken uit lijst van vooraf bepaalde vakken.
Formeel is een ``wisselvak'' gedefinieerd als een keuzevak met een bijbehorend
lijstje van benoemde alternatieven.
De intentie van het wisselvakkensysteem kan als volgt worden geformuleerd:
\begin{itemize}
  \item Er zijn aan het \NIII\ twee standaard studieprogrammas. E\'en voor
  informatiekunde en \'e\'en voor informatica.
  \item Deze programma's kunnen gezien worden als twee ``extremen'' die
  het spectrum van het \NIII onderwijs opspannen.
  \item Het wisselvakkensysteem is er primair voor bedoeld om studenten in
  staat te stellen om zich van de ene extreem in de richting van de andere
  extreem te bewegen.
  \item Het is dus in principe niet de bedoeling dat studenten zich buiten
  het opgespannen spectrum begeven. De enige uitzondering daarop vormen
  vakken die specifiek nodig zijn voor eventuele externe specialisaties van
  de masterfase. Denk bijvoorbeeld aan een masterspecialisatie
  Medische Informatiekunde.
\end{itemize}
Het NIII bepaald jaarlijks een lijst van alternatieven per
wisselvak. Daarnaast mag 6 ECTS van de 30 ECTS aan wisselvakruimte
overgehevelt worden naar de ruimte voor verbredingsvakken.

Voor het collegejaar 2003 is er geen specifieke verdeling gemaakt
per wisselvak. De lijst van alternatieven voor het collegejaar
2003 is:
\begin{itemize}
  \item Semantiek \& logica
  \item Geheugen, distribute \& netwerken
  \item Abstractie \& compositie
  \item Analyse van algoritmen
  \item Vertalerbouw
\end{itemize}
Het zal duidelijk zijn dat elk alternatief slechts \'e\'en maal
gebruikt mag worden.
}
\Ragged{\chapter{Master Informatiekunde 2003 voor Academische Bachelors}
\label{h:VWOMasterProgramma}

Studenten die een master informatiekunde doen, zullen zich nader willen
specialiseren op deelgebieden van het informatiekunde vakgebied, of ten aanzien
van de precieze rol die men na afloop van de studie ambi\"eerd.  Dit betekent
ook dat keuzevrijheid een belangrijk ontwerpcriterium is voor het programma van
de master. Tegelijkertijd is het noodzakelijk om een balans tussen de
verschillende aspecten van het vakgebied te bewaren. Daarnaast is het niet de
bedoeling om per definitie een brede masteropleiding op te zetten, maar de
\NIII\ Informatiekunde master juist een helder profiel mee te geven op basis
van de sterke punten van het \NIII\ binnen het vakgebied:
informatiearchitectuur. Het idee is dan ook om ons vooralsnog te profileren met
\'e\'en specialisatie voor de master: ``Informatiearchitectuur'', maar de
studenten wel veel vrijheid te bieden ten aanzien van het aanbrengen van
variaties.

\section{Basisstramien}
Om voldoende vrijheid te bieden bij het inrichten van eventuele andere
specialisaties, en om studenten ook een eigen keuze te laten maken, is voor het
volgende basisstramien gekozen:
\begin{itemize}
   \item Specialisatievak gericht op de rol die de student na
   de studie wil vervullen: 6 ECTS

   Afhankelijk van de voorkeur: architect, manager of onderzoeker,
   dient de student te kiezen voor:
   \begin{itemize}
     \item Software Engineering 3 (architect of manager)
     \item Onderzoekslaboratorium (onderzoeker)
   \end{itemize}

   \item Inhoudelijke specialisatie: 21 ECTS

         Deze dient als volgt verdeeld te worden:
         \begin{itemize}
            \item Grondslagen: 3 ECTS
            \item Proces van verandering \& bestendiging: 6 ECTS
            \item Subject van verandering \& bestendiging: 12 ECTS

                  Hierbij dient een duidelijke balans te zijn tussen de
                  menselijke, organisatorische, informationele, technologische
                  en systemische aspecten.
         \end{itemize}

   \item ICT \& Samenleving 2: 3 ECTS

   \item Vrije keuze: 12 ECTS

   \item Scriptie: 18 ECTS (eventueel kan een deel van de vrije keuzeruimte
         gebruikt worden om een extra grote scriptie te schrijven).
\end{itemize}
Binnen deze parameters staat het studenten vrij een eigen wending aan hun
master's programma te geven.  Echter, het uiteindelijke programma wat een
student volgt moet uiteraard ter toetsing voorgelegd worden aan de
examencommissie.

Het bovenstaande stramien van de master leidt tot het volgende
programma-template:
\begin{center} \small
 \begin{tabular}{|llr|}
   \hline
   Sem. & Vak & ECTS\\
   \hline
   \hline
   4.1      & Software Engineering 3 {\it of}\ Onderzoekslaboratorium      &  6 \\
            & Vak 1                                                        &  6 \\
            & Vak 2                                                        &  6 \\
            & Vak 3                                                        &  6 \\
            & Vak 4                                                        &  6 \\
   \hline
            &                                                              & 30 \\
   \hline
   \hline
   4.2      & Vak 4                                                        &  6 \\
            & Vak 5                                                        &  3 \\
            & ICT \& Samenleving 2                                         &  3 \\
            & Scriptie                                                     & 18 \\
   \hline
            &                                                              & 30 \\
   \hline
   \hline
 \end{tabular}
\end{center}
Er zal een lijst van keuzevakken voor de master komen. Deze vakken
zullen niet alleen uit de Informatica en Informatiekunde
voortkomen, maar ook uit de verbredingsrichtingen, de
menswetenschappen en de organisatiewetenschappen. Daarnaast zullen
er een aantal specialisatierichtingen voor de master worden
gedefinieerd, die studenten als basis kunnen gebruiken.

\section{Toegangseisen}
Als toegangseis voor de \'e\'en-jarige master geldt dat studenten
een relevante academische bacheloropleiding hebben afgerond. Dit
zal doorgaans een informatica, informatiekunde, of
bedrijfsinformatiekunde bacheloropleiding zijn. Hierbij is het
belangrijk dat de studenten een redelijke balans aan vakken hebben
gevolgd wat betreft de benoemde aspecten van het informatiekunde
vakgebied.

De uiteindelijke toelating van een student tot de Master-fase zal
worden beoordeeld door een toelatingscommissie.
De toelatischecommissie zal hierbij mede gebruik maken van:
\begin{itemize}
  \item Algemene instroomeisen voor de master (zie hieronder).
  \item Balans in de aspecten van het vakgebied.
  \item Specifieke instroomeisen zoals die voortkomen uit
  de instroomeisen van de vakken uit de gekozen specialisatie.
\end{itemize}

Middels de vaardigheden en eindtermen uit
hoofdstuk~\ref{h:wat-doelstellingen} kunnen we de algemene instroomeisen
nader concretiseren. In hoofdstuk~\ref{h:wat-doelstellingen} waren
de vaardigheden en eindtermen opgesplitst in de volgende aspecten:
\begin{enumerate}
   \item Algemeen
   \item Proces van verandering \& bestendiging
   \item Subject van verandering \& bestendiging
   \item Verbredingsgebieden
   \item Onderzoek \& reflectie
\end{enumerate}
Het is zeker niet zo dat de eindtermen van de bachelorfase
onverkort als ingangseisen van de masterfase zullen gelden. Dit
zou namelijk in strijd zijn met de doelstellingen van de
bachelor/master structuur. Echter, om de instroomeisen voor de
\NIII\ Informatiekunde Master te formuleren maken we wel gebruik
van de vaardigheden zoals benoemd in
hoofstuk~\ref{h:wat-doelstellingen}.

Voor de masterfase van de Informatiekunde opleiding gelden als
instroomeisen:
\begin{center} \footnotesize
     \begin{tabular}{l|l}
     Vaardigheid                               & Instroomeis\\
     \hline
     \hline
     \multicolumn{2}{c}{\it Algemeen}\\
     \hline
     Academisch                                & \U \\
     Zelf leren                                & \U \\
     Kennis ontsluiten                         & \U \\
     Kennis inzetten                           & \U \\
     Reflectie op leren                        & \B \\
     Reflectie op handelen                     & \B \\
     \hline
     \multicolumn{2}{c}{\it Proces van verandering \& bestendiging}\\
     \hline
     Probleem oplossend                        & \U \\
     Uitvoeren van veranderen \& bestendigen   & \U \\
     Aanbesteden van veranderen \& bestendigen & \B \\
     Besturen van veranderen \& bestendigen    & \B \\
     Analyseren \& modelleren                  & \U \\
     Belangen behartigen                       & \B \\
     Onderhandelen                             & \B \\
     Leven met vaagheden                       & \B \\
     Communicatief                             & \U \\
     Balans tussen product en proces           & \U \\
     \hline
     \multicolumn{2}{c}{\it Subject van verandering \& bestendiging}\\
     \hline
     Gezichtspunten                            & \U \\
     Integrale visie                           & \U \\
     \hline
     \multicolumn{2}{c}{\it Verbredingsgebieden}\\
     \hline
     Inwerken in verbredingssgebieden          & \B \\
     Reflectie over verbredingsgebieden        & \B \\
     \hline
     \multicolumn{2}{c}{\it Onderzoek \& reflectie}\\
     \hline
     Onderzoeksvraag                           & \U \\
     Besturen van onderzoek                    & \B \\
     Uitvoeren van onderzoek                   & \U \\
     \hline
     \end{tabular}
\end{center}

\section{Specialisatie: Informatiearchitectuur}
Omdat we onze master willen profileren als een opleiding die studenten
voorbereid op een rol\footnote{Denk hierbij aan een rol als
informatiearchitect, informatiemanager, of onderzoeker op het gebied van de
Informatiearchitectuur.} in het vakgebied van de informatiearchitectuur, hebben
we het volgende specialisatie ``Informatiearchitectuur'' opgesteld:
\begin{center} \small
 \begin{tabular}{|llrllr|}
   \hline
   Sem. & Vak & ECTS & Status & Afdeling & Invoering\\
   \hline
   \hline
   4.1  & {\it Software Engineering 3}               & 6 & B & \NIII    & \\
        & Informatiearchitectuur                     & 6 & N & \AfdIRIS & 03/04\\
        & Kwaliteit van Informatiesystemen           & 6 & N & \AfdST   & 03/04\\
        & Security Protocols                         & 6 & N & \AfdITT  & 03/04\\
        & Visualiseren \& Communiceren               & 6 & V & \AfdST   & 03/04\\
   \hline
        &                                            &30 &   &          & \\
   \hline
   \hline
   4.2  & Informatie- \& Communicatietheorie         & 6 & B & \AfdGS   & 03/04\\
        & Systeemtheorie: Structuur en Co\"ordinatie & 3 & N & \AfdIRIS & 03/04\\
        & {\it ICT \& Samenleving 2}                 & 3 & B & \FNWI    & \\
        & Scriptie                                   &18 & B & -        & \\
   \hline
        &                                            &30 &   &          & \\
   \hline
   \hline
        & TOTAAL Master                              &60 &   &          & \\
   \hline
 \end{tabular}
\end{center}
Wederom geldt voor de \emph{cursiefgedrukte} vakken dat ze samen
gegeven worden met informatica.
Het ligt in de verwachting dat, zodra het Informatiekunde onderzoek in
Nijmegen op gang komt, deze specialisatie nog verder verfijnd zal
worden.
Tevens ligt het in de bedoeling om voor de master nog specifieke
keuzevakken in te voeren.

Voor de studenten van cohort 2000,
die dus in principe in 2003 in de masterfase moeten instromen, is
een afwijkend programma samengesteld. Dit omdat die
generatie studenten de vakken ``Visualiseren'' en ``ICT \&
Samenleving 2'' reeds gevolgd hebben. Deze studenten zullen
krijgen het volgende basisprogramma aangeboden:
\begin{center} \small
 \begin{tabular}{|llrllr|}
   \hline
   Sem. & Vak & ECTS & Status & Afdeling & Invoering\\
   \hline
   \hline
   4.1  & {\it Individueel project SO3}              & 6 & B & \NIII    & \\
        & Informatiearchitectuur                     & 6 & N & \AfdIRIS & 03/04\\
        & Kwaliteit van Informatiesystemen           & 6 & N & \AfdST   & 03/04\\
        & Cognitie en Representatie                  & 6 & N & \AfdIRIS & 03/04\\
        & Computational Intelligence                 & 6 & V & \AfdIRIS & 03/04\\
   \hline
        &                                            &30 &   &          & \\
   \hline
   \hline
   4.2  & SO4                                        & 6 & B & \NIII    & 03/04\\
        & ICT Management                             & 3 & N & \NIII    & 03/04\\
        & Systeemtheorie: Structuur en Co\"ordinatie & 3 & N & \AfdIRIS & 03/04\\
        & Scriptie                                   &18 & B & -        & \\
   \hline
        &                                            &30 &   &          & \\
   \hline
   \hline
        & TOTAAL Master                              &60 &   &          & \\
   \hline
 \end{tabular}
\end{center}

In termen van het basisstramien geldt voor de bovenstaande invulling:
\begin{itemize}
   \item Specialisatievak gericht op de rol die de student na
         de studie wil vervullen:
         ``Software Engineering 3'' of ``Onderzoekslaboratorium'' (6 ECTS)

   \item Inhoudelijke specialisatie: 21 ECTS

         Deze dient als volgt verdeeld te worden:
         \begin{itemize}
            \item Grondslagen: ``Systeemtheorie: Structuur en Co\"ordinatie'' (3 ECTS)
            \item Proces van verandering \& bestendiging:
                  ``Kwaliteit van Informatiesystemen'' (6 ECTS)
            \item Subject van verandering \& bestendiging:
                  ``Informatiearchitectuur'' (6 ECTS) en
                  ``Security Protocols'' (6 ECTS)
         \end{itemize}

   \item ICT \& Samenleving 2: 3 ECTS

   \item Vrije keuze: ``Visualiseren \& Communiceren'' (6 ECTS) en
         ``Informatie- \& Communicatietheorie'' (6 ECTS)

   \item Scriptie: 18 ECTS
\end{itemize}

Het kan niet genoeg worden benaderukt dat studenten binnen
de spelregels van het basisstramien hun eigen Master specialisatie
samenstellen.
\begin{itemize}
   \item Een student die zich meer aangetrokken voelt tot het doen van onderzoek
   in het vakgebied van de informatiearchitectuur, zal wellicht het vak
   ``Software Engineering 3'' willen vervangen door ``Onderzoekslaboratorium''.
   \item Een student die meer wil weten van beheer/bestendiging van systemen zal misschien
   het ``Visualisatie \& communicatie'' willen vervangen door een (nog te ontwikkelen!)
   keuzevak ``ICT Management''.
\end{itemize}

Naast de master ``Informatiearchitectuur'' zal er in de komende jaren
nagedacht moeten worden over alternatieve specialisaties. Denk aan:
\begin{itemize}
   \item Medische Informatiekunde
   \item Informatiemanagement
   \item Computer en recht
\end{itemize}
}
\Ragged{\chapter{Master Informatiekunde 2003 voor HBO Bachelors}
\label{h:HBOMasterProgramma}
Hieronder staat het beoogde programma voor studenten die na het voltooien
van een HBO Bachelor Informatica of Informatiekunde in 2003 het master
Informatiekunde curriculum gaan volgen.
Dit programma bestaat uit een aantal schakelvakken en uiteraard
een regulier masterprogramma.

Bij het benoemen van schakelvakken maken we een onderscheid tusen
studenten met een HBO Bachelor Informatica en een HBO Bachelor Informatiekunde.
Als we uitgaan van het programma-template van de reguliere master:
\begin{center} \small
 \begin{tabular}{|llr|}
   \hline
   Sem. & Vak & ECTS\\
   \hline
   \hline
   4.1      & Software Engineering 3 {\it of} Onderzoekslaboratorium      &  6 \\
            & Vak 1                                                       &  6 \\
            & Vak 2                                                       &  6 \\
            & Vak 3                                                       &  6 \\
            & Vak 4                                                       &  6 \\
   \hline
            &                                                             & 30 \\
   \hline
   \hline
   4.2      & Vak 5                                                       &  6 \\
            & Vak 6                                                       &  3 \\
            & ICT \& Samenleving 2                                        &  3 \\
            & Scriptie                                                    & 18 \\
   \hline
            &                                                             & 30 \\
   \hline
   \hline
 \end{tabular}
\end{center}
dan ziet het programma-template er voor HBO-ers er als volgt uit:
\begin{center} \small
 \begin{tabular}{|llr|}
   \hline
   Sem. & Vak & ECTS\\
   \hline
   \hline
   1.1  & {\sc Formeel Denken (schakelvak)}                      & 3\\
        & {\sc Soft-Systems Methodology}                         & 0/4\\
        & {\sc Informatiesystemen}                               & 6\\
        & Vak 1                                                  & 6\\
        & Vak 2                                                  & 6\\
        & Vak 3                                                  & 6\\
   \hline
        &                                                        & 27/31\\
   \hline
   \hline
   1.2  & {\sc Beweren \& Bewijzen}                              & 6\\
        & {\sc Onderzoeksvaardigheden}                           & 3\\
        & {\sc Mens-Machine Interactie}                          & 3\\
        & {\sc Datastructuren}                                   & 6/0\\
        & Vak 5                                                  & 6\\
        & Vak 6                                                  & 3\\
        & ICT \& Samenleving 2                                   & 3\\
   \hline
        &                                                        & 24/30\\
   \hline
   \hline
   2.1  & {\sc Software Engineering 3} {\it of} {\sc R\&D Laboratorium} & 6\\
        & Vak 4                                                  & 6\\
        & Scriptie                                               &18\\
   \hline
        &                                                        &30\\
   \hline
        & TOTAAL Master                                          &90\\
   \hline
 \end{tabular}
\end{center}
De in {\sc hoofdletters} aangegeven vakken zijn schakelvakken.
Het is de bedoeling dat HBO instromers met een informatica achtergrond
het vak ``Soft-Systems Methodology'' volgen, terwijl HBO instromers
met een bedrijfskundige en/of informatiekundige achtergrond het vak
``Datastructuren'' moeten volgen.

Voor de master ``Informatiearchitectuur'' leidt dit uiteindelijk
tot de volgende concrete invulling:
\begin{center} \small
 \begin{tabular}{|llr|}
   \hline
   Sem. & Vak & ECTS\\
   \hline
   \hline
   1.1  & {\sc Formeel Denken (schakelvak)}                   & 3\\
        & {\sc Soft-Systems Methodology}                      & 0/4\\
        & {\sc Informatiesystemen}                            & 6\\
        & Informatiearchitectuur                              & 6\\
        & Visualiseren \& Communiceren                        & 6\\
        & Kwaliteit van Informatiesystemen                    & 6\\
   \hline
        &                                                     &27/31\\
   \hline
   \hline
   1.2  & {\sc Beweren \& Bewijzen}                           & 6\\
        & {\sc Onderzoeksvaardigheden}                        & 3\\
        & {\sc Mens-Machine Interactie}                       & 3\\
        & {\sc Datastructuren}                                & 6/0\\
        & Informatie- \& Communicatietheorie                  & 6\\
        & Systeemtheorie: Structuur en Co\"ordinatie          & 3\\
        & ICT \& Samenleving 2                                & 3\\
   \hline
        &                                                     &30/24\\
   \hline
   \hline
   2.1  & {\sc Software Engineering 3}                        & 6\\
        & Security Protocols                                  & 6\\
        & Scriptie                                            &18\\
   \hline
        &                                                     &30\\
   \hline
        & TOTAAL Master                                       &90\\
   \hline
 \end{tabular}
\end{center}

Wegens capaciteitsbeperkingen bij de invoering van het nieuwe
Informatiekunde curriculum (onder andere leidende tot een
stapsgewijze invoering van nieuwe vakken), zal het programma in
bovenstaande vorm nog niet voor de HBO instroom van 2003 kunnen
gelden. Zij zullen het volgende programma doorlopen:
\begin{center} \small
 \begin{tabular}{|llr|}
   \hline
   Sem. & Vak & ECTS\\
   \hline
   \hline
   1.1  & {\sc Formeel Denken (schakelvak)}                   & 3\\
        & {\sc Soft-Systems Methodology}                      & 0/4\\
        & {\sc Informatiesystemen}                            & 6\\
        & Informatiearchitectuur                              & 6\\
        & Visualiseren \& Communiceren                        & 6\\
        & Kwaliteit van Informatiesystemen                    & 6\\
   \hline
        &                                                     &27/31\\
   \hline
   \hline
   1.2  & {\sc Beweren \& Bewijzen}                           & 6\\
        & {\sc Onderzoeksvaardigheden}                        & 3\\
        & {\sc Mens-Machine Interactie}                       & 3\\
        & {\sc Datastructuren}                                & 6/0\\
        & Informatie Retrieval 1                              & 6\\
        & Systeemtheorie: Structuur en Co\"ordinatie          & 3\\
        & ICT \& Samenleving 2                                & 3\\
   \hline
        &                                                     &30/24\\
   \hline
   \hline
   2.1  & {\sc SO 3}                                          & 6\\
        & Security Protocols                                  & 6\\
        & Scriptie                                            &18\\
   \hline
        &                                                     &30\\
   \hline
        & TOTAAL Master                                       &87/85\\
   \hline
 \end{tabular}
\end{center}
}
\Ragged{\chapter{Bachelor Informatiekunde 2002}
\label{h:Bachelor2002}

Studenten die in 2002 zijn ingestroomd hebben in 2003 hun eerste jaar
er in principe al opzitten. Dit deel van hun programma nemen we
dan ook voor de volledigheid zonder veranderingen over:
\begin{center} \small
 \begin{tabular}{|llr|}
   \hline
   Jaar & Vak & Sp.\\
   \hline
   \hline
   1  & Formeel Denken                              & 4 \\
      & Beweren en Bewijzen                         & 4 \\
      & Communicatie 1 (Schriftelijke vaardigheden) & 1 \\
      & Communicatie 2 (Mondelinge vaardigheden)    & 1 \\
      & Introductie Informatica en Informatiekunde  & 1 \\
      & Programmeren voor Informatiekundigen 1      & 4 \\
      & Programmeren voor Informatiekundigen 2      & 4 \\
      & Inleiding Bedrijfscommunicatie              & 3 \\
      & Introductie Mens-Machine Interactie         & 4 \\
      & Informatiesystemen in hun context           & 4 \\
      & Organisatie en Informatievoorziening        & 4 \\
      & Ori\"entatiecollege Toepassingsgebieden     & 6 \\
      & Integratieproject Informatiekunde           & 2 \\
   \hline
      &                                             &42 \\
   \hline
 \end{tabular}
\end{center}
Beschrijvingen van de vakken van de oude curricula van voor 2003 zijn
te vinden in appendix~\ref{h:OudeVakken}.

Bij het opstellen van het vervolgprogramma voor jaargang 2002 is
zoveel mogelijk uitgegaan van het nieuwe curriculum. Echter, om
de invoering van het pakket aan nieuwe vakken, zoals die nodig zijn voor
de masteropleiding en de vernieuwde bacheloropleiding, over meerdere
jaren uit te smeren, is het streven om het vervolgprogramma van
de bestaande jaargangen zoveel mogelijk op te bouwen middels
bestaande vakken.

Hieronder staat het beoogde programma voor de vervolgjaren voor studenten
die in 2002 zijn begonnen.
Enkele opmerkingen vooraf:
\begin{itemize}
   \item De {\it cursiefgedrukte} vakken zullen zoveel mogelijk met
         Informatica samen gegeven worden.
   \item De vakken in {\sc hoofdletters} zijn vakken die uit de
         oude curricula van voor Curriculum 2003 komen.
   \item De ``Verbredingsvakken'' dienen ingevuld te worden met vakken uit
         \'e\'en of hooguit twee Verbredingsgebieden van de informatiekunde.
         Deze zullen per jaar nader bepaald worden in samenwerking met
         externe faculteiten.
   \item Vakken met een ``\Ja'' in de kolom ``Wisselvak'', zijn door de
         studenten in te wisselen voor andere vakken.
         Dit biedt de studenten de ruimte voor persoonlijke profilering binnen
         de voor Informatica en Informatiekunde relevante vakgebieden.
         Echter, elk ingewisseld vak moet vervangen worden door een vak
     wat zich richt op dezelfde thema's van het vakgebied.
   \item Voor elke bacheloropleiding geldt dat er 6 ECTS volledig vrije
         keuze dient te zijn.
         Informatiekunde studenten mogen daarom \'e\'en van de
         ``Verbredingsvakken'' vervangen door een willekeurig ander vak
         van een willekeurige andere bacheloropleiding van de Universiteit.
\end{itemize}

\begin{center} \small
 \begin{tabular}{|llrr|}
   \hline
   Sem. & Vak & ECTS & Wisselvak\\
   \hline
   \hline
   2.1      & Verbreding                                           & 4/5 & \\
            & {\sc Conceptueel Modelleren}                         & 6   & \\
            & Soft-Systems Methodology                             & 4   & \Ja \\
            & Methoden van organisatieverandering                  & 5   & \\
            & {\sc Inleiding Computer Architectuur}                & 3   & \\
            & {\sc Software Technologie 1}                         & 3   & \\
            & Keuzeruimte                                          & 3   & \\
   \hline
            &                                                      &28/29& \\
   \hline
   \hline
   2.2      & Verbreding                                           & 5/4 & \\
            & {\sc Requirements Engineering in het Medisch Domein} & 6   & \\
            & {\sc Functioneel Specificeren}                       & 6   & \\
            & {\sc Software Technologie 2}                         & 6   & \Ja\\
            & Research \& Development 1                            & 6   & \\
            & Kansrekening voor Informatiekundigen                 & 3   & \\
   \hline
            &                                                      &32/31& \\
   \hline
   \hline
   3.1      & Verbreding                                           & 6 & \\
            & Statistiek                                           & 3 & \\
            & {\it Software Engineering 1}                         & 3 & \\
            & Kennis- \& Informatiemanagement                      & 6 & \Ja\\
            & Introductie CEM                                      & 6 & \\
            & {\it Mens-Machine Interactie}                        & 3 & \Ja\\
            & {\it Security}                                       & 3 & \Ja\\
   \hline
            &                                                      &30 & \\
   \hline
   \hline
   3.2      & Verbreding                                           & 6 & \\
            & Systeemtheorie: Ontwerp \& Evolutie                  & 3 & \\
            & {\it Software Engineering 2}                         & 6 & \Ja\\
            & {\it ICT \& Samenleving 1}                           & 3 & \\
            & Intelligente Systemen                                & 6 & \\
            & Scriptie                                             & 6 & \\
   \hline
            &                                                      &30 & \\
   \hline
 \end{tabular}
\end{center}
}
\Ragged{\chapter{Bachelor Informatiekunde 2001}
\label{h:Bachelor2001}

Studenten die in 2001 zijn ingestroomd hebben in 2003 hun tweede
jaar er in principe al opzitten. Dit deel van hun programma nemen
we dan ook voor de volledigheid zonder veranderingen over. Het
curriculum voor studenten die in 2001 zijn begonnen kende drie
varianten met betrekking tot de verbredingsgsgebieden:
\begin{enumerate}
   \item taaltechnologie,
   \item medische informatiekunde,
   \item informatiemanagement.
\end{enumerate}
Beschrijvingen van de vakken van de oude curricula van voor 2003 zijn
te vinden in appendix~\ref{h:OudeVakken}.

\begin{center} \small
 \begin{tabular}{|llr|}
   \hline
   Jaar & Vak & Sp.\\
   \hline
   \hline
   1  & Introductie Informatica                          & 1 \\
      & Beweren en Bewijzen                              & 4 \\
      & Inleiding programmeren A - 1e deel               & 2 \\
      & Inleiding programmeren A - 2e deel               & 2 \\
      & Informatiesystemen in hun context                & 4 \\
      & Integratieproject Informatiekunde                & 3 \\
      & Practicum psychologische functieleer             & 2 \\
      & Inleiding cognitieve ergonomie                   & 4 \\
      & Syntactische Analyse                             & 3 \\
      & Inleiding Algemene Fonetiek                      & 3 \\
      & Inleiding Informatie- en Communicatietechnologie & 3 \\
      & Statistiek 1                                     & 3 \\
      & Inleiding Medische Informatiekunde               & 4 \\
      & Organisatie en Informatievoorziening             & 4 \\
   \hline
      &                                                  &42 \\
   \hline
   \hline
   2  & Introductie CEM                                  & 4 \\
      & Informatica en Samenleving 1                     & 2 \\
      & Software Technologie 1                           & 2 \\
      & Programmeren voor Informatiekundigen 2           & 4 \\
      & Inleiding Computer Architectuur                  & 2 \\
      & Functioneel Specificeren                         & 4 \\
      & Conceptueel Modelleren                           & 4 \\
      & Het softwareontwikkelproces                      & 3 \\
      & Architectuur en Alignment 1                      & 4 \\
      & Communicatieve Aspecten van Informatiesystemen   & 3 \\
      & Verdieping Cognitieve Ergonomie                  & 2 \\
      & Verbreding                                       & 8 \\
   \hline
      &                                                  &42 \\
   \hline
 \end{tabular}
\end{center}

De 8 studiepunten ``Verbreding'' zoals boven vermeld, mogen door
studenten uit deze jaargang op \'e\'en van de drie volgende wijzen
worden ingevuld:
\begin{center}
  \begin{tabular}{lll}
     \multicolumn{2}{l}{Taaltechnologie:}\\
     &Computerlingu\"{\i}stiek en Corpustaalkunde    & 3\\
     &Taaltechnologie 1                              & 2\\
     &Taaltechnologie 2                              & 2\\
     \\
     \multicolumn{2}{l}{Informatiemanagement:}\\
     &Diverse keuzevakken Informatiemanagement       & 8\\
     \\
     \multicolumn{2}{l}{Medische Informatiekunde:}\\
     &Inleiding Medisch Handelen                     & 4\\
     &Requirements Engineering in het Medisch Domein & 4\\
 \end{tabular}
\end{center}

Hieronder staat het beoogde programma voor de vervolgjaren voor studenten
die in 2002 zijn begonnen.
Enkele opmerkingen vooraf:
\begin{itemize}
   \item De {\it cursiefgedrukte} vakken zullen zoveel mogelijk met
         Informatica samen gegeven worden.
   \item De vakken in {\sc hoofdletters} zijn vakken die uit de
         oude curricula van voor Curriculum 2003 komen.
   \item De ``Verbredingsvakken'' dienen ingevuld te worden met vakken uit
         \'e\'en verbredingsgebied van de informatiekunde.
         Deze zullen per jaar nader bepaald worden in samenwerking met
         externe faculteiten. Merk op: het gekozen verbredingssgebied \emph{mag}
         hetzelfde zijn als het verbredingsgebied wat in het tweede jaar
         is gekozen.
   \item Vakken met een ``\Ja'' in de kolom ``Wisselvak'', zijn door de
         studenten in te wisselen voor andere vakken.
         Dit biedt de studenten de ruimte voor persoonlijke profilering binnen
         de voor Informatica en Informatiekunde relevante vakgebieden.
         Echter, elk ingewisseld vak moet vervangen worden door een vak
         wat zich richt op dezelfde thema's van het vakgebied.
   \item Voor elke bacheloropleiding geldt dat er 6 ECTS volledig vrije
         keuze dient te zijn.
         Informatiekunde studenten mogen daarom \'e\'en van de
         ``verbredingsvakken'' vervangen door een willekeurig ander vak
         van een willekeurige andere bacheloropleiding van de Universiteit.
\end{itemize}

\begin{center}
 \begin{tabular}{|llrr|}
   \hline
   Sem. & Vak & ECTS & Wisselvak\\
   \hline
   \hline
   3.1      & Verbreding                                           & 4 & \\
            & Informatieverzorging                                 & 3 & \\
            & Visualiseren \& Communiceren                         & 6 & \\
            & Informatie- \& Kennismanagement                      & 4 & \Ja \\
            & Methoden van organisatieverandering                  & 5 & \\
            & Soft-Systems Methodology                             & 4 & \\
            & {\it Security}                                       & 3 & \Ja \\
   \hline
            &                                                      &29 & \\
   \hline
   \hline
   3.2      & Verbreding                                           & 4 & \\
            & Opslaan en Terugvinden                               & 6 & \\
            & {\it Mens-Machine Interactie}                        & 3 & \Ja \\
            & {\sc Software Technologie 2}                         & 6 & \Ja \\
            & Intelligente Systemen                                & 6 & \\
            & Scriptie                                             & 6 & \\
   \hline
            &                                                      &31 & \\
   \hline
 \end{tabular}
\end{center}
}
\chapter{Toetsing van Inrichtingsprincipes}
\label{h:hoe-principes}

\section{Opleiding}
\Principe{Focus op verbanden} Er dient in de opleiding veel
aandacht te zijn voor verbanden tussen gezichtspunten, vakgebieden
\& verbredingsgebieden.
\Status{
  In diverse vakken, zoals de Software Engineering vakken, Domein Modellering,
  de twee Systeem Theorie vakken en het Informatiearchitectuur vak, wordt
  aandacht besteed aan de verschillende gezichtspunten en de wederzijdse
  verbanden.

  In het vak R\&D 2 en de bachelor scriptie wordt daarnaast nog eens extra
  aandacht besteed aan de relaties met de verbredingsgebieden.

  Binnen Introductie Informatica \& Informatiekunde en de Software Engineering
  vakken wordt expliciet aandacht besteed aan de synergetische relatie tussen
  het Informatica en het Informatiekunde vakgebied.
}

\Principe{Geen verzuiling van de gezichtspunten} De 4+1 gezichtspunten dienen
eveneens expliciet in de opleiding naar voren te komen. Echter, hierbij
dient verzuiling te worden voorkomen!  De nadruk moet liggen op een
integrale visie op informatiesystemen vanuit de gezichtspunten, en de
onderlinge impact van die gezichtspunten.
\Status{
  De gezichtspunten (Mens, Organisatie, Informatie, Technologie, Informatiesystemen)
  komen in verschillende vakken naar voren. Er zijn ook diverse vakken waarin
  meerdere gezichtspunten, en met name hun onderlinge relaties, naar voren komen.
}

\Principe{Processen ook inhoud van studie} De ontwikkelings-, bestendigings-,
en aanbestedingspro\-ces\-sen dienen zelf ook inhoudelijk ontwerp van studie te
zijn. Zowel in het onderwijs als in het onderzoek.
\Status{
  In de Software Engineering vakken worden de ontwikkelprocessen zelf ook als
  subject van studie behandeld, waarbij de studenten ook leren redeneren over
  het te volgen ontwikkelproces.
}

\Principe{Taakgericht} Er moeten in de opleiding vakken zijn die zowel vanuit
theoretisch als praktisch perspectief het toekomstige takenpakket benaderen,
waarbij informatica en informatiekunde studenten geacht worden nauw samen te
werken.
\Status{
  Met name het prakticum deel (GIPHouse) van de Software Engineering vakken
  komt hieraan tegemoet. Daar zullen Informatica en Informatiekunde studenten
  samen optrekken binnen systeemontwikkelingsprojecten.
}

\Principe{Brede denkers} In de opleiding dient er ook aandacht besteed te
worden aan holistische en niet-verbale denkprocessen.
\Status{
  Hier zijn nog geen expliciete vakken voor aan te wijzen in het curriculum.
  Enkele vakken zullen echter gaan expirimenteren met nieuwe onderwijsvormen
  waar aandacht gegeven kan worden aan dergelijke denkprocessen.

  De vakken waar met dergelijke onderwijsvormen ge\"expirimenteerd zal worden
  zijn: Domein Modelleren en de beide systeemtheorie vakken.
}

\Principe{Ingebouwde dynamiek} De inrichting van de opleiding dient zo
gekozen te zijn dat vakken up-to-date (moeten en) kunnen blijven zonder dat
dit gezien moet worden als een curriculum wijziging.
\Status{
  De vakbeschrijvingen, en de namen, van de vakken uit het curriculum zijn
  gesteld in generieke termen, waarbij duidelijk is aangegeven dat er gebruik
  gemaakt zal worden van hedendaagse technologie/technieken. Hierbij worden
  de actuele technologie\"en/technieken wel als voorbeeld genoemd.
}

\Principe{Aandacht voor trends} Hoewel het voor een academische opleiding
essenti\"eel is zich vooral te focussen op de onderliggende theorie\"en,
dient er in de opleiding toch aandacht te zijn voor hedendaagse
trends in het vakgebied en hun relatie naar de dieperliggende theorie\"en.
Op technologisch vlak zal zich dit onder andere uiten in aandacht voor
actuele ontwikkelingen, zoals web-services, XML, middleware, etc.
\Status{
  Zie voorgaande principe. In de vakbeschrijvingen (en de vakinhoud) wordt waar
  relevant verwezen naar (c.q. gebruik gemaakt van) hedendaagse
  technologie/technieken.
}

\Principe{Verbredingsgebieden} De opleiding moet stil staan bij de
diversiteit aan verbredingsgebieden. De focus moet hierbij liggen
op het kunnen reflecteren over de \emph{verschillen} en de
\emph{overeenkomsten} tussen de diverse verbredingsgebieden,
evenals het kunnen inwerken in nieuwe verbredingsgebieden.

De specifieke verbredingsgebieden, zoals die op de KUN reeds
worden onderzocht en gedoceerd (medische informatiekunde, taal \&
spraaktechnologie), dienen hierbij een illustrerende rol te
hebben. \Status{
   Er is in de bachelor ruimte voor totaal 24 ECTS aan vakken uit
   verbredingsgebieden. De student mag maximaal twee verbredingsgebieden
   uitkiezen. In R\&D 2 dient het werkstuk gerelateerd te zijn aan het
   verbredingsgebied uit het tweede jaar, terwijl de bachelorscriptie
   gerelateerd dient te zijn aan \'e\'en van de gekozen verbredingsgebieden.
}

\Principe{Brede beroepsorientering} In de bachelor-fase zal er geen
expliciete voorsortering zijn op een specifieke beroepsrichting. In de
bachelor-fase zal er voor alle studenten aandacht zijn voor:
\begin{itemize}
   \item Onderzoeksvaardigheden
   \item Doceer- \& presentatievaardigheden
   \item Praktijkgerichte vaardigheden
\end{itemize}
\Status{
   De mix van: Introductie CEM, R\&D 1, R\&D 2, Software Engineering 1 en 2, voldoet
   hier naar onze mening aan.
}

\Principe{Uitvoeren \& besturen} In de opleiding zal er zowel aandacht zijn
voor uitvoerende, besturende, als beleidsmatige aspecten van het vakgebied.
\Status{
   Deze aspecten komen met name in de Software Engineering vakken naar voren,
   waarbij in de bachelorfase de focus ligt op de uitvoerende aspecten en in de
   masterfase op de sturende/beleidsmatige aspecten.
}

\Principe{Doorstroomprogramma} Voor zij-instromers dienen er
doorstroomprogramma beschikbaar te zijn. Uitgaande van een relevante
vooropleiding mag dit programma niet meer dan een ${1}\over{2}$ jaar aan
studietijd kosten.
\Status{
   Zie de instroomprogramma's voor HBO instromers. Indien er een substantieel
   aanbod van instromers uit andere richtingen ontstaat, zal ook daar een specifiek
   instroomprogramma voor worden ontwikkeld.
}

\Principe{Makkelijk schakelen} De \NIII\ informatica en informatiekunde
opleidingen dienen zodanig opgezet te worden dat het schakelen tussen de
twee studies zo min mogelijk impact heeft op het studieverloop van de
studenten.
\Status{
   Na afloop van het vak Introductie Informatica \& Informatiekunde, wat
   als blok-cursus aan het begin van semester 1.1 wordt gegeven, kan een
   student nog zonder vertraging overstappen tussen beide studies.

   Na afloop van het eerste semester moet een student rekenen op ongeveer
   6 ECTS vertraging, terwijl overschakelen na een voltooid eerste jaar
   ongeveer 12 ECTS zal kosten.
}

\Principe{Profilering van de Master} Voor de \NIII\ Master Informatiekunde
zal in eerste de specialisatie op Informatiearchitectuur expliciet
geprofileerd worden.  Wellicht dat er op termijn nog specialisaties bijkomen,
maar het is belangrijk ons eerst goed te specialiseren in \'e\'en
specialisatierichting.
\Status{
   Zie de Informatiearchitectuur specialisatie van de master.
}

\Principe{Cliff-hanger} De bachelor-fase dient zodanig ingericht te zijn dat
er als vanzelf een `honger naar meer' ontstaat bij de studenten. De
Master-fase dient te voldoen aan die honger.
\Status{
   Er zijn in het derde jaar van de bachelorfase diverse momenten waarop
   de KUN specialisaties mbt de master kunnen worden gepositioneerd.
}

\Principe{Engels} De engelse en nederlandse taal als volgt gebruiken:
\begin{itemize}
   \item Leerstof -- Bachelor: optie, Master: Engels
   \item Tentamens -- Bachelor: Nederlands, Master: Engels
   \item Voorlichtingsmateriaal -- Bachelor: Nederlands, Master: beide
   \item Offici\"ele reglementen -- Bachelor: Nederlands, Master: beide
   \item Colleges -- Bachelor: optie, Master: Engels
\end{itemize}
\Status{
   Dit zal zsm worden ingevoerd. Zeker voor nieuw te ontwikkelen master
   vakken zal het gebruik van Engels voor de leerstof verplicht worden gesteld.
}

\section{Onderwijs}
\Principe{Ondervinden van ICT} De snelle opmars van ICT moet ook doorklinken
in de opleiding. Studenten dienen daarom dan ook ICT aan den lijve te
ondervinden, bijvoorbeeld door het inzetten van moderne ICT in het onderwijs
zelf.
\Status{
   Hier moet nog een \NIII\ brede specifieke inrichting aan gegeven worden.
   Los daarvan kunnen we wel verwijzen naar het Notebook project, het draadloze
   netwerk op de \FNWI\ campus, en het toenemend gebruik van communicatiemiddelen
   zoals IRC, Jabber en ICQ.
}

\Principe{Aandacht voor creativiteit en verandering} Studenten dienen
gestimuleerd te worden om zich te ontwikkelen tot creatieve mensen die
openstaan voor, en kunnen omgaan met veranderingen.
\Status{Zie principe: ``Brede Denkers''.}

\Principe{Rechter hersenhelft} Naast de typische linkerhersenhelft
benaderingswijze dient het onderwijs ook de rechter\-hersenhelft manier van
denken, informatie verwerken en leren te stimuleren, echter zonder dat de
wetenschappelijke werkwijze die iedere (toekomstige) academicus zich eigen
moet maken, verloren gaat.
\Status{Zie principe: ``Brede Denkers''.}

\Principe{Leren onder eigen verantwoordelijkheid} Studenten dienen zelf
verantwoordelijk gesteld te worden voor het leerproces. Een zelfstandigheid
welke in de loop van de opleiding geleidelijk zal moeten groeien (groeiende
zelfsturing).
\Status{Zie principe: ``Brede Denkers''.}

\Principe{Aandacht voor leerprocessen} In de opleiding dient aandacht besteed
te worden aan leerprocessen. Hierbij is het belangrijk om studenten te
\emph{helpen ontdekken} hoe zij \emph{zichzelf}  kunnen ontplooien, met
andere woorden, hoe zij hun eigen (en eventueel andermans) capaciteiten
optimaal kunnen benutten en onder welke omstandigheden.
\Status{Zie principe: ``Brede Denkers''.}

\section{Toetsing en beoordeling}
\Principe{Grotere projecten} Elk studiejaar kent een of meerdere grotere
projecten waarin een aantal samenhangende thema's aan bod komt.\footnote{Te
overwegen valt bijvoorbeeld om in de propedeuse met \een\ of twee van zulke
projecten te defini\"eren, en het aantal te laten toenemen in latere jaren.} De
inhoud van verschillende samenhangende vakken wordt aan deze projecten
opgehangen. De projecten hebben een concrete vraag of probleemstelling als
insteek, waarin een specifiek stuk kennis en inzicht centraal staat. Aan het
einde van het project geeft de student blijk van het feit dat hij deze kennis
en inzichten verworven heeft (bijvoorbeeld door middel van verslaglegging).
\Status{De informatiekunde opleiding zal hierbij de innovaties bij de
informatica opleiding volgen. Voor informatiekunde geldt momenteel dat
het opzetten van een inhoudelijk complete en goed gestructureerde
de voorrang heeft.}

\Principe{Initiatief bij de student} Studenten worden aangemoedigd aan het
begin van een project vanuit hun eigen optiek na te denken over de manier
waarop zij de beoogde inhoudelijke leerdoelen zouden willen bereiken, en
daarnaast voor zichzelf een aantal procesgerichte leerdoelen te formuleren.
Hierbij kunnen studenten (al naar gelang hun achtergrond) verschillende
trajecten volgen.
\Status{De informatiekunde opleiding zal hierbij de innovaties bij de
informatica opleiding volgen. Voor informatiekunde geldt momenteel dat
het opzetten van een inhoudelijk complete en goed gestructureerde
de voorrang heeft.}

\Principe{Samenwerkend leren} Studenten worden gestimuleerd om in het kader van
de projecten samen te werken. Bij aanvang van een project kunnen zij zelf
onderling afstemmen welke taken zij zouden willen vervullen. Hierdoor kan de
ene student in een bepaald project andere vaardigheden verwerven als de andere
student. Door ervoor te zorgen dat rollen rouleren, wordt bewerkstelligd dat studenten
ook steeds verschillende vaardigheden leren, en niet alleen de dingen doen die
hen makkelijk afgaan.
\Status{De informatiekunde opleiding zal hierbij de innovaties bij de
informatica opleiding volgen. Voor informatiekunde geldt momenteel dat
het opzetten van een inhoudelijk complete en goed gestructureerde
de voorrang heeft.}

\Principe{Ondersteunende vaklijnen} Ter ondersteuning van de grotere, probleem-
en kennisgedreven projecten zijn er een aantal vaklijnen waarin typisch
vaardigheden centraal staan. Te denken valt hierbij bijvoorbeeld aan:
programmeervaardigheden, communicatieve vaardigheden, specificeervaardigheden
(Beweren \& Bewijzen; Formeel Denken). Door de ondersteunende lijnen te
verbinden aan de context van de lopende projecten, wordt de rol van deze
vaardigheden duidelijk.
\Status{De informatiekunde opleiding zal hierbij de innovaties bij de
informatica opleiding volgen. Voor informatiekunde geldt momenteel dat
het opzetten van een inhoudelijk complete en goed gestructureerde
de voorrang heeft.}

\section{Kwaliteitsbeheersing}

\Principe{Synergie onderwijs \& onderzoek} Er dient synergie nagestreefd te
worden tussen onderwijs en onderzoek.  Concreet, dient het onderwijs moet
gefundeerd zijn op een gemeenschappelijke visie op het wetenschapsgebied,
het onderwijs en de beroepspraktijk, en de onderdelen goed op elkaar
afgestemd zijn qua inhoud en studeerbaarheid.
\Status{
  Diverse vakken zijn sterk gerelateerd aan onderzoeksinteresses binnen
  het \NIII. Zie ook het visiedocument~\cite{Visie2003}.
}

\Principe{Voorlichting}
Potenti\"ele studenten, collega wetenschappers en beroepsbeoefenaars dienen over
de opleiding te worden voorgelicht middels gedegen, eerlijke en uitdagende
voorlichting. Dit begint met het hebben van een duidelijke visie op het
vakgebied, en de rol die afgestudeerde informatiekundigen hierin kunnen
vervullen.
\Status{
  Dit is in voorbereiding. Voor informatiekunde zal er iemand aangewezen
  worden die zich expliciet gaat bezighouden met de inhoudelijke kant van
  de voorlichtingsactiviteiten.
}

\Principe{Dialoog met de praktijk} Er dient een dialoog te worden aangegaan met
de praktijk, middels: gastdocenten, bijzondere hoogleraren, gemeenschappelijke
onderzoeksprojecten, gebruik van praktijkcasussen in het onderwijs, en een
raad van advies van vertegenwoordigers van de drie beroepsvelden.
\Status{
  Behalve de raad van advies, voldoen we hieraan. De raad van advies zal
  ook zo snel mogelijk worden opgezet.
}

\subsection{Randvoorwaarden}

\Principe{Ba-Ma structuur} De opleiding dient conform het bachelor \& master
stelsel te worden opgezet
\Status{
  Gedaan.
}

\Principe{Instroom in Master} Er dienen in het master-deel van de opleiding
instroom mogelijkheden te zijn voor bachelors van andere universitaire
studies. Het master-deel van de opleiding dient tevens toegankelijk te zijn
voor studenten uit het buitenland.
\Status{
   Er zijn instroomprogramma's benoemd. In de master is gekozen voor Engels
   als standaardtaal.
}

\Principe{In lijn met certificering} Waar mogelijk en relevant zal de
informatiekunde opleiding zoveel mogelijk worden afgestemd op de eisen die
voortvloeien uit de certificering van architecten in de ICT wereld
\Status{
   Certificeringseisen zijn nog niet voorhanden.
}

\Principe{Filosofie \& ethiek} Er dienen zowel in de Master als in de Bachelor 3
ECTS besteed te worden aan vakken die zijn gericht op filosofische en
ethische verdieping, middels reflectie op het eigen vakgebied.
\Status{
   Hier wordt middels de vakken ICT \& Samenleving 1 en 2 aan voldaan.
}

\Principe{Koppeling onderwijs \& onderzoek} Er dient een consequente koppeling
tussen onderwijs en onderzoek te zijn.
\Status{
   Diverse vakken komen direct voort uit de onderzoeksinteresses van de vier
   \NIII\ afdelingen.
}

\Principe{Aansluiting bij zwaartepunten} De opleiding moet aansluiten aan bij c.q.\
integreert zwaartepunten van het facultair onderzoek.
\Status{
}

\Principe{Afgestemd op beroepsperspectief} De opleiding moet inhoudelijk zijn
afgestemd op de beroepsperspectieven en -profielen van de afgestudeerde.
\Status{
   De opleiding is gebaseerd op een visie van het vakgebied, uitgaande van de
   werkzaamheden van een informatiekundige in de beroepspraktijk. Dit beeld
   is middels een extern klankbord getoetst aan de praktijk.
}

\Principe{Transparante opbouw} De opleiding dient transparant te zijn qua
opbouw. Met andere woorden, er dient een samenhangende, cumulatieve,
inhoudelijke opbouw door de opleiding heen te zijn, die de ontwikkeling van
de student weerspiegelt.
\Status{
   De opleiding is zoveel mogelijk opgezet vanuit het beeld van de werkzaamheden
   van een afgestudeerde informatiekundige. Op basis van dat beeld zijn componenten
   binnen de opleiding onderscheiden.
}

\Principe{Wenselijke inhoud} De inhoud van de opleiding moet actueel, aantrekkelijk
en uitdagend zijn.
\Status{
   Naar onze mening voldoet de opleiding hieraan.
}

\Principe{Rol propedeuse} De propedeuse dient een \emph{selecterende}.
\emph{ori\"enterende} en \emph{verwijzende} functie te hebben. Studenten
dienen daarom in het eerste jaar ook een tijdige terugkoppeling te krijgen
t.a.v. hun prestaties, bijvoorbeeld middels een studieadvies.
\Status{
   Naar onze mening voldoet voorgestelde propedeuse hieraan.
}

\Principe{Breed; doch diepe focus} Het Bachelor-deel van de opleiding dient
breed ge\"orienteerd te zijn, maar met een sterke vakdisciplinaire
component.
\Status{
   Naar onze mening voldoet de bachelorfase hieraan.
}

\Principe{Academisch vormend} De Bachelorfase dient tevens academisch
vormend te zijn, hetgeen de academische bacheloropleiding moet onderscheiden
van een HBO bachelor.
\Status{
   Diverse vakken in de bachelorfase zijn duidelijk op academisch niveau
   gepositioneerd. Vakken zoals Formeel Denken, Beweren \& Bewijzen, R\&D 1,
   R\&D 2 en Systeemtheorie: Ontwerpe \& Evolutie, benadrukken dit onderscheid
   nog eens expliciet.
}

\Principe{Aansluiten op voorkennis} Het opleidingsniveau sluit inhoudelijk
aan op het voorkennis- en abstractieniveau van de vwo-eindprofielen.
\Status{
   Voor zover we dit van te voren kunnen beoordelen voldoet onze
   opleiding hieraan.
}

\Principe{Smaakt naar meer} Het derde jaar van de bachelor dient te worden
opgezet als aanzet tot het doen van een master, en niet als uitstroommoment.
\Status{
   Zie principe ``Cliff-hanger''.
}

\Principe{Afstudeervarianten} Er zijn drie afstudeervarianten
\begin{itemize}
   \item Onderzoek
   \item Communicatie \& educatie
   \item Management \& toepassing
\end{itemize}
\Status{
   Voor de 4-jarige Informatiekunde opleiding wordt het niet als haalbaar
   beschouwd om drie beroepsgeorienteerde afstudeervarianten op te zetten.
}

\Principe{Volgen in innovatie} Het informatiekunde curriculum dient zich te
conformeren aan de uitkomsten van de facultaire onderwijsinnovatie.
Echter, het zorgen dat de (inhoudelijk) juiste vakken worden aangeboden
binnen het informatiekunde curriculum heeft een \emph{hogere} prioriteit dan
het inzetten van de juiste onderwijsvorm.
\Status{
  Bij enkele vakken zal al wel ge\"expirimenteerd worden met nieuwe onderwijsvormen.
  Zie principe ``Brein denken''.
}

\Principe{Essenti\"ele kwaliteiten}
De inrichting van de opleiding dient zodanig te zijn dat deze
(in volgorde van prioriteit) een verantwoord kwaliteitsniveau heeft met betrekking
tot:
\begin{enumerate}
   \item Studeerbaarheid
   \item Roosterbaarheid
   \item Robuustheid
   \item Migreerbaariheid
\end{enumerate}
\Status{
  Bij het ontwerpen van het nieuwe curriculum is steeds rekening gehouden met
  de impact hiervan op de bestaande cohorts. De bijstellingen van de lopende
  curricula zijn daarom ook bijgesloten in dit document.

  Daarnaast is zorgvuldig gekeken naar de roosterbaarheid van de combinatie(!)
  van resulterende curricula, over beide opleidingen heen.
  Door het geven van volledig ingevulde studieprogramma's (met default keuzes
  voor eventuele keuzevakken) is gepoogd de studeerbaarheid zoveel mogelijk
  aan te tonen. Een kanttekening moet gemaakt worden mbt de extern te volgen
  vakken. Omdat het \NIII\ daar geen controle over heeft is het onmogelijk om
  daar voor de zomer van 2003 zekerheid over te verkrijgen.
}

\Principe{Regie in eigen handen} Informatiekunde onderwijs zoveel mogelijk onder
directe (roostering \& inhoudelijke) controle van het \NIII\ georganiseerd
te worden. Dit geldt met name voor onderwijs dat tot de kern van het
vakgebied behoord.
\Status{
  Voor zover de personele en financi\"ele middelen dit toestaat worden vakken
  door `eigen personeel' gegeven.
}

\Principe{Gedeelde verantwoordelijkheid} Als onderwijs niet onder directe
\NIII\ controle georganiseerd kan worden, dan dienen de studenten hiervan op
de hoogte te zijn. Ook van studenten wordt een actieve houding gevraagd om in de
loop van het studiejaar de roosterbaarheid en studeerbaarheid optimaal te houden.
Denk hierbij heel concreet aan het vroegtijdig signaleren van clashes in de
roostering.
\Status{
  In de vakbeschrijvingen staat expliciet vermeld wat externe vakken zijn en
  wat interne vakken zijn. Daarnaast zal ook de studiegids hier aandacht
  aan besteden.
}

\Principe{Samen; met wederzijds respect} Waar mogelijk moet het informatica en
informatiekunde onderwijs gezamenlijk aangeboden worden.
Dit moet juist niet alleen uit efficiency overwegingen gedaan worden.
De beide vakgebieden moeten wel in hun waarde gelaten worden.
\Status{
  In de opleiding zitten diverse vakken die (deels) door informatica
  en informatiekunde studenten samen gevolgd worden.
}

\Principe{Semester als standaard} Het \NIII\ gebruikt semesters, onderbroken
met een `rustperiode' ten behoeve van mid-semester tests, als standaard.
Kwartaalvakken zijn toegestaan.
\Status{
   Is zo ingevoerd
}

\Principe{Standaardomvang} Het \NIII\ gebruikt 3 en 6 ECTS als standaardomvang
voor vakken.
\Status{
   Is zo ingevoerd. Wel moet opgemerkt worden dat vakken die gevolgd worden bij
   externe faculteiten soms roet in het eten kunnen gooien. In dergelijke gevallen
   kan desgewenst via een persoonlijke opdracht een gat van 1 ECTS opgevuld worden.
   Typisch zal in zo'n opdracht de relatie tussen het gevolgde vak en de
   informatiekunde nader uitgediept worden.
}

\Principe{Wisselvakken} Er wordt naar gestreefd om 30 ECTS aan wisselvakken te hebben
in de bachelorfase. Dit geldt voor beide \NIII\ opleidingen.
\Status{
  Zie de inrichting van de bachelorfase. Daar is 30 ECTS aan ruimte voor
  wisselvakken ingericht.
}

\appendix
\Ragged{\chapter{Vakbeschrijvingen}
\label{h:Vakken}

Deze appendix bevat de vakbeschrijvingen van de vakken die met ingang van het
curriculum 2003 ingevoerd zullen worden.


\Vak{Algoritmiek}{6 ECTS}{Herst}{\AfdST}{Sjaak Smetsers}{
   2003 --
}{
  \Hernoeming{
  - Programmeren voor Informatiekundigen 1
  }
}{
\item Bij het programmeren instrueer je een computer hoe hij een bepaald
taak moet uitvoeren. Omdat een computer geen benul heeft van je
bedoelingen valt dat niet mee. In dit eerste programmeervak leer je de
basisconstructies kennen waar mee je een zogenaamd imperatief
computerprogramma kunt samenstellen. We zullen de programmeertaal C++
gebruiken. Hoe je dit gereedschap op de goede manier gebruikt om
programma's te maken of aan te passen is belangrijker dan de
programmeertaal zelf.

\Onderwerpen{
\begin{itemize}
  \item taalbeschrijvingen in de vorm van syntaxdiagrammen
  \item controlstructuren: opeenvolging, keuzes, voorwaardelijke
        herhalingen, herhalingen-met-teller
  \item functies en procedures
  \item objecten, variabelen en constanten, globale versus lokale
        objecten
  \item parameteroverdracht: call-by-value, call-by-reference
  \item datastructuren: typesynoniemen, enumeratietypes, rijen,
        structuren en klassen, bestanden
  \item systematische programmaontwikkeling door het opsplitsen van
        problemen in deelproblemen (top-down programma ontwikkeling);
  \item eenvoudige complexiteitsanalyses van algoritmen;
  \item standaardalgoritmen voor zoeken en sorteren;
  \item recursie, recursief sorteren.
\end{itemize}}
}{
\item Inzicht in de werking en opbouw van programma's te geven door
      studenten zelf programma's te laten maken.
\item Gegeven algoritmen kunnen implementeren.
\item De werking van een gegeven programma doorgronden.
\item Voor eenvoudige problemen zelf systematisch een algoritme
      ontwikkelen en de geschiktheid hiervan aannemelijk maken.
\item Globale afschattingen maken van de complexiteit van algoritmen en
      programma's.
\item De kwaliteit van een programma's beoordelen
      (zowel door redeneren als door testen).
\item De correcte werking van een programma verifi\"eren.
\item De geschiktheid van een implementatie valideren.
\item Programma's ontwikkelen die aanpasbaar zijn
      (duidelijke structuur, goede naamgeving, abstractie via
      typesynoniemen, functies en klassen).
}

\Vak{Beweren \& Bewijzen}{6 ECTS (was 4 sp)}{Lente}{\AfdITT}{Hanno Wupper}{
   2000 --
}{
   \Bestaand
}{
   \item Hoe bereikt men helderheid? Wanneer is een bewering waar? Wanneer doet
   een ICT-systeem wat het moet doen? We beschouwen verschillende
   toepassingsgebieden van taal, juridische wetten bijvoorbeeld, en contracten.

   \item Voor informatici belangrijke speciale gevallen zijn specificaties (als
   contract) en algoritmen (uitvoeringsvoorschriften, speciale gevallen van een
   speciaal geval van wetten).

   \item We gaan uit van uitspraken in natuurlijke taal.
   Deze gaan we analyseren en beperken tot constructies die we echt begrijpen,
   en formaliseren, d.w.z. in een notatie gieten met een goed gedefinieerde
   betekenis.

   \item Vervolgens gaan we bestuderen, aan welke regels deze formele
   uitspraken onderhevig zijn en hoe men tot aantoonbaar ware uitspraken kan
   komen.

   \item Er zal exemplarisch gebruik gemaakt worden van SQL.  Doel daarvan is
   dat de deelnemers ervaring opbouwen in het gebruik van het retrieval-deel
   van SQL.

   \item Merk op: er zijn drie bachelor vakken die aspecten van SQL belichten
   de docenten van deze vakken zullen dit onderling nader afstemmen. Het gaat
   hierbij om: ``Beweren \& Bewijzen'', ``Domain Modelleren'' en
   ``Opslaan \& Terugvinden''.

   \Onderwerpen{
      Realiteit, abstractie, modellen, contracten, natuurlijke en formele
      talen, syntaxis en semantiek, typering, propositie- en predikatenlogica,
      waarheidstabellen, natuurlijke deductie, specificatie, correctheid van
      systemen, Chinese dozen (hi\"erarchische decompositie), proof tools,
      relationele algebra, SQL, state-based systems.
   }
}{
   \item inconsistenties en incorrectheden aanwijzen in niet
   deugende uitspraken
   \item heldere, consistente en correcte uitspraken
   formuleren
   \item de correctheid van eigen beweringen beredeneren
   \item oplossingen systematisch kunnen afleiden c.q. een
   systematische afleiding presenteren
   \item actief en constructief meewerken aan het
   verhelderen van onduidelijke uitspraken
   \item teksten en discussies structureren d.m.v.
   begripsdefinities
   \item propositie- en predicatenlogica en exemplarische
   andere theorie\"en relateren.
}

\ExternVak{Cognitie}{6 ECTS}{Lente}{
   2004 --
}{
   \item Introductie cognitiewetenschap A (COGW.KICO111)
}

\Vak{Cognitie en Representatie}{6 ECTS}{Herfst}{\AfdIRIS}{Janos
Sarbo}{
   2003 --
}{
   \Hernoeming{Kennis Representatie}
}{
  \item De term 'kennis' wordt vaak geassocieerd met (a) een idee, iets
wat we bedenken, of (b) een observatie, iets wat we ervaren.
Ruwweg zijn deze twee complementair.
  \item De term representatie
correspondeert, met name in de informatika, met formalisering.
Maar hoe kom je tot geformaliseerde kennis?
  \item In dit vak staat de
tweede betekenis van kennis centraal. We stellen de vraag: Wat is
er in de 'reele' wereld, en hoe kan men dat systematisch
specificeren? We zoeken naar een formele methode, en vinden dat in
de tekenleer (semiotic).
  \item Toepassingen hiervan in de logica en taal
worden bestudeerd.
}{
  \item Wat zijn signs en welke aspecten
(verschillen) kunnen door deze gesignalleerd worden?
  \item Hoe onstaan
primitieve en complexe tekens?
  \item Hoe kunnen we problemen als
fenomenen `zien' door middel van tekens?
  \item Hoe is dit gerelateerd
aan de logica, de natuurlijke taal en semantiek?
  \item Hoe vindt je
tekens in een tekst?
  \item Hoe kunt je deze tekens reduceren tot een
teken dat de inhoud samenvat.
  \item Tekens zijn `reeele' concepten.
  \item Wat zijn dan `formele' concepten, en waar ligt het verschil tussen de
twee.
}

\Vak{Computational Intelligence}{6 ECTS}{Herfst}{\AfdIRIS}{Peter
Lucas}{
   2003 --
}{\Nieuw}{
  \item Al bij de ontwikkeling van de eerste kennissystemen in
       de Kunstmatige Intelligentie bleek het noodzakelijk te zijn
        rekening te houden met onzekerheid in probleemdomeinen.
  \item Zonder een goede aanpak van het vastleggen en redeneren met onzekere
kennis zijn veel problemen, zoals trouble-shooting van software en
hardware, en medische diagnostiek en behandeling, niet goed aan te
pakken.
  \item In Windows XP heeft Microsoft bijvoorbeeld diverse
kennissystemen ingebed die met onzekere kennis kunnen omgaan, en
helpen bij het achterhalen van de oorzaak van problemen; ook zijn
diverse onderzoeksgroepen actief bezig met de ontwikkeling van
kennissystemen die artsen kunnen helpen bij diagnostiek en
behandelingskeuze.
  \item Daarnaast is de afgelopen jaren meer nadruk
komen te liggen op het representeren van en redeneren met modellen
van apparaten en biologische mechanismen bij het ontwikkelen van
kennissystemen.
  \item Autofabrikanten zoals Fiat hebben al
model-gebaseerde software ontwikkeld die autobezitters helpt bij
trouble-shooting, en ook Philips is druk bezig in
consumentenelectronica computational intellegence in te bouwen.
  \item Model-gebaseerd redeneren en Bayesiaanse netwerken zijn twee
verwante collecties methoden en technieken die de basis vormen van
dit soort systemen.
  \Onderwerpen{Principes van kennissystemen,
representatie van onzekerheid in kennissystemen, Bayesiaanse
netwerken (representatie en redeneeralgoritmen),
toepassingsklassen van Bayesiaanse netwerken, automatisch leren
van Bayesiaanse netwerken, model-gebaseerd representeren en
redeneren.}
}{
  \item Inzicht opdoen in recente ontwikkelingen in
de Kunstmatige Intelligentie, in het bijzonder in de mogelijkheden
van Bayesiaanse netwerken en model-gebaseerde methoden bij de
ontwikkeling van kennissystemen.
}

\Vak{Datastructuren}{6 ECTS}{Lente}{\AfdST}{Pieter Koopman}{
   2003 --
}{
   \Hernoeming{
   - Programmeren voor Informatiekundigen 2
   }
}{
\item In het vak \emph{datastructuren} worden volgens dezelfde systematische methoden
als in het vak \emph{algoritmen} complexere algoritmen ontworpen en gerealiseerd.

\Onderwerpen{\begin{itemize}
\item backtracking; branch and bound
\item recursieve datatypen: lijsten, queues, stacks, grafen, bomen
\item virtual functions en inheritance
\item graphical user interfaces (GUI's)
\item modules, templates en exceptions
\end{itemize}}
}{
\item Ingewikkeldere algoritmen begrijpen, afleiden en implementeren.
\item Toepassingsmogelijkheden van zoekalgoritmen (als back-tracking).
      herkennen, het geschiktste algoritme kunnen kiezen en implementeren.
\item Recursieve datastructuren (lijsten en bomen) gebruiken en ontwikkelen.
\item Basisprincipes van OO begrijpen, herkennen en toepassen.
\item Abstraheren via modulen en klassen.
\item Complexere bibliotheken gebruiken.
\item Gegeven programma's kunnen begrijpen en uitbreiden.
}

\Vak{Domeinmodellering}{6 ECTS}{Herfst}{\AfdIRIS}{-}{
   2003 --
}{
   \Herschikking{
   - Informatiesystemen in hun Context\\
   - Architectuur \& Alignment
   }
}{
   \item In dit vak zul je kennis maken met het
   modelleren van domeinen (een stuk van ``de werkelijkheid die we
   waarnemen'').

   \item Dit zullen we in eerste instantie doen aan de hand van
   concrete voorbeelden waarbij studenten middels de UML technieken
   aspecten van organisaties, informatiesystemen, etc, dienen te modelleren.
   Er zal ook gebruik gemaakt worden van enkele oefeningen met SQL en
   met Prolog. SQL zal hierbij gebruikt worden om een UML
   klassediagrammen te concretiseren in termen van een implementatie, en
   Prolog zal gebruikt worden om kennis te maken met de semantiek van
   logische eigenschappen van domeinen (bijvoorbeeld de semantiek van
   constraints in UML).

   \item In dit vak wordt ook beoogd om de studenten kennis te laten maken met
   het feit dat eenzelfde domein, vanuit verschillende waarnemers en
   perspectieven gemodelleerd kan worden.

   \item Gaandeweg het semester zullen we de overstap maken naar een wat
   abstracter denkniveau en overgaan naar het denken in termen van systemen obv
   systeemtheorie.

   \item Merk op: er zijn drie bachelor vakken die aspecten van SQL belichten
   de docenten van deze vakken zullen dit onderling nader afstemmen. Het gaat
   hierbij om: ``Beweren \& Bewijzen'', ``Domain Modelleren'' en
   ``Opslaan \& Terugvinden''.
}{
   \item Het primaire doel van dit vak is kennis te maken met de vaardigheid van
   modelleren, en de redenen om te modelleren.
   \item Secondaire doelen zijn:
   \begin{itemize}
     \item Kennismaking met UML, Prolog, SQL, elementen uit de systeemtheorie
     \item Bewust van het onderscheid tussen:
           \begin{itemize}
              \item het gemodelleerde domein,
              \item de waarnemer,
              \item het beeld van het domein zoals dit leeft bij de waarnemer en
              \item de beschrijving van het domein in een taal (natuurlijke taal of
                    een formele taal).
           \end{itemize}
     \item Bewust van de invloed die de gebruikte beschrijvingstaal kan hebben op het
           beeld van een waarnemer van het domein. Met een hamer in de hand is alles
           een spijker ...
     \item Begrip van de plaats van modelleren binnen de systeemontwikkeling.
   \end{itemize}
}

\Vak{Formeel Denken}{6 ECTS (was 4 sp)}{Herst}{\AfdGS}{Henk Barendregt}{
   2002 --
}{
   \Bestaand
}{
   \item In dit college leer je met precisie om te gaan met de objecten die
   voor de informatica van belang zijn.

   \item Om een idee te krijgen van de toepasbaarheid van deze formele
   (wiskundige) methode zullen we een aantal exemplarische voorbeelden
   bekijken.

   \item De volgende begrippen zijn bijvoorbeeld van belang in de wereld van
   de informatica \& informatiekunde: 'Werkelijkheid', '(formele) taal',
   'betekenis', 'waarschijnlijkheid' en `beweging'.

   \item De hulpmiddelen die we zullen gebruiken om met precisie met deze
   onderwerpen om te gaan komen uit de wiskunde: logica, algebra, analyse,
   kansrekening.

   \item Het gaat er niet zozeer om alle technieken te beheersen, maar meer om
   te weten dat ze er zijn. Toch zal er in het college flink geoefend worden.
}{
   \item Kunnen beargumenteren wat de rol is/kan zijn van formele talen
   in de wereld van de informatici en informatiekundigen.

   \item Zelf formele talen kunnen inzetten bij het analyseren, modelleren,
   ontwerpen van enkele \'e\'envoudige voorbeelddomeinen.
}

\Vak{Formeel Denken (Schakelvak)}{3 ECTS}{Herst}{\AfdGS}{Henk
Barendregt}{
   2003 --
}{
   \Bestaand
}{
   \item In dit college leer je met precisie om te gaan met de objecten die
   voor de informatica van belang zijn.

   \item Om een idee te krijgen van de toepasbaarheid van deze formele
   (wiskundige) methode zullen we een aantal exemplarische voorbeelden
   bekijken.

   \item De volgende begrippen zijn bijvoorbeeld van belang in de wereld van
   de informatica \& informatiekunde: 'Werkelijkheid', '(formele) taal',
   'betekenis', 'waarschijnlijkheid' en `beweging'.

   \item De hulpmiddelen die we zullen gebruiken om met precisie met deze
   onderwerpen om te gaan komen uit de wiskunde: logica, algebra, analyse,
   kansrekening.
}{
   \item Kunnen beargumenteren wat de rol is/kan zijn van formele talen
   in de wereld van de informatici en informatiekundigen.

   \item Zelf formele talen kunnen inzetten bij het analyseren, modelleren,
   ontwerpen van enkele \'e\'envoudige voorbeelddomeinen.
}

\Vak{Gedistribueerde software systemen}{6 ECTS}{Herfst}{\AfdST}{Peter Achten}{
   2004 --
}{
   \Herschikking{
   - Internet Exploitatie\\
   - Software Technologie 1
   }
}{
  \item
   De technologische vooruitgang en toenemende mate van beschikbaarheid van
   computer-hardware en netwerktechnologie stimuleert de ontwikkeling van
   gedistribueerde software systemen.
   Bovendien dwingt de toenemende complexiteit van applicaties ontwikkelaars er
   toe software in componenten op te bouwen.

   \item Gedistribueerde software systemen onderscheiden zich op een aantal punten van
   `traditionele' software systemen:
   \begin{itemize}
   \item
    het eindproduct is een samenstelling van bestaande (door derden ontwikkelde)
    componenten en zelf-ontwikkelde componenten. Hierbij is een component niet
    een bibliotheek, maar een zelfstandig executerende (deel)applicatie,
   \item
    software componenten communiceren met elkaar middels vastgelegde afspraken
    (protocol),
   \item
    de software componenten zijn niet in dezelfde programmeertaal ontwikkeld,
   \item
    het systeem draait op een netwerk: het aantal processoren waarvan een systeem
    gebruikt maakt kan gedurende de levensduur van een systeem vari\"eren,
    evenals de kenmerken van deze processoren (homogene versus heterogene netwerken,
    processor-capaciteit, snelheid van netwerkverbindingen),
   \item
    software is veel gevoeliger voor timing eigenschappen vanwege het intensieve en
    niet altijd duidelijk aanwijsbare gebruik van inter-processor-communicatie.
    Dit leidt tot een aanzienlijke toename in de complexiteit in het redeneren over en
    testen en debuggen van gedistribueerde software systemen.
   \end{itemize}

   \item In dit vak leren studenten zelf gedistribueerde software systemen te ontwikkelen
   en maken kennis met een aantal bestaande protocollen om de componenten met elkaar
   te laten communiceren.
   Deze protocollen zullen op een relatief laag niveau staan (bijvoorbeeld TCP/IP) of
   worden zelf ontwikkeld om een goed inzicht te krijgen (en in aanraking te komen met)
   de valkuilen die optreden bij het ontwikkelen van gedistribueerde software systemen.

   \item De software systemen die in dit vak gemaakt worden zijn met de programmeertaal Java
   gemaakt. De benodigde kennis van Java wordt in dit vak aangeboden. De student wordt
   geacht de inleidende programmeervakken succesvol afgerond te hebben.
   Dit vak bereidt voor op het vak ``Integratie van Softwaresystemen''.
}{
  \item
   Op te leveren gedistribueerde applicaties in kaart kunnen brengen, en een ontwerp
   hiervoor kunnen maken.
  \item
   Kleine gedistribueerde software systemen zelfstandig kunnen realiseren in de
   programmeertaal Java.
  \item
   Kunnen bepalen welke technologie\"en en architecturen in welke situaties gebruikt
   dienen te worden.
  \item
   Op illustratief niveau in staat zijn om toonaangevende gedistribueerde
   technologie\"en te gebruiken.
}

\Vak{ICT Infrastructuren}{3 ECTS}{Herfst}{\AfdITT}{-}{
   2004 --
}{
   \Herschikking{
   - Inleiding Computer Architectuur
   }
}{
   \item Om een volwaardige gesprekspartner te zijn voor informatici moet een
   informatiekundige de ontwikkelingen in de ICT met voldoende kennis van zaken
   kunnen volgen om ze naar waarde te kunnen schatten.

   \item Dit vereist een achtergrondkennis van bijvoorbeeld
   computerarchitectuur, besturingssystemen,
   datacommunicatie en netwerken. Dit vak is een eerste introductie tot deze
   gebieden.

   \Onderwerpen{
      Computers van binnen, machinetaal, operating systems, netwerken,
      protocollen.
   }
}{
   \item Na het volgen van dit vak kunnen de studenten in abstracte termen
   aangeven hoe een computer werkt.
   \item Tevens zijn ze in staat in te schatten welke problemen er optreden (en
   hoe men deze kan oplossen) als men meerdere computers met elkaar wil laten
   samenwerken.
}

\Vak{ICT Management}{3 ECTS}{Lente}{\NIII}{Paul Frederiks}{2003 --}%
{\Nieuw}{
  \item strategie ontwikkeling,
  \item Business Balanced Score Card (BBSC),
  \item kwaliteit(systemen) en audits,
  \item grondslagen van beheer,
  \item onderdelen van beheer,
  \item ITIL processen,
  \item Service Level Agreements (SLA),
  \item beveiliging en risicobeheersing
}{
  \item kennismaking met management van ICT organisaties,
  \item inzicht verschaffen in beheerorganisaties en beheerprocessen
}

\Vak{ICT \& Samenleving 1}{3 ECTS (was 2 sp.)}{Lente}{\FNWI}{-}{
   2001 --
}{
   \Bestaand
}{
   \item Historische, filosofische en ethische aspecten van de Informatica \&
   Informatiekunde.
   \item De menselijke maat; de positie van de mens in de digitale samenleving.
   \item Zijn we als mens slaaf geworden van de digitale technologie?
   Misschien niet fysiek; maar dan wel mentaal?
}{
   \item Herkennen van, analyseren van en redeneren over vraagstukken rond de
   gevolgen van de Digitale Revolutie voor wetenschappelijk en menselijk
   handelen
}

\Vak{ICT \& Samenleving 2}{3 ECTS (was 2 sp.)}{Lente}{\FNWI}{-}{
   2003 --
}{
   \Bestaand
}{
   \item Historische, filosofische en ethische aspecten van de Informatica \&
   Informatiekunde.
   \item De ethiek van de architect van de digitale wereld.
   \item De menselijke maat; de positie van de mens in de digitale samenleving.
}{
   \item Herkennen van, analyseren van en redeneren over vraagstukken rond de
   gevolgen van de Digitale Revolutie voor wetenschappelijk en menselijk
   handelen
}

\Vak{Informatie- \& Communicatietheorie}{6 ECTS}{Lente}{\AfdGS}{Dick
van Leijenhorst}{
   2004 --
}{
   Nieuw vak
}{
   \item Kennismaking met theorie\"en tav communicatie en informatie.  De
   cursus geeft een overzicht van theoretische aspecten van
   communicatiekanalen. Daarnaast wordt een aanzet gegeven tot een formeel
   model voor communicatie.
   \item Informatietheorie van Shannon.
   Met name gaat het om:
   \begin{itemize}
     \item Wat is informatie? Kolmogorow tegenover statistisch.
     \item Een stuk statistische ondergrond van de informatietheorie.
     \item De drie hoofdonderwerpen in de concrete Shannon-informatietheorie:
     \begin{enumerate}
        \item Cryptografie
        \item Errorcorrectie
        \item Datacompressie
     \end{enumerate}
     Van elk een behandeling in concrete voorbeelden plus een bespreking van de
     Shannon hoofdstellingen.
   \end{itemize}
   \item Communicatietheorie op basis van Minzberg, Winograd, Flores, Searle en
   Habermas. Uitgaande van cognitieve modellen van communicatie, wordt een
   overzicht gegeven van de diverse aspecten die onderdeel uit moeten maken van
   een formele theorie over communicatie.
}{
   \item Kunnen aanwenden van inzichten uit de informatietheorie van Shannon.
   \item Kunnen aanwenden van inzichten uit communicatietheorie\"en,
   zoals speech-act theory van Habermas.
}

\Vak{Informatiearchitectuur}{6 ECTS}{Herfst}{\AfdIRIS}{Daan Rijsenbrij}{
   2003 --
}{
   \Nieuw
}{
   \item De kwaliteit van de informatievoorziening is voor moderne organisaties
   een succesfactor van belang. In veel gevallen is die voorziening het
   resultaat van ongeleide 'organische' groei. Vaak is dan tegelijk met de
   toename van het aantal technische en informatorische voorzieningen het
   geheel uitgegroeid tot een onontwarbare kluwen digitale onderdelen.
   \item De kosten zijn toegenomen en de inflexibiliteit eveneens terwijl het
   reactievermogen van de organisatie juist groter moet worden en daardoor
   extra eisen stelt aan de souplesse van de informatievoorziening.
   \item In dit vak wordt deze situatie nader geanalyseerd en ontleed, en worden
   concepten ontwikkeld om de structurele veroudering en verstarring tegen
   te gaan.
   \item Derhalve wordt veel aandacht besteed aan 'functies en constructies'
   van de informatievoorziening en aan beheersing van de complexiteit.
   \item Architectuur en informatie-infrastructuur zijn daarbij belangrijke begrippen.
}{
   \item (Kennis) Een eerste kennismaking met:
   \begin{itemize}
      \item het concept architectuur en enkele van de definities die hieraan
      gegeven worden, met nadruk op de rol van architectuur als communicatie-
      en onderhandelingsmiddel.
      \item de rol van digitale architectuur in ondernemingen en organisaties
      mbt het bereiken van een betere afstemming tussen de bedrijfsvoering en
      de ICT (business-ICT alignment).
      \item adaptiviteitscriteria die organisaties in staat stellen om beter in
      te spelen op de immer veranderende, nauwelijks voorspelbare omgeving.
      belangrijke raamwerken en patronen met betrekking tot architectuur, zoals
      service ori\"entatie, etc.
   \end{itemize}
   \item (Cognitieve vaardigheden) In staat zijn om:
   \begin{itemize}
      \item in te schatten welke impact architectuurkeuzes hebben.
      \item relevante wetenschappelijke/professionele literatuur op het gebied
      van architectuur te duiden en de relevantie ervan te bepalen voor een
      gegeven probleemsituatie.
   \end{itemize}
   \item (Professionele vaardigheden) In staat zijn om in een praktische situatie:
   \begin{itemize}
      \item de noodzaak te duiden van een goede afstemming tussen
      bedrijfsvoering en ICT.
      \item te beredeneren welke adaptiviteitscriteria relevant zouden kunnen
      zijn.
      \item een eerste inschatting te maken welke raamwerken en patronen met
      betrekking tot architectuur relevant zijn.
   \end{itemize}
}

\Vak{Informatieverzorging}{3 ECTS}{Herfst}{\NIII }{
  Paul Frederiks}{2003 --}{\Nieuw}{
  \item organisatiedoelen en beleid
  \item hoofdstromen en processen,
  \item informatie als object van management,
  \item informatiestrategie en -planning,
  \item kwaliteit en informatiesystemen,
  \item doelmatigheid bij informatiesystemen
}{
  \item bewustwording cre\"eren voor de waarde van informatie,
  \item belang van informatiestrategie en -planning
}

\Vak{Informatiesystemen}{6 ECTS}{Herfst}{\AfdIRIS}{Patrick van Bommel}{
   2004 --
}{
   \Herschikking{
   - Informatiesystemen in hun context\\
   - Functioneel Specificeren\\
   - Conceptueel Modelleren\\
   - Informatiesystemen 1\\
   - Informatiesystemen 2
   }
}{
   \item Dit vak richt zich primair op
   \begin{itemize}
     \item het aanleren van modelleervaardigheden (werkwijze)
           voor de ontwikkeling van informatiesystemen,
     \item de formele betekenis van de opgeleverde modellen
   \end{itemize}

   \item Hierbij worden een aantal modelleermethoden gebruikt om vanuit
   statisch en dynamisch perspectief (1) de context van een
informatiesysteem,
   (2) het informatiesysteem zelf, en (3) het geautomatiseerde deel van
het
   informatiesysteem te modelleren.

   \item De behandelde methoden, geselecteerd op basis van de kwaliteit
van hun
   gedocumenteerde werkwijze, zijn ORM (statische aspecten) en Testbed
   (dynamische aspecten). Tijdens de colleges zal een brug geslagen
worden naar
   andere notaties, waaronder ER en UML, en zal de relatie met
Object-Oriented
   modellen besproken worden.

   \item Om goed te modelleren is het niet alleen nodig om te kunnen
gaan met
   de syntactische constructen. Een goed begrip van de achterliggende
formele
   semantiek is ook noodzakelijk. Daarom wordt ook aandacht besteed aan
de
   formele betekenis van de diverse modellen.

   \item De informatiekunde variant van dit van besteed extra aandacht
aan
   modelleren en validatie, terwijl de informaticavariant van dit vak
juist
   extra aandacht geeft aan technische aspecten, zoals implementatie en
   verificatie.
}{
   \item Versterking van modelleervaardigheden voor de ontwikkeling van
   informatiesystemen.
   \item Begrip hebben van de formele semantiek van informatiemodellen
en deze
   kunnen terugvertalen naar praktische situaties.
   \item De toepassing van geavanceerde relationele algebra in de
context van
    ORM.
   \item Verbanden tussen ORM, UML en object-georienteerde modellen
   begrijpen.
}

\Vak{Integratie van Softwaresystemen}{6 ECTS}{Lente}{\AfdST}{Marko van Eekelen}{
   2003 --
}{
   \Herschikking{
   - Software technologie 2\\
   - Ontwikkeling van grote softwaresystemen
   }
}{
\item
In dit vak staat de architectuur, ontwikkeling, en onderhoud van
grote software systemen centraal. Kenmerkend van een dergelijk
systeem is dat deze bestaat uit een integratie van een groot aantal
samenwerkende componenten die op verschillende periodes geprogrammeerd
zijn door verschillende mensen met verschillende doelstellingen en
in verschillende programmeertalen.
\item
Deze problematiek wordt ook wel eens
``Enterprise Application Integration'' genoemd.
Het draait hierbij om het integreren van applicaties teneinde de
bedrijfsprocessen binnen \'e\'en organisatie, maar ook die tussen
meerdere organisaties lopen (bijvoorbeeld klanten en leverancier)
beter te integreren.
\item
Dit vak beschouwt het doel en de noodzaak van
Enterprise Application Integration, en de fundamentele eigenschappen
van de technologische middelen om dit te ondersteunen.
Dit zal concreet gemaakt worden door een aantal specifieke voorbeeld
technologie\"en nader te onderzoeken. De nadruk zal liggen op de
relevante software technologische aspecten.
\Onderwerpen{
CBD (component based development),
de architectuuraspecten van grote systemen, herbruikbaarheid,
inheritance, sharing van data en code, middleware, interfaces,
Corba, Com, dCom, XML, dynamic link libraries, dynamics, J2EE,
client-server architectuur, plug-ins, .Net, mobile code.
}
}{
\item
gebruik makend van een overkoepelend kader de structuur
van grote software systemen in kaart kunnen brengen
\item
op illustratief niveau in staat zijn om gebruik te maken van
toonaangevende technologie\"en voor het ontwikkelen van
grote software systemen
\item
gebruik makend van een overkoepelend kader een
ontwerp voor de integratie van softwaresystemen kunnen maken
\item
gebruik makend van een overkoepelend kader kunnen bepalen
welke technologie\"en en architecturen in welke situaties
gebruikt dienen te worden
\item
een mini-prototype van een groot software systeem
(waarin een aantal van de meest relevante aspecten aan de orde komen)
zelfstandig kunnen realiseren
met behulp van de daarvoor meest geschikte technologie\"en
}

\Vak{Intelligente Systemen}{6 ECTS}{Herfst}{\AfdIRIS}{Peter Lucas}{
   2004 --
}{
   \Nieuw
}{
   \item Dit vak biedt een inleiding op de kunstmatige intelligentie,
   in het bijzonder de onderwerpen probleemoplossers, kennisrepresentatie,
   automatisch redeneren, kennisacquisitie en knowledge engineering.
   \item Wat is kunstmatige intelligentie?
   \item Geschiedenis: GPS, MYCIN, DENDRAL, ATP.
   \item Probleemoplossen en zoeken in toestandsruimten:
   uitputtend zoeken, backtracking, heuristisch zoeken op basis van A*,
   hill climbing, tabu search, simulated annealing.
   \item Kennisrepresentatie en automatisch redeneren: produktieregels en
   redeneren, logica en resolutie, Herbrand universum en basis, object
   representatie in de AI, redeneren met onzekere kennis, produktieregels en
   onzekerheid, Bayesiaanse netwerken.
   \item Knowledge engineering: methodology van de ontwikkeling van
   kennissystemen, machinaal leren.
   \item AI programmeren in een daarvoor geschikte taal.
}{
   \item Kennismaken met de belangrijkste onderwerpen in de kunstmatige
   intelligentie (AI), zoals probleemoplossen, toestandsruimte, heuristisch
   zoeken, kennisrepresentatie en automatisch redeneren, machinaal leren,
   intelligent agent, kennisacquisitie, knowledge engineering.
   \item Inzicht opdoen in de toepassing van algoritmische, logische en
   wiskundige methoden in de kunstmatige intelligentie.
   \item Ervaring opdoen met de ontwikkeling van een kennissysteem voor een
   concreet domein.
   \item Ervaring opdoen in de ontwikkeling van AI- programma's.
}

\Vak{Introductie CEM}{6 ECTS}{Herfst}{\FNWI}{-}{
   2002 --
}{
   \Bestaand
}{
   \item Communicatie- en leertheorie\"en;
   \item Management en organisatie (capita selecta);
   \item Effectief schrijven;
   \item Het houden van een voordracht;
   \item Strategie en management van Research en Development;
   \item Project- \& procesmanagement;
   \item Organiseren van leerprocessen in interactie;
   \item Leren probleemoplossen
}{
   \item Inzicht in een aantal relevante communicatietheorie\"en, leertheorie\"en
   en in een aantal concepten en instrumenten op het gebied van management en
   organisatie (procesmanagement, projectmanagement).
   \item Toepassen van een aantal communicatieve vaardigheden om kennis zowel
   mondeling als schriftelijk over te kunnen brengen in multidisciplinaire
   samenwerkingssituaties.
   \item Toepassen van een aantal managementvaardigheden gericht op het
   multidisciplinair samenwerken (onderhandelen, conflicthantering).
   \item Reflecteren op hun in praktijk gebrachte vaardigheden.
   \item Ervaring met en inzicht in een aantal aspecten van het toekomstig
   beroepsmatig functioneren en in hun mogelijkheden zich in deze
   beroepssituaties verder te ontwikkelen.
 }

\Vak{Introductie Informatica \& Informatiekunde}{3 ECTS}{Herfst}{\NIII}{-}{
   2001 --
}{
   \Onderhoud{
   - Uitbreiding informatica-deel met programmeertalen, programmeren en\\
   \mbox{~~~}gegevensopslag.\\
   - Uitbreiding informatiekunde-deel met architectuuraspecten.\\
   - Actualisering om aan te sluiten op nieuwe vakken.
   }
}{
   \item Een kennismaking met de vakgebieden Informatica en Informatiekunde aan
   de hand van het stapsgewijs ontleden en bekijken van een relevante
   probleemstelling.

   \item Tevens is dit vak bedoeld om, alvorens je studie `echt' aanvangt,
   je keuze voor informatiekunde of informatica nog even voor jezelf
   te herbevestigen. Dit is de laatste kans om zonder extra studietijd
   te schakelen tussen beide studies.

   \item Verder maak je kennis met de NIII-onderwijsomgeving. Zowel de
   infrastructuur als de mensen die je daarin tegenkomt (docenten,
   co\"ordinatoren,etc.).

   \Onderwerpen{
      netwerken, communicatie, protocollen, representatie en opslag,
      programmeren, natuurlijke versus kunstmatige taal, inbedding van
      softwaresystemen in organisaties, mens-machine interactie,
      organisatie van systeemontwikkelingsprojecten, besturingssystemen,
      specificaties, logische schakelingen.
   }
}{
   \item De karakteristieke fenomenen in de Informatica en Informatiekunde
   herkennen en verklaren, en wetenschappelijke onderzoeksvragen eraan
   relateren;
   \item Deze fenomenen plaatsen aan de hand van stapsgewijze ontrafeling van
   complexe technische (en organisatorische) systemen tot elementaire
   onderdelen;
   \item Deze fenomenen plaatsen aan de hand van casu\"{\i}stiek van
   systeemontwikkelingsprojecten;
   \item De wisselwerking tussen informatica en informatiekunde expliciteren
   \item Elementaire constructies en analyses uitvoeren (en op basaal niveau
   verantwoorden).
}

\Vak{Kansrekening}{3 ECTS}{Herfst}{\NIII}{-}{
   2004 --
}{
   \Bestaand
}{
   \item Dit vak heeft als doel een inleiding te geven in de kansrekening. De
   kansrekening richt zich op de bestudering van verschijnselen en processen
   waarvan het verloop (of de afloop) niet van tevoren te voorspelen is, maar als
   het ware door ``toeval'' bepaald wordt.
   \item In dit college laten we zien hoe de begrippen toeval en kans wiskundig
   geformaliseerd worden, zodat een wiskundige analyse mogelijk is van
   problemen waarin het toeval een belangrijke rol speelt.
   \item Denk hierbij aan toepassingen zoals: data-mining, expert systemen,
   probabilistische algoritmes, prestatie van netwerken, genetica, herkennen van
   beelden, het voorspellen van de aandeelkoersen, populatie-groei enz.
   \Onderwerpen{
      vaasmodel, algemeen model, voorwaardelijke kansen, verwachting, binomiale
      verdeling, Poissonverdeling, normale verdeling
   }
}{
   \item In praktische situaties herkennen van structuren uit de kanstheorie.
   \item Berekenen van kansen met combinatorische middelen.
   \item Hanteren van kanstheoretische begrippen rond kansruimten, stochasten en
   kansverdelingen.
}

\ExternVak{Kennis \& Informatiemanagement}{4 \'a 6 ECTS}{Herfst}{
   2002 --
}{
  \item Kennismanagement (NSM.BCU244)
}

\Vak{Kwaliteit van Informatiesystemen}{6 ECTS}{-}{\AfdST}{Jan Tretmans}{
   2003 --
}{
   \Nieuw
}{
   \item Dat nieuw gebouwde informatiesystemen goed moeten zijn behoeft geen
   discussie. Maar wat is goed? Voldoen aan de functionele requirements
   is allang niet goed meer. De non-functionele requirements zijn soms
   zelfs belangrijker dan de functionele requirements.

   \item Hoe kun je de kwaliteit van een systeem (softwaresysteem of informatiesysteem)
   benoemen en expliciet maken? Hoe kun je deze kwaliteit testen? Kan dat al
   tijdens de ontwikkeling van het systeem? Dit zijn typische vragen waar in
   het kader van dit van specifieker naar gekeken zal worden.
}{
   \item Verschillende kwaliteitsattributen waar een informatiesysteem aan moet voldoen
   kunnen benoemen.
   \item Kunnen redeneren over de mogelijke impact van kwaliteitsattributen op het
   ontwerp van een informatiesysteem.
   \item Een strategie kunnen opstellen om te bepalen (testen) of een informatiesysteem
   voldoet aan de vooraf gestelde kwaliteitseisen.
}

\Vak{Mens-Machine Interactie}{3 ECTS}{Lente}{\AfdIRIS}{Gert Veldhuijzen
van Zanten}{
   2003 --
}{
   \Nieuw
}{
  \item The purpose of this course is to learn to design usable user-interfaces.
\item Human-machine interaction is concerned with
      the performance of tasks by both humans and machines,
      the structure of the communication between human and machine,
      human capabilities to use machines (including the learnability of interfaces),
      algorithms and programming of the interface,
      engineering concerns in designing and building interfaces, and
      the process of specification, design and implementation of interfaces.

    \Onderwerpen{
        Human capabilities, human problem solving, interaction design,
        task analysis and modeling, usability testing.
    }
}{
   \item After successful completion of the course, students will be able
   \begin{itemize}
     \item to design and construct user-interfaces that support
           efficient and effective usage, user satisfaction, fun and ease of use,
     \item to assess the design of user-interfaces, recognize fallacies and
     explain why users find certain systems hard to use, and
     \item to critically evaluate product designs and prototypes with regard to learnability,
         ease of use, and user capabilities and limitations.
   \end{itemize}

 \item Furthermore, students will be familiar with
   \begin{itemize}
     \item the basic ergonomic principles, such as consistency,
           conformance to (implicit) standards, and the concept of memory load,
     \item some theories of perception and cognition as a foundation for
           human capabilities and limitations, and with
       \item some aspects of information and communication technology
         relevant for user-interface design.
   \end{itemize}
}

\Vak{Onderhandelen \& veranderen}{6 ECTS}{Lente}{\NIII}{Erik
Proper}{
  2004 --
}{
  \Nieuw
}{
  \item In het informatiekunde vakgebied speelt onderhandelen met de verschillende
  belanghebbenden van een nieuw informatiesysteem een belangrijke rol.
  Hierbij gaat het dan in eerste instantie niet zozeer om het
  behartigen van een belang van de informatiekundige \emph{zelf}, maar veeleer om het
  adequaat kunnen inschatten en behartigen van de belangen van verschillende
  \emph{stakeholders}, en deze optimaal op elkaar af te stemmen middels een vertaling
  naar een adequaat ontwerp.

  \item De invoering van een ICT-product in een organisatie brengt veranderingen te
  weeg. Veranderingen in een organisatie, de technologie, de informatievoorziening,
  maar ook veranderingen voor de mensen die met het product moeten werken.
  Veranderingsprocessen in mens,
  organisatie en technologie hebben hun eigen dynamiek, die vaak verschillend
  is, en kunnen met weerstand gepaard gaan. Met deze factoren dient met name
  bij de invoering van een product rekening gehouden te worden.

  \item De informatiekundige zal derhalve met veel verschillende partijen in
  contact komen en hierbij een belangrijke brugfunctie moeten vervullen door
  met nieuwe openingen te komen. Voor het vervullen van deze rol dient hij over
  goede onderhandelings-vaardigheden te beschikken. Daarnaast moet hij in staat
  zijn een goede inschatting te maken van de factoren die bij de verschillende
  stakeholders meespelen.

  \item In dit vak maken studenten kennis met de theorie en praktijk van
  verschillende aspecten van het onderhandelingsproces. Onderwerpen die aan bod
  komen zijn o.a.\ de machtsbalans, mimima en maxima, het verschil tussen
  standpunten en belangen, het ontwikkelen van flexibiliteit.

  \item De inhoud van dit vak heeft een sterke binding met lopend onderzoek binnen
  net \NIII\ op het gebied van de informatiekunde. Op basis van de uitkomsten van
  dit onderzoek zal de inhoud van dit vak ook ge-actualiseerd worden. Met name op
  het gebied van de formele theoretische onderbouwing van onderhandelingsprocessen
  en hun invloed op het ontwerp van zowel het systeem als de daarvoor benodigde
  invoeringsstrategie.
}{
  \item Na afloop is een student in staat om in een concrete situatie te
  redeneren over het onderhandelings- en veranderingsdomein.

  \item Na afloop is een student in staat verschillende
  onderhandelingsstrategie\"en te duiden, en de toepasbaarheid in een gegeven
  onderhandelingssituatie te beargumenten.

  \item Na afloop is een student in staat om in afgebakende situaties
  onderhandelingen te voeren tussen verschillende stakeholders van een
  te ontwikkelen en in te voeren systeem.
}

\Vak{Onderzoekslaboratorium}{6 ECTS}{Herfst}{\AfdITT}{Jozef Hooman}{
   2001 --
}{
   \Bestaand
}{
   \item In dit vak komt naast het zelf plannen en uitvoeren van onderzoek ook het
   begeleiden van medestudenten aan bod.
   \item Ook het leiden van voordrachten en onderzoeksbijeenkomsten speelt nu
   een rol.
   \item Tenslotte is er ook aandacht voor het beoordelen van het onderzoek van
   anderen.
}{
   \item Het kunnen plannen en uitvoeren van onderzoek.
   \item Het begeleiden van medestudenten bij het onderzoek.
   \item Het leiden van een voordracht met discussie.
   \item Het leiden van een onderzoeksbijeenkomst.
   \item Het beoordelen van onderzoek van medestudenten.
}

\Vak{Onderzoeksvaardigheden}{3 ECTS}{Herfst}{\NIII}{-}{
   2002 --
}{
   \Bestaand
}{
   \item In dit worden doorstroomstudenten ingeleid in de manier waarop academisch onderzoek
   wordt verricht binnen de context van de Informatiekunde. Dit wordt gekoppeld aan een klein
   eigen onderzoek waarin de onderdelen van zo'n onderzoek aan bod komen.
   \item Het doel van dit vak is dat studenten inzicht hebben in de academisch onderzoek, en
   dit hebben getoond door succesvol een onderzoek volgens de regels der kunst uit te voeren.
}{
   \item Hoe je het probleemgebied afbakent,
   \item Hoe je de probleemdefinitie opstelt: wat is het probleem; wie heeft het probleem,
   \item Hoe je een academisch onderzoek op moet zetten,
   \item Hoe je een onderzoeksvraag moet formuleren,
   \item Hoe je de juiste informatiebronnen selecteert en kritisch bestudeert (onder andere
   het lezen van wetenschappelijke publicaties),
   \item Hoe je werk van anderen kritisch beoordeelt,
   \item Hoe je de juiste onderzoeksmethoden daarbij selecteert,
   \item Hoe deze onderzoeksmethode moet worden uitgevoerd,
   \item Hoe je de onderzoeksresultaten opschrijft en presenteert,
   \item Hoe je een en ander vertaalt naar een planning.
}

\Vak{Opslaan \& Terugvinden}{6 ECTS}{Herfst}{\AfdIRIS}{Franc Grootjen}{
   2004 ---
}{
   \Herschikking{
   - Informatiesystemen in hun Context\\
   - Functioneel Specifiseren\\
   - Conceptueel Modelleren
   }
}{
   \item Dit vak beoogt studenten kennis te laten kennis met een breed scala aan
   technieken en hun achterliggende theorie, voor het opslaan en ontsluiten van
   informatie.
   \item Aan bod zullen komen: basis opslagtechnieken (B-bomen, hashing), het
   indexed-sequential bestand, de tabel, de geneste tabel, SQL databases,
   Object-Oriented databases, XML gebaseerde opslag, diverse querytalen, opslag
   van ongestructureerde informatie, tools om exploratief te zoeken in grote
   hoeveelheden informatie, etc.
   \item Merk op: er zijn drie bachelor vakken die aspecten van SQL belichten
   de docenten van deze vakken zullen dit onderling nader afstemmen. Het gaat
   hierbij om: ``Beweren \& Bewijzen'', ``Domain Modelleren'' en
   ``Opslaan \& Terugvinden''.
}{
   \item Na afloop van dit vak kan een student verschillende technieken voor
   de opslag en ontsluiting van gegevens duiden.
   \item Na afloop is een student tevens in staat om voor een gegeven situatie een
   beargumenteerde keuze maken.
   \item Verder zijn studenten in staat om voor enkele technieken programma's
   te schrijven waarin deze worden toegepast.
}

\ExternVak{Organisatiekunde}{6 ECTS}{Herfst}{
   2003 --
}{
   \item Inleiding in de organisatietheorie (NSM.BIN114)
   \item Organisatie en Informatievoorziening (MIK)
   \item Human resource management en organisational behaviour (NSM.BCU202)
}

\Vak{Orientatiecollege Toepassingen}{3}{Herst}{\NIII}{Janos Sarbo}{
   2001 --
}{
   \Bestaand
}{
   \item In dit college maak je kennis met de toepassingsgebieden van de
   Informatiekunde waar je je in het vervolg van de opleiding in kunt
   specialiseren. Diverse toepassingsgebieden passeren de revue, zoals Medische
   Informatiekunde, Taal- en Spraaktechnologie, Juridisch Kennisbeheer,
   Informatiemanagement e.a.
   \item Aan de hand van verschillende praktijkvoorbeelden krijg je een beeld van het
   ruime en gevarieerde vakgebied waar jij als toekomstig Informatiekundige je
   werk zult kunnen vinden.
   \item Er zal met name ruim aandacht gegeven worden aan een orientatie op de
   specialisatierichtingen uit de masterfase.
   \item Het college wordt mede verzorgd door gastdocenten
   uit het bedrijfsleven; indien mogelijk zullen ook excursies of studiebezoeken
   worden gepland.
   \item De inhoud van het college kan aan de actualiteit worden aangepast.
   Maar: resultaten behaald in het verleden bieden geen garantie voor de
   toekomst.
}{
   \item Studenten kunnen na afloop een, vanuit hun eigen interesses,
   beargumenteerd keuze maken voor het kiezen van een invulling van
   de toepassingsvakken in het tweede en derde jaar van de bachelorfase.
   \item Studenten kunnen na afloop reflecteren over typische informatiekundige
   aspecten van diverse toepassingsgebieden; zowel vanuit theoretisch als
   praktisch perspectief.
}

\Vak{R\&D 1}{6 ECTS}{Lente}{\NIII}{Docenten uit het eerste jaar}{
   2004 --
}{
   \Onderhoud{
   - Vergroten van het huidige integratieproject tot 6 ECTS.\\
   - Van blok-vak tot semester vak.\\
   - Focus op ontwikkelproject.\\
   - Combineren van praktisch werk ihkv de overige vaksen.
   }
}{
   \item Voor een wat groter probleem worden vaardigheden uit verschillende
   vakken uit het eerste jaar ingezet.
   \item Voor dit probleem wordt een compleet systeemontwikkeltraject doorlopen
   beginnend bij specificeren van de verlangde eigenschappen tot en met het
   documenteren van de geproduceerde oplossing. Het gaat hierbij om een systeem
   waarin menselijke, organisatorische \emph{en} technologische aspecten een
   rol spelen.
   \item Indien noodzakelijk voor het welslagen van het project kan er enig
   aanvullend onderwijs verzorgd worden.
   \item Daarnaast zal in dit vak ook expliciet aandacht besteed worden aan het
   aanleren van vaardigheden op het gebied van communicatie en presentatie.
}{
   \item Methoden en technieken uit het eerste jaar combineren in een
   ontwikkelproject.
   \item In projectvorm werken volgens gegeven richtlijnen.
   \item In een groter team (circa 4 studenten) werken.
   \item De resultaten schriftelijk en mondeling presenteren.
}

\Vak{R\&D 2}{6 ECTS}{Lente}{\NIII}{Docenten uit het tweede jaar}{
   2005 --
}{
   \Nieuw
}{
   \item In dit vak worden de verschillende algemene en inhoudelijke
   vaardigheden in de afzonderlijke vaksen uit de eerste twee jaar expliciet
   samengebracht. Belangrijk onderdeel hiervan zijn ook de gekozen
   toepassingsgebieden.
   \item In dit vak ga je in teamverband een concrete
   systeemontwikkelingsbehoefte van een klant volledig in kaart brengen, de
   behoeften analyseren, onderzoek doen naar de achtergronden, een ontwerp
   maken en deze implementeren.  Hierbij zullen de informatiekunde studenten
   een opdracht uitvoeren binnen het door hun gekozen toepassingsgebied.
   \item Je zult bij het oplossen moeten herkennen of toepassing van een of
   meer van de verschillende technieken in de diverse vaksen mogelijk,
   zinvol en effectief is.
   \item Verder leer je gestructureerd projectmatig werken en het presenteren
   van je resultaten.
}{
   \item Methoden en technieken uit de eerste twee jaar combineren voor het
   oplossen van een groter vraagstuk binnen het door de student gekozen
   toepassingsgebied.
   \item In projectvorm werken.
   \item In een klein team (2-3 personen) werken.
   \item Reflecteren op het product volgens zelf geformuleerde kwaliteitscriteria.
   \item Reflecteren op het proces volgens gegeven criteria; uitvoeren van
   `peer assessment'.
   \item De resultaten schriftelijk en mondeling presenteren.
   \item Reflectie over toepassingsgebieden.
}

\Vak{Requirements Engineering}{6 ECTS}{Lente}{\AfdIRIS}{Stijn Hoppenbrouwers}{
   2004 --
}{
   \Nieuw
}{
   \item Een belangrijk deelproces van systeemontwikkeling is het goed helder
   krijgen van de behoeften. Dit is het ``requirements engineering'' proces.
   Het belangrijkste doel van dit proces is er voor te zorgen dat een eventueel
   te ontwikkelen systeem ook het goede systeem is. Centraal in dit proces
   staat de vergaring, specificatie, validatie en de bewaking van de
   systeemeisen.
   \item In dit vak leren de studenten hoe ze moeten komen tot een goede
   requirementsspecificatie, hoe ze die moeten valideren, en hoe ze deze bij de
   daadwerkelijke realisatie en invoering van een systeem kunnen bewaken.
}{
   \item De student is in staat in een gegeven organisatorische context een
   informatievoorzienings-probleem te analyseren en samen met de
   belanghebbenden in die organisatie een programma van eisen voor een nieuw
   (in te voeren of te wijzigen) informatiesysteem(deel) opstellen.
   \item De student is in staat om op basis van enerzijds de eisen mbt. een
   informatiesysteem en anderzijds een gereed produkt (elders gebouwd maatwerk
   informatiesysteem danwel een 'getuned' multipurpose softwarepakket) de
   bruikbaarheid van dat produkt aan te beoordelen
   en te beargumenteren.
}

\Vak{Lerende \& Redenerende Systemen}{6 ECTS}{Lente}{\AfdIRIS}{Theo van der Weide}{
   2006 --
}{
   \Nieuw
}{
   \item In dit vak zullen onderwerpen uit de computationele intelligentie,
   zoals representaties bij automatisch leren (beslisbomen,
   classificatieregels, neurale netwerken) aan de orde komen, naast onderwerpen
   uit de information retrieval met betrekking tot kennisextractie uit concrete
   bronnen (text mining, concepttralies, extentionele semantiek).

   \item Daarnaast zal aandacht besteed worden aan probleemklasse-specifieke
   redeneervormen, zoals case-based retrieval and reasoning, model-gebaseerd
   redeneren, spatieel redeneren, en redeneren met ontwerpspecificaties.
}{
   \item In staat om theorie\"en uit de computationele intelligentie en information
   retrieval te be\"oordelen op hun toepasbaarheid in een concrete situatie.
   \item Redeneren over basiseigenschappen van deze theorie\"en.
   \item Te duiden wat de plaats van mens \& organisatie is ten opzichte van
   deze theorie\"en.
}

\Vak{Security}{3 ECTS}{Herfst}{\AfdITT}{-}{
  2002 --
}{
  \Bestaand
}{
  \item Security is widely recognised as being of great importance in all areas of
  computer science: networks, operating systems, databases etc.
  \item Security is about regulating access to assets. Crucial questions are:
  Who are you? and: Should you be doing that? Authentication (of people and
  computers) and access control are basic aspects of computer security.
  \item Cryptography provides a mathematical toolset that for realising key
  security goals, via appropriate protocols.
  \item The course introduces the basic notions and techniques in the area of
  computer security. It surveys cryptography, but does not go into the
  underlying mathematics.
}{
  \item To understand different security goals and their threats
  \item To be able to find a balance between between technical, organisational
  and legal security measures.
  \item To be able to appreciate the various differences between symmetric
  (secret key) and asymmetric (public key) cryptography
  \item To be able to formalise a security protocol and to analyse it with a
  modelchecker (FDR + Casper)
  \item To be able to use Java's security API
}

\Vak{Security protocols}{3 ECTS}{Herfst}{\AfdITT}{-}{
  2003 --
}{
  \Nieuw
}{
  \item Security is widely recognised as being of great importance in all areas of
  computer science: networks, operating systems, databases etc.
  \item Security is about regulating access to assets. Crucial questions are:
  Who are you? and: Should you be doing that? Authentication (of people and
  computers) and access control are basic aspects of computer security.
  \item Cryptography provides a mathematical toolset that for realising key
  security goals, via appropriate protocols.
  \item The course introduces some advanced notions and techniques in the area of
  computer security. The emphasis lies on security protocols.
  \item Formal methods, like model
  checking will be introduced to assess protocols in order to detect possible
  attacks
}{
  \item To develop a suitable level of paranoia, needed for designing security
  sensitive computer applications;
  \item To learn some basic techniques for evaluating specific security
  designs.
}

\ExternVak{Soft-Systems Methodology}{4 \'a 6 ECTS}{Herfst}{
   2003 --
}{
   \item Sociotechniek (NSM.BCU242)
   \item Systeemtheorie (NSM.??)
}

\ExternVak{Statistiek}{3 ECTS}{Herfst}{
   2004 --
}{
   \item Statistiek 1 (CogW.PSST110)
   \item Statistiek 2 (CogW.PSST210)
}

\Vak{Software Engineering 1}{3 ECTS}{Herfst}{\NIII}{Franc Grootjen}{
   2003 --
}{
   \Onderhoud{Explicieter aandacht voor de rol van informatiekunde.}
}{
   \item GIP en theorie, gericht op de uitvoering van systeemontwikkelprocessen.
   Hierbij dient ook het SO-proces expliciet onderwerp van studie te zijn.
   Vragen zoals ``wanneer een evolutionaire ontwikkelstrategie gebruiken''
   dienen hierbij expliciet aan bod te komen.

   \item Het einddoel van het vak is dat de student wordt opgeleid tot information system
   engineer, met de nadruk op de ontwikkeling van standaard systeem ontwikkelprojecten.

   \item In het GIP zullen de informatiekunde studenten typisch een belangrijke
   rol vervulling bij het bewaken van de belangen van de verschillende
   belanghebbenden. Dit komt niet alleen tot uiting bij het vaststellen
   van de requirements, maar blijft ook bij de verdere systeemontwikkeling
   van cruciaal belang.
}{
   \item Inhoudelijk heeft de student na het volgen van dit vak kennis
   (methoden en technieken) om standaard systeem ontwikkelprojecten als
   software engineer uit te voeren.
   \item Daarnaast leert de student binnen een projectgroep te werken
   (GiP-House).
}

\Vak{Software Engineering 2}{6 ECTS}{Lente}{\NIII}{Franc Grootjen}{
   2004 --
}{
   \Onderhoud{Explicieter aandacht voor de rol van informatiekunde.}
}{
   \item GIP en theorie, gericht op de uitvoering van systeemontwikkelprocessen.
   Hierbij dient ook het SO-proces expliciet onderwerp van studie te zijn.
   Vragen zoals ``wanneer een evolutionaire ontwikkelstrategie gebruiken''
   dienen hierbij expliciet aan bod te komen.

   \item Binnen dit vak wordt de student opgeleid tot senior information
   system engineer. In dit vak gaat de student de ervaring (zoals geleerd in
   ``Software Engineering 1'') inzetten ten behoeve van
   creativiteit. Het gaat nu meestal om innovatieve en/of experimentele
   projecten, waarbij nieuwe diensten en producten gemaakt worden en de laatste
   ontwikkelingen van het vakgebied worden toegepast.

   \item In het GIP zullen de informatiekunde studenten typisch een belangrijke
   rol vervulling bij het bewaken van de belangen van de verschillende
   belanghebbenden. Dit komt niet alleen tot uiting bij het vaststellen
   van de requirements, maar blijft ook bij de verdere systeemontwikkeling
   van cruciaal belang.
}{
   \item Aan het eind van dit vak heeft de student de kennis (methoden en
   technieken) om innovatieve systeem ontwikkelprojecten uit te voeren.
   \item Tevens leert de student om in projectgroepen - als onderdeel van een
   systeem ontwikkel team - innovatieve trajecten uit te voeren.
}

\Vak{Software Engineering 3}{6 ECTS}{Herfst}{\NIII}{Patrick van Bommel}{
   2003 --
}{
   \Onderhoud{Explicieter aandacht voor de rol van informatiekunde.}
}{
   \item GIP en theorie gericht op sturing bij systeemontwikkelprocessen.
   Dit kan gaan om inhoudelijke sturing (informatiearchitect) en om procesmatige
   sturing (projectleider).

   \item In dit vak wordt de student opgeleid tot projectleider of informatiearchitect
   van standaard systeem ontwikkelprojecten. Hierbij is de student betrokken in het
   management van het practicum van ``Software Engineering 1''.

   \item In het GIP zullen de informatiekunde studenten typisch een belangrijke
   rol vervulling bij het bewaken van de belangen van de verschillende
   belanghebbenden. Dit komt niet alleen tot uiting bij het vaststellen
   van de requirements, maar blijft ook bij de verdere systeemontwikkeling
   van cruciaal belang.
}{
   \item Aan het eind van dit vak heeft de student de vaardigheden om
   standaard systeemontwikkel project inhoudelijk of procesmatig te leiden.
}

\Vak{Systeemtheorie: Ontwerp \& Evolutie}{3 ECTS}{Herfst}{\AfdIRIS}{Erik Proper}{
   2004 --
}{
   \Herschikking{Architectuur \& Alignment}
}{
   \item Doel van dit vak is om vanuit systeemtheoretisch perspectief te kijken
   naar systeemontwikkeling en -evolutie, en de belangrijkste processen,
   stakeholders, perspectieven, etc, vanuit dat perspectief te duiden.

   \item Vanuit dit algemene raamwerk worden, ten aanzien van het ontwerp \&
   evolutie, bruggen geslagen vanuit de denkwereld van systemen in het algemeen
   naar informatiesystemen in het bijzonder.
}{
   \item Na afloop zijn studenten in staat om vanuit een systeemtheoretisch kader
   aspecten van specifieke systeemontwikkelingsaanpakken te duiden en te relateren.
   \item Na afloop beheersen de studenten een generieke terminologie ten aanzien van
   (architectuurgedreven) systeemontwikkeling, op basis waarvan men verschillende
   systeemontwikkelingsaanpakken onderling kan relateren.
}

\Vak{Systeemtheorie: Structuur \& Co\"ordinatie}{3 ECTS}{Lente}{\AfdIRIS}{Erik Proper}{
   2004 --
}{
   \Nieuw
}{
   \item Doel van dit vak is studenten bewust te maken van algemene systemische
   patronen tav structuur en co\"ordinatie. Qua inhoud zal dit vak een
   mengeling zijn tussen de wereld van design patterns en systeem theorie.

   \item Er zal in dit vak ook bewust buiten de traditionele ICT context
   gekeken worden.  Denk aan structuren en co\"ordinatiepatronen uit de
   biologische wereld.  In de bouwwereld is het al jarenlang gebruikelijk om
   fysieke patronen uit de biologische wereld te gebruiken voor fysieke
   constructies.

   \item Recentelijk zijn dergelijke ontwikkelingen ook waar te nemen in de
   ICT- en organisatiewereld, waarbij sociale structuren en
   co\"ordinatiepatronen uit de biologische wereld worden gecopi\"eerd.
}{
   \item Na afloop van dit vak is de student bekent met een aantal
   basisbegrippen uit de cybernetica, systeemdynamica en de general systems
   theorie.
   \item Kennis van met relevante systeemtheoretische raamwerken, zoals het
   viable systems model.
   \item Inzicht in de theorie van architectuurpatronen, en de rol die deze
   spelen in de structuur en co\"ordinatie van/in systemen.
   \item In staat zijn om zelfstandig na te denken over het omgaan met
   complexiteit in systemen (en organisaties).
   \item In staat zijn om in een bepaalde practische situatie een relevant
   systeemtheoretisch raamwerk te selecteren en in te gebruiken.
   \item In staat zijn om in een bepaalde practische situatie
   architectuurpatronen te herkennen en toe te passen.
}

\Vak{Visualiseren \& Communiceren}{6 ECTS}{Herfst}{\AfdST}{Susan
Even}{
   2004 --
}{
   \Onderhoud{
   - Verandering van naam. Was: Visualisatie\\
   - Meer aandacht voor de rol van visualisatie in communicatie
   }
}{
  \item Hoe krijg je beelden uit je hersenen?
In deze cursus bestudeer je softwaretechnieken die het mogelijk maken
je innerlijke creatieve beelden vorm te geven op een andere manier
dan met taal.

  \item Je maakt virtuele objecten en virtuele omgevingen met behulp van
een state-of-the-art visualisatie tool voor computer-aided design.

\item Je maakt kennis met verschillende 3D ontwerptechnieken:
polygonal modelling, solid modelling, surface modelling,
organic modelling, booleans, patches, splines, en nurbs.

\item Hierbij oefen je met deze kennis door je eigen objecten en omgevingen
te ontwerpen.

\item We gebruiken ``principles of clean design'':
abstraction, parameterization, composition, application,
parallelism, en multiple channels.
Herbruikbaarheid komt overal voor.
}{
\item
Je kunt 3D vormen herkennen, analyseren, namaken, ver- en hervormen.
\item
Je kunt top-down en bottom-up virtuele dingen ontwerpen en bouwen.
\item
Je kunt systematisch een 3D vorm bestuderen om de constructiemethode
erachter te ontdekken (reverse engineering)
of te verzinnen (forward engineering).
\item
Je kunt verschillende ontwerpmethodes, tools, en technieken toepassen
om eenzelfde vorm te realiseren.
Hier kun je de voor- en nadelen van toelichten.
\item
Je kan een gegeven 3D model kritisch bestuderen
en er verbeteringen in aanbrengen.
\item
Je toont expert-user gedrag: Je bent in staat software manuals
grondig door te lezen om nieuwe combinaties van technieken
bij elkaar te stoppen, je eigen maken,
en vervolgens in te zetten in het ontwerpproces.
\item
Je kunt niet-precieze beschrijvingen precies maken,
door een beschrijving in natuurlijke taal te vertalen naar
een 3D ontwerp. Dit houdt zowel de globale ontwerpstrategie in
als de specifieke constructie-stappen (vormgeving).
\item
Je beseft hoe precisie zich manifesteert als kwaliteit
in het ontwerpproces en je bent in staat precies te werken
indien dat nodig is.
}
}
\Ragged{\chapter{Vakbeschrijvingen van oude vakken}
\label{h:OudeVakken}

Deze appendix bevat de vakbeschrijvingen van de vakken
die met ingang van het curriculum 2003 uitgefaseerd
zullen worden.

De inhoud van de onderstaande vakbeschrijvingen is tot standgekomen
op basis van de informatie zoals deze op het web of in de studiegids
voorhanden was. Spijtig genoeg is de curriculum commissie niet in
staat geweest om de beschrijvingen van alle vakken te achterhalen.

\OudVak{Architectuur \& Alignment}{6 ECTS}{Lente}{\AfdIRIS}{Erik Proper}{
  2002 -- 2005
}{
  Bestaand vak, zal worden opgedeeld over:\\
  - Domeinmodellering\\
  - Systeemtheorie: Ontwerp \& Evolutie
}{
  Onderscheiden van niveaus en manieren waarop een informatiesysteem (of een
  portfolio van informatiesystemen) en zijn context in kaart gebracht kan
  worden (architecturen), waaronder informatiearchitectuur,
  bedrijfsarchitectuur, communicatie-architectuur en kennisarchitectuur, maar
  ook (relaties met) applicatie-architectuur, systeemarchitectuur,
  software-architectuur.

  Specifieke onderwerpen die aan bod komen zijn: Systemen, paradigma's voor het
  vormgeven van systemen, soorten systemen (zoals business, informatie,
  applicatie, etc.), systemic-alignment, kwaliteitseisen voor systemen,
  architectuur, aspect-architecturen, alignment technieken,
  architectuurprincipes.
}{
  \item Kunnen beschrijven van de vele factoren die van invloed zijn op de kwaliteit
  en relevantie van bepaalde architecturen, en het invullen en beoordelen van
  deze factoren voor specifieke situaties.

  \item Kunnen beschrijven van relaties tussen diverse architecturen en de manieren
  waarop deze tot stand gebracht en onderhouden kunnen worden (methodiek), en
  het correct toepassen daarvan.
}

\OudVak{Communicatie 1 (Schriftelijke vaardigheden)}{1 sp}{Herfst}{Extern}{-}{
   2001 -- 2002
}{
   Bestaand vak
}{
   Niet voorhanden
}{
   \item -
}

\OudVak{Communicatie 2 (Mondelinge vaardigheden)}{1 sp}{Lente}{Extern}{-}{
   2002 -- 2003
}{
   Bestaand vak
}{
   Niet voorhanden
}{
   \item -
}

\OudVak{Communicatieve Aspecten van Informatiesystemen}{3 sp}{Lente}{\AfdIRIS}{Stijn Hoppenbrouwers}{
   2001 -- 2003
}{
   Bestaand vak, zal worden opgedeeld over:\\
   - Requirements Engineering\\
   - Informatie \& Communicatie Theorie
}{
   De cursus geeft een overzicht van theoretische en practische aspecten van
   informatiesystemen die te maken hebben met het gebruik van die systemen als
   communicatiekanaal of 'medium'. Het gaat daarbij niet zozeer om 'open'
   gebruik van ICT-gebaseerde communicatiekanalen (zoals audio, video, e-mail)
   maar om meer voorgestructureerde vormen van communicatie: via databases,
   transactiesystemen, electronische formulieren, etc. De nadruk ligt hierbij
   op organisatie-ondersteunende informatiesystemen (kantoorautomatisering). De
   cursus werpt vanuit een taal- en communicatieperspectief licht op het
   (functioneel) ontwerp en gebruik van dergelijke informatiesystemen. Daarbij
   staan de voor- en nadelen van het voorstructureren van talige communicatie
   centraal.
}{
   \item Het kunnen doorzien en analyseren van organisatie-ondersteunende
   informatiesystemen vanuit een taal- en communicatieperspectief
   \item Het kunnen toepassen van theoretische verworvenheden op het gebied
   van taal en communicatie op de analyse van informatiesystemen
   \item Het kunnen aangeven van beperkingen en mogelijkheden m.b.t.
   informatiesystemen als taaldrager/communicatiemedium
   \item Het kunnen laten meewegen van taal-gerelateerde argumentenh bij het
   ontwerpen van informatiesystemen
   \item Het kunnen positioneren van taalgebaseerde methoden en technieken
   als ondersteuning van taal- en communicatie-gerelateerd ontwerp en
   gebruik van informatiesystemen.
}

\OudVak{Conceptueel Modelleren}{6 ECTS (was 4 Sp)}{Lente}{\AfdIRIS}{Franc Grootjen}{
   2001 -- 2004
}{
   Bestaand vak, zal worden opgedeeld over:\\
   - Opslaan \& Terugvinden\\
   - Informatiesystemen
}{
   Om goed te modelleren is het niet alleen nodig goed om te kunnen gaan met de
   syntactische constructen. Een goed begrip van de achterliggende formele
   semantiek van de modellen is ook noodzakelijk. Daarom wordt er in dit vak
   ook aandacht besteed aan de formele betekenis van de diverse modellen.

   Tijdens de analyse en ontwikkeling van software intensieve systemen is het
   cruciaal om (deel) resultaten te kunnen representeren, vastleggen en
   overdragen.  Deze vaardigheid is het hoofdonderwerp van het vak
   ``Functioneel Specificeren''.  De studenten leren zowel verbaal als
   schriftelijk, met tekst of plaatjes, informeel (zonder strenge regels) als
   uiterst formeel (bijvoorbeeld wiskundig) zich uit te drukken.

   Als leidraad wordt tijdens het college een moderne object geori\"enteerde
   specificatie methode (UML) gebruikt om inzicht te krijgen in de
   verschillende facetten van het specificeren.
}{
   \item -
}

\OudVak{Functioneel Specificeren}{6 ECTS (was 4 sp)}{Herfst}{\AfdIRIS}{Erik Proper}{
   2001 -- 2003
}{
   Bestaand vak, zal worden opgedeeld over:\\
   - Domeinmodellering\\
   - Informatiesystemen
}{
   In 2003 zal dit vak alvast gaan opschuiven in de richting van het vak
   ``Informatiesystemen'' zoals dit in 2004 zal worden ingevoerd.

   Dit vak richt zich primair op het aanleren van modelleervaardigheden
   (werkwijze) voor de ontwikkeling van informatiesystemen. Hierbij worden een
   aantal modelleer methoden gebruikt om vanuit statisch en dynamisch
   perspectief de context van een informatiesysteem, een informatiesysteem
   zelf, en het geautomatiseerde deel van een informatiesysteem te modelleren.

   De te gebruiken voorbeeld methoden zijn geselecteerd op basis van de
   kwaliteit van hun gedocumenteerde werkwijze. Dit vertaald zich naar een
   keuze voor ORM (Statische aspecten) en Testbed (Dynamische aspecten).
   Tijdens de colleges zal echter ook een brug geslagen worden naar de UML
   notatie.
}{
   \item -
}

\OudVak{Het Software Ontwikkelproces}{3 sp}{Herfst}{\NIII}{Gert Veldhuijzen van Zanten}{
   2002 -- 2003
}{
   Bestaand vak
}{
   Dit vak is je kennismaking met het studenten software-huis (GIP -house).
   Je vervult daarin zelf een rol en wordt begeleid door managers die SO3/SO4
   (Informatica) of SOW1/SOW2 (Informatiekunde) volgen.

   Het GIP-house beoogt een realistische simulatie van een groot software-huis
   te zijn.

   Specifieke onderwerpen in dit vak zijn:
   software proces modellen, requirements analyse, design, documentatie,
   implementatie.
}{
   \item Samen werken aan grotere softwareprojecten.
   \item Software modellen (her)kennen en gebruiken.
   \item Theoretisch en praktisch inzicht in softwareproces
   \item In open probleemstelling oplossingen herkennen en realiseren.
}{}

\OudVak{Informatiesystemen in hun Context}{4 sp}{Herfst}{\AfdIRIS}{Ger Paulussen}{
   2001 -- 2002
}{
   Bestaand vak
}{
   In dit vak gaan we vooral in op het ontwerp en gebruik van eenvoudige
   informatiesystemen in het kader van organisaties waarin mensen en machines
   samenwerken.
   In het vervolg van de opleiding zul je bij meerdere vakken gebruik maken
   van de bij dit vak opgedane kennis en vaardigheden.

   We richten ons in deze cursus specifiek op:
   \begin{itemize}
      \item specificeren van commando's op
      vierde-generatie-gegevensbank-structuren in een concrete
      vierde-generatie-taal (SQL; Structured Query Language) en realiseren van
      deze commando's op een corresponderend gegevensbanksysteem
      \item specificeren van organisatiestructuren door middel van concrete
      schematechnieken (SADT, ISAC)
      \item gebruiken van een concrete informatie-analyse-methode (ORM; Object
      Role Modeling) voor het specificeren van conceptuele schemata van niet al
      te grote organisaties, tegen het einde van de cursus gebruiken we daarbij
      een informatie-analyse-tool
      \item omzetten van conceptuele schemata naar
      vierde-generatie-gegevensbank-structuren
      \item toepassen van een aantal basisvereisten bij het ontwerpen van een
      GUI (Grafische User Interface)
      \item opstellen van een proces-model voor objecten binnen een organisatie
      \item specificeren van data-structuren voor gebruik op het WWW (World
      Wide Web) via XML
   \end{itemize}
}{
   \item Het maken van een model van een organisatie.
   \item Het uitvoeren van een informatieanalyse op basis van dat model,
   leidende tot een database.
   \item Het kunnen manipuleren maar ook beschermen van de in die database
   opgeslagen gegevens. De nadruk ligt daarbij op het modelleren van gegevens
   en in mindere mate op dat van processen.
}

\OudVak{Inleiding Algemene Fonetiek}{3 sp}{-}{Taalwetenschappen}{H. Strik}{
   2001--2002
}{
   Bestaand vak
}{
   Taal manifesteert zich in geschreven en gesproken vorm. Fonetiek houdt zich
   bezig met verschillende facetten van gesproken taal: spraak. Centraal hierin
   staat de zogenaamde spraakketen: de productie van spraak (spreken), het
   spraaksignaal zelf en zijn eigenschappen (akoestiek), en de perceptie van
   spraak (luisteren).

   De centrale onderwerpen zijn: spraakproductie, spraakperceptie, en
   akoestische analyse en beschrijving van spraak. Deze stof wordt behandeld
   a.d.h.v. de bijbehorende hoofdstukken uit het boek 'Algemene Fonetiek'. Deze
   hoofdstukken uit het boek vormen de tentamenstof. Op het hoorcollege worden
   aanvullende multimediale middelen (beeld, geluid, film, demonstraties, etc.)
   gebruikt om de onderwerpen te illustreren en van extra uitleg te voorzien.
}{
   \item Je bezit kennis over de (experimentele) fonetiek in brede zin, en meer
   specifiek over de verschillende onderdelen van de spraakketen (productie,
   akoestiek en perceptie).
   \item Tevens heb je geleerd dat er een nauwe samenhang is tussen de
   onderdelen van de spraakketen, en wat die samenhang is.

}

\OudVak{Inleiding Cognitiewetenschap}{4 sp}{-}{Cognitiewetenschappen}{Louis Vuurpijl}{
   2001 -- 2003
}{
   Bestaand vak
}{
   Bij het ontwerpen van computer-software dient systematisch rekening te
   worden gehouden met de eigenschappen van perceptuele en cognitieve
   vaardigheden bij de menselijke gebruiker. Over de vraag, hoe dit doel moet
   worden gerealiseerd, is gedurende de laatste decennia een grote hoeveelheid
   onderzoek verricht. Het boek van Dix et al (1998) geeft een heldere
   inleiding over dit vakgebied dat zich op de grens van technologie en
   cognitiewetenschap begeeft.

   De perceptuele en cognitieve vaardigheden van de mens (horen, zien, voelen,
   motoriek, geheugen, redeneren, communicatie) en de faciliteiten en
   beperkingen van de computertechnologie
   (input/output/netwerk/rekenkracht/geheugen) worden in het kader van MMI
   behandeld. Onderwerpen waaraan in dit college verder aandacht wordt besteed:
   interactiestijlen en mens-machine dialogen; een overzicht over succesvolle
   interactieve systemen (WIMP,direct manipulation, multi-modaliteiten, hyper-
   en multi-media, virtual reality, WWW, mobile en ubiquitous computing); het
   ontwerp proces, de software life cycle, usability engineering, design en
   prototyping; informatie- en datavisualisatietechnieken, spraak- en
   pen-gestuurde gebruikersinterfaces.
}{
   \item Volgens Dix is mens-machine interactie: "the study of people, computer
   technology and the ways these influence each other". Het bestuderen van MMI
   aspecten heeft als doel computertechnologie "beter bruikbaar te maken'' voor
   mensen.
   \item De student zal de beperkingen van de mens en de technologie met
   betrekking tot de interactie tussen beide gaan leren kennen.
   \item Deze beperkingen zullen door de student geplaatst moeten kunnen worden
   in de context van de in het college behandelde theorien en case-studies.
}

\OudVak{Inleiding Informatie- en Communicatietechnologie}{3 sp}{Herfst}{\AfdIRIS}{Franc Grootjen}{
   2001 -- 2003
}{
   Bestaand vak
}{
   Dit vak biedt een eerste orientatie op het zuster-vakgebied Informatica.
   Hierbij staat het verkrijgen van inzicht, niet het verwerven van technische
   vaardigheden zoals het gebruik van een specifieke tekstverwerker of
   besturingssysteem.

   Specifieke onderwerpen zijn:
   \begin{description}
      \item[Netwerken --] Waar komen tegenwoordig overal computers voor? Hoe
      zijn ze aan elkaar gekoppeld tot een wereldomspannend netwerk? Wat doet
      dat netwerk allemaal, en welke problemen levert het op?

      \item[Computers --] Hoe zit een computer in elkaar? Uit welke onderdelen
      bestaat hij en op welke manier communiceert hij met de buitenwereld? Hoe
      werkt een computer eigenlijk, en welke problemen levert dat op?

      \item[Bedrijfssystemen --] Hoe is het mogelijk dat een computer al zijn
      complexe taken tegelijk aan kan? Wat gebeurt achter de schermen als een
      gewone gebruiker met een simpele opdracht een applicatie-programma
      opstart? Waarom komen de gegevens op de harde schijf niet in de war?
      Welke programmatuur moet een computer meekrijgen om \"uberhaupt te kunnen
      worden gebruikt, en welke problemen levert dat op?

      \item[Talen --] Welke rol speelt taal in het omgaan met computers? Hoe
      kan een machine taal begrijpen? Hoe kan een machine vertalen van een taal
      naar een andere? Wat voor talen kunnen machines aan en welke niet, en wat
      voor problemen levert dat op?

      \item[Processoren --] Hoe zit de `central processing unit' van een
      computer in elkaar? Wat is een programma, wat is software en hardware?
      Waaruit bestaat een computer eigenlijk? Hoe kan men een machine maken die
      opdrachten uitvoert die de maker niet kon voorzien, en welke problemen
      levert dat op?
   \end{description}
}{
   \item Doel van het vak is het verkrijgen van begrip, inzicht en overzicht
   met betrekking tot  de fenomenen uit de ICT. De deelnemers ervaren een
   breed, representatief spectrum van fenomenen die de informatica
   wetenschappelijk bestudeert, zodat ze ze later kunnen herkennen, benoemen en
   in context plaatsen.
}

\OudVak{Inleiding Medische Informatiekunde}{4 sp}{-}{Medische Informatiekunde}{Peter de Vries Robb\'e}{
   2001--2002
}{
   Bestaand vak
}{
   Dit vak richt zich op het kennismaken met het handelen van de individuele
   arts. Beslissingen over individuele pati\"enten vinden plaats in het kader van
   de pati\"entenzorg. Voor het nemen van dit type beslissingen gaan we uit van
   de gegevens die reeds over de pati\"ent bekend zijn en maken we gebruik van de
   algemene medische kennis die aanwezig is in boeken en ervaring. Resultaten
   van de besluitvorming over een individuele pati\"ent worden als
   pati\"entgegevens opgeslagen in het medisch dossier. Zowel voor het omgaan met
   algemene medische kennis als met pati\"entgegevens is het van belang dat het
   duidelijk is wat medische termen betekenen en hoe deze kunnen worden
   opgeslagen.  Bij het evalueren van het handelen in de praktijk wordt
   uigegaan van aanwezige kennis en wordt gebruik gemaakt van de gegevens van
   pati\"enten die voldoen aan zogenaamde inclusiecriteria voor de betreffende
   studies. De resultaten van die evaluatie-studies worden weer verwerkt in
   onze kennis. De resultaten van studies kunnen ook een directe bijdrage
   leveren aan het beslissingsproces bijvoorbeeld in de vorm van (behandel)
   protocollen.

   Tijdens de cursus zult u onder meer praktische klinische bijeenkomsten
   bijwonen waarin pati\"entcasussen besproken worden. Verder zult u kennis maken
   met de relatie tussen het medisch beslissingsproces en het opstellen en
   invoeren van (behandel)protocollen en andere formele beslissingsprocedures
   en met het modelleren van kennis van meetinstrumenten voor het systematisch
   opzetten van evaluatiestudies.
}{
   \item Na deze cursus heeft de student een beeld gekregen van de
   informatieprocessen  die zich voordoen bij het medisch handelen en bij het
   uitvoeren van medisch wetenschappelijk  onderzoek.
   \item U kunt zich een voorstelling maken van de rol van de informatiekundige
   bij het analyseren van de medisch inhoudelijke processen.
}

\OudVak{Inleiding Programmeren A (deel 1 en 2)}{4 sp}{Herfst}{\AfdST}{Sjaak Smetsers}{
   2001--2002
}{
   Bestaand vak
}{
   Bij het programmeren wil je een computer zover krijgen dat hij een bepaalde
   opdracht uitvoert. Je moet daartoe in de eerste plaats aan jezelf en
   vervolgens aan de computer duidelijk moeten maken wat er gedaan moet worden.
   Het blijkt eenvoudiger te zijn om aan mensen een bedoeling duidelijk te
   maken dan aan een computer: mensen zijn veel flexibeler dan computers.
   Mensen begrijpen wat je bedoelt en voeren dat uit. Computers hebben geen
   idee van bedoeling, ze doen alleen wat precies opgedragen is. De
   beschrijving van wat de computer moet doen dient dan ook uiterst precies te
   zijn. Zo'n beschrijving noemen we een algoritme. Een algoritme wordt
   opgeschreven in een formalisme dat de computer 'begrijpt' of, beter gezegd
   'kan uitvoeren'. We noemen zo'n formalisme een programmeertaal.

   Het leren van fundamentele begrippen als 'algoritme' en
   samenstellingsmechanismen hiervoor, de zogenaamde besturingsstructuren
   (conditionals, herhalingen en functies) en datastructuren (basistypes en
   operaties, structuren/klassen en rijen).  Abstract, dwz
   programmeertaal-onafhankelijk specificeren van algoritmen uitgaande van een
   concrete probleemstelling.

   Het omzetten van zo'n specificatie in een realisatie in de vorm van een
   concreet (C++) programma gebruikmakend van de meest geschikte data- en
   bestruringsstructuren.  Het kunnen redeneren over zowel een abstract als
   concreet algoritme, dwz je moet in staat zijn om jezelf maar ook andere te
   kunnen overtuigen van de juistheid van jouw oplossing.  Het gebruik van
   functies als abstractiemiddel voor (deel)algoritmes. Verschillende vormen
   van parameteroverdracht weten te onderscheiden en de voor- en nadelen
   hiervan kunnen aangeven.  Begrijpen en kunnen hanteren van recursie.  Kunnen
   omgaan met enkele datastructuren zoals rijen en klassen en combinaties
   hiervan.  Implementeren van enkele zoek- en sorteermethodes, zoals binair en
   recursief zoeken, backtracking, eenvoudig iteratief sorteren en recursief
   sorteren.
}{
   \item In dit vak staat het systematisch ontwikkelen van algoritmen centraal,
   waarbij gebruik wordt gemaakt van de programmeertaal C++. De nadruk bij het
   ontwerpen ligt op het gebruik van abstractie: algoritmes zullen stapsgewijs
   worden gespecificeerd waarbij details langzamerhand nader worden ingevuld
   (verfijning). Deze methodiek heet ook wel Top-Down programmering. De taal
   C++ dient hierbij als hulpmiddel en vormt zeker niet het 'n doel. Het vak
   mag dan ook niet beschouwd worden als een programmeercursus C++.
}

\OudVak{Inleiding Bedrijfscommunicatie}{3 sp}{Herfst}{Letteren}{Carel Jansen}{
   2001 -- 2002
}{
   Bestaand vak
}{
   In dit vak maak je kennis met een scala aan begrippen, modellen en theorie\"en
   die relevant zijn voor de studie en de praktijk van professionele
   (bedrijfs)communicatie. Aan de orde komen (onder meer) een model voor de
   beschrijving en verklaring van communicatief gedrag, het verband tussen imago en
   identiteit van een organisatie, de relatie tussen organisatiestructuur en interne
   communicatie, de relatie tussen marktkenmerken en externe communicatie, en de
   consequenties van dit alles voor het onderwerp van documenten.

   Verdere informatie over dit vak, inclusief studiemateriaal, is te vinden via de
   website van de opleiding Bedrijfscommunicatie, http://www.let.kun.nl.ciw-bc.
}{
   \item -
}

\OudVak{Inleiding Computer Architectuur}{2 sp}{Lente}{\AfdIRIS}{Franc Grootjen}{
   2001--2003
}{
   Bestaand vak
}{
   Om een volwaardige gesprekspartner te zijn voor informatici moet een
   informatiekundige de ontwikkelingen in de ICT met voldoende kennis van zaken
   kunnen volgen om ze naar waarde te kunnen schatten. Dit vereist een achtergrondkennis
   van bijvoorbeeld computerarchitectuur, besturingssystemen, datacommunicatie en
   netwerken. Dit vak is een eerste introductie tot deze gebieden.

   Specifieke onderwerpen:
   Computers van binnen, machinetaal, operating systems, netwerken, protocollen.
}{
   \item Na het volgen van dit vak kunnen de studenten in abstracte termen
   aangeven hoe een computer werkt en zijn in staat in te schatten welke
   problemen er optreden (en hoe men deze kan oplossen) als men meerdere
   computers met elkaar wil laten samenwerken.
}

\OudVak{Integratieproject Informatiekunde}{2 sp}{Lente}{\AfdIRIS}{Wil Dekkers}{
   2001 -- 2003
}{
   Bestaand vak\\
   Zal in 2003 al vooruitlopen op R\&D Lab 1 en 2.
}{
   Het Integratieproject Informatiekunde belicht de samenhang tussen de
   eerstejaarsvakken informatiekunde aan de hand van een aantal basale thema's
   die centraal zijn voor het veld, maar die in de verdere studie niet zo gauw
   expliciet belicht worden. Begrippen als perspectief, model, functionaliteit,
   automatisering etc. worden nader bekeken in context van de steeds
   terugkerende vraag: wat is informatiekunde en wat doet een
   informatiekundige? E.e.a. is beschouwend-analytisch opgezet en wordt
   ge\"illustreerd d.m.v. voorbeelden uit de eerstejaars lesstof. De collegereeks
   wordt afgesloten met een groepsproject.

   Specifieke onderwerpen:
   Informatiekunde, perspectieven, domeinen, stakeholders,  functionaliteit,
   structuren, modellen, automatiseren, methoden, tools, technieken,
   (tegenstrijdige) belangen.
}{
   \item Men de eerstejaars vakken informatiekunde goed kan plaatsen ten
   opzichte van het vakgebied, en de relevantie ervan voor dat vakgebied
   helder is
   \item Men een aantal informatiekundige basisbegrippen genuanceerd kan
   hanteren
   \item Men basaal inzicht heeft in de algemene processen en structuren
   waarmee een informatiekundige te maken heeft
   \item Men in staat is een basale informatiekundige kijk op toe te passen
   op bijvoorbeeld een casus
   \item Men deze kijk ook op een heldere en doelgerichte manier kan
   verwoorden
}

\OudVak{Introductie Mens-Machine Interactie}{4 sp}{Lente}{CogW}{-}{
   2001 -- 2003
}{
   Bestaand vak
}{
   Niet voorhanden
}{
   \item -
}

\ExternVak{Methoden voor Organisatieverandering}{5 ECTS}{Herfst}{
   2003
}{
   \item Methoden voor Organisatieveranderin (NSM)
}

\OudVak{Organisatie en Informatievoorziening}{4 sp}{Lente}{MIK}{Hans ten Hoopen}{
   2001--2003
}{
   Bestaand vak
}{
   In deze cursus staan kennis en vaardigheden met betrekking tot het inrichten
   en onderhouden van de informatievoorziening in organisaties centraal. Eerst
   zal worden stilgestaan bij de rol die informatie in organisaties speelt en
   bij wat het `voorzien' in informatie inhoudt. Vervolgens komt aan de orde
   hoe de informatievoorziening kan worden ingericht en onderhouden. Dat zal
   worden behandeld aan de hand van de interventiecyclus: diagnose, ontwerp,
   verandering en evaluatie van de informatievoorziening. Daarbij zal blijken
   dat bij het inrichten van de informatievoorziening niet alleen informatie-
   en communicatietechnologie een rol speelt, maar dat ook de structuur van de
   organisatie van invloed is. Nadat methoden voor elk van de stappen van de
   interventiecyclus aan bod zijn gekomen, zal de student in staat zijn om een
   informatieplan op te stellen, waarin stapsgewijs duidelijk gemaakt wordt hoe
   de informatievoorziening van een organisatie kan worden ingericht en
   onderhouden.  Hier wordt als voorbeeld het management van de zorgverlenende
   instelling en de daaruit voortvloeiende konsekwenties voor de
   informatievoorziening behandeld.  Inzicht wordt verworven in hoe
   beleidsvraagstukken worden aangepakt op het niveau van een zorginstelling,
   maar ook op zowel het interne niveau van de concrete zorguitvoering
   (case-management) als op het niveau van de zorgsector of het gehele stelsel.
   Vaardigheden m.b.t. de planning van de informatievoorziening worden
   aangeleerd, d.w.z. hoe te onderkennen welke informatiesystemen nodig zijn,
   gegeven doelen, aktiviteiten en processen in een sector of organisatie,
   welke informatie hoe gedefinieerd relevant is.
}{
   \item De student kan enkele eenvoudige organisatiekenmerken
   onderscheiden.
   \item De student kan aangeven welke conditionerende en
   afhankelijkheidsrelaties van het maatschappelijke veld zijn te
   onderscheiden voor een organisatie voor gezondheidszorg.
   \item De student kan primaire functies onderscheiden voor het management
   welke voortvloeien uit kenmerken van de organisatie en uit condities
   gesteld vanuit de maatschappelijke omgeving.
   \item De student kan gegeven de functies, taken en operationele
   aktiviteiten van het management in een instelling vaststellen welke
   informatiebehoefte dat management heeft.
   \item De student kan methoden en technieken inzetten bij het vaststellen
   van informatiegebieden en van een globale systeemarchitectuur
   \item De student leert hulpbronnen (referentie informatiemodellen)
   selecteren tbv. informatiebehoeftenvaststelling en informatieplanning.
   \item De student kan een globaal Corporate datamodel opstellen om daarmee
   de eenheid van taal in een instelling te bevorderen.
}

\OudVak{Practicum Psychologische Functieleer}{2 sp}{-}{Cognitieweteneschappen}{B. Hofstede}{
   2001--2002
}{
   Bestaand vak
}{
   Onderzoek op het gebied van motoriek, perceptie, taal of cognitiewetenschap.
}{
   \item Je kunt een eenvoudig experiment op het gebied van de cognitieve psychologie
   bedenken, opzetten, uitvoeren, analyseren, interpreteren, presenteren en
   rapporteren volgens A.P.A.-normen; Je kunt een eenvoudig computermodel op
   het gebied van de cognitiewetenschap draaien, analyseren, aanpassen,
   interpreteren, presenteren en rapporteren volgens A.P.A.-normen.
}

\OudVak{Programmeren voor Informatiekundigen 1}{4 sp}{Herst}{\AfdST}{??}{
   2001 -- 2002
}{
   Bestaand vak, wordt hernoemd tot:\\
   - Algoritmiek
}{
   Zie ``Algoritmiek''
}{
   \item -
}

\OudVak{Programmeren voor Informatiekundigen 2}{4 sp}{Lente}{\AfdST}{??}{
   2001 -- 2002
}{
   Bestaand vak, wordt hernoemd tot:\\
   - Datastructuren
}{
   Zie ``Datastructuren''
}{
   \item -
}

\OudVak{Requirements Engineering in het Medische Domein}{6 ECTS (was 4 sp)}{Herfst}{MIK}{
   Hans ten Hoopen
}{
   2001 -- 2003
}{
   Bestaand vak
}{
   Het vaststellen van een programma van eisen in het medisch domein begint
   veelal met het verkrijgen van inzicht in de werkprocessen in de
   gezondheidszorg. Veel werkers in de zorg zijn zodanig met concrete
   pati\"enten bezig, dat zij moeilijk ook hebben voor de meer generieke
   aspecten van de zorg. Dat betekent dat uit de concrete voorbeelden die in
   het contact met zorgverleners worden verzameld door de informatiekundige de
   meer algemene procedures moeten worden afgeleid. Dat maakt requirements
   engineering in de gezondheidszorg tot een taak die veel inhoudelijk begrip
   vraagt van de informatiekundige.

   Op grond van de werkprocessen worden vervolgens de voor die processen
   noodzakelijke informatiebehoeften bepaald die vervolgens de basis vormenvan
   de te ontwerpen informatiesystemen. Dit blok geeft de student vaardigheden
   om een sturende \'en uitvoerende rol te spelen in de aanvangsfase van een
   ontwikkelingstraject voor een informatiesysteem in een
   gezondheidszorgorganisatie. Met aanvangsfasen worden bedoeld de diverse
   probleem- verhelderings-, analyse- en functionele oplossingsspecificatie
   fase. De nadruk ligt daarbij op die zaken waarbij kennis van zorgverlening
   en -organisatie een essenti\"ele rol speelt. Het relatief steeds eenvoudiger
   te beheersen \& uit te voeren, en ook meer technische deel van het
   systeemontwikkelingstraject, n.l. de feitelijke bouw van informatiesystemen
   (ontwerp en programmering), maakt bewust geen deel uit van dit blok.

   In het eerste deel van het leertraject wordt o.b.v. concrete problemen in
   zorgorganisaties geleerd hoe met eindgebruikers via systeem- of proces- en
   informatie-analyse gekomen wordt tot het opstellen van functionele
   specificaties voor een informatiesysteem(deel).
   Daarna wordt de essentie van projectmanagement in de kontekst van
   systeemontwikkeling behandeld. In een volgend deel wordt geleerd hoe om te
   gaan met de voor analisten belangrijke domeinspecifieke kennis:
   referentie-informatiemodellen voor de gezondheids-zorg. Tenslotte wordt het
   geheel geplaatst in de kontekst van modellen voor het
   systeem-ontwikkelingsproces als geheel. Gedurende het gehele leertraject zal
   tenminste
}{
   \item  De student is in staat in een concrete zorgorganisatie een
   informatievoorzienings-probleem te analyseren en kan een project formuleren
   en uitvoeren om samen met de eindgebruikers in die organisatie een programma
   van eisen voor een nieuw (in te voeren of te wijzigen)
   informatiesysteem(deel) op te stellen.
   \item De student is in staat om op basis van enerzijds de eisen mbt. een
   informatie-systeem en anderzijds een gereed produkt (elders gebouwd maatwerk
   informatie-systeem danwel een 'getuned' multipurpose softwarepakket) de
   bruikbaarheid van dat produkt aan te geven.
}

\OudVak{Software Technologie 1}{2 sp}{Herfst}{\AfdST}{-}{
   2000 -- 2002
}{
   Bestaand vak, zal overgaan in:\\
   - Gedistribueerde Software Systemen
}{
   In dit vak maken studenten kennis met Object Georienteerd programmeren aan
   de hand van de programmeertaal JAVA. De behandelde begrippen en technieken
   worden ondersteund door een practicum. In dit practicum zullen we een
   incrementeel aan een internet-georienteerde applicatie werken, dat zal
   leiden tot een interactief netwerk spel.
}{
   \item -
}

\OudVak{Software Technologie 2}{6 ECTS (was 4 sp)}{Lente}{\AfdST}{-}{
   2000 -- 2004
}{
   Bestaand vak, zal overgaan in:\\
   - Integratie van Software Systemen
}{
   In dit vak wordt een aantal moderne softwaretechnologieen besproken die van
   belang zijn bij ontwerp, architectuur en implementatie van grote, heterogene
   systemen. Zulke systemen zijn vaak opgebouwd uit "componenten" die via
   "middleware" met elkaar communiceren; denk aan Corba, .net, IDL, XML,
   dynamics, etc. Van studenten wordt een actieve bijdrage gevraagd tijdens het
   college.
}{
   \item -
}

\OudVak{Statistiek voor Informatiekunde}{2 Sp}{Herfst}{Sociale Wetenschappen}{??}{
   2000 -- 2003
}{
   Wordt vervangen door ``Statistiek''.
}{
   Zie ``Statistiek''.
}{
   \item -
}

\OudVak{Syntactische Analyse}{3 sp}{-}{Taalwetenschappen}{Peter Coppen}{
   2001 -- 2003
}{
   Bestaand vak
}{
   De cursus syntactische analyse vervult een brugfunctie tussen de
   traditionele zinsontleding van de middelbare school en de moderne
   ontleedmethoden  van de generatieve grammatica. De cursus is gericht op het
   opfrissen dan wel verwerven van praktische analytische vaardigheden.

   In kwartaal 1: de traditionele zinsontleding, redekundig en taalkundig. In
   kwartaal 2: de belangrijkste principes van de generatieve grammatica:
   basisstructuur, afleiding, naamval en thematische rol, agreement, beknopte
   bijzinnen, anaforische relaties, domeinbeperkingen op afleidingen.
}{
   \item
   Aan het eind van periode 1 kun je Nederlandse zinnen analyseren volgens de
   methoden van de traditionele zinsontleding. Je kent de benoemingen in
   woordsoorten en zinsdelen, en kunt die benoemingen aanbrengen op aangewezen
   woorden en woordgroepen. Deze vaardigheid wordt in de eerste tentamenperiode
   met een deeltentamen getoetst.
   \item
   In periode 2 maak je kennis met de generatieve grammatica, de taalkundige
   theorie zoals die in de tweede helft van de twintigste eeuw ontwikkeld is.
   Je leert hoe zinnen kunnen worden afgeleid van basisstructuren, en je kunt
   structurele relaties in de eindstructuren aanwijzen en benoemen.
}

\OudVak{Systematisch Bouwen van Eenvoudige Systemen}{4 sp}{Herfst}{\AfdIRIS}{Ger Paulussen}{
   2001 -- 2003
}{
   Bestaand vak
}{
   In deze cursus gebruiken we een aantal methoden en technieken om
   bestuurlijke informatiesystemen op een gestructureerde manier te ontwerpen
   en te realiseren. Het ontwerpen gebeurt via de System Development
   Methodology II (SDM) en we gebruiken daarbij o.a. informatieanalyse (een
   NIAM-versie), de Systems Analysis and Design Technique (SADT),
   risico-analyse en functiepunt-analyse (FPA). De realisatie gebeurt via SQL
   (Structured Query Language) en programmeren in een relationeel
   gegevensbanksysteem met tools (MS Access/Windows).
}{
   \item -
}

\OudVak{Verdieping Cognitieve Ergonomie}{4 sp}{Herfst}{Cognitiewetenschappen}{Louis Vuurpijl}{
   2001--2002
}{
   Bestaand vak
}{
   Deze cursus bouwt voort op de cursus Inleiding in de cognitieve ergonomie.
   Er wordt aan de hand van artikelen dieper ingegaan op de analyse
   en op het formeel modelleren van mens-computer-interactie (HCI). Terwijl het
   met computers niet moeilijk is om de gebruiker te bestoken met grote
   hoeveelheden 'bits' in de vorm van multimediale informatie, blijkt het nog
   steeds een hele klus om nuttige informatie van de gebruiker de machine in te
   krijgen. Een nieuwe vorm van interactie, multi-modaliteit, maakt gebruik van
   de mogelijkheid om de verschillende output channels die de gebruiker heeft
   (o.a. handschrift en spraak) te combineren. Binnen het 'multiple agent'
   paradigma is het mogelijk de gebruiker hierin te ondersteunen met
   kennisgebaseerde technieken.
}{
   \item -
}

\OudVak{Visualisatie}{6 ECTS (was 4 sp)}{Herfst}{\AfdST}{Susan Even}{
   2001 -- 2003
}{
   Bestaand vak
}{
   Het doel van de cursus is tweevoudig: visualisatie als hulpmiddel in het
   overbrengen van informatie aan de gebruiker; en visualisatie als analytisch
   gereedschap. Dit laaste wordt onderzocht door middel van computer-aided
   design. Ook bekijken we ruimtelijke en tijdsaspecten van data en data
   analyse, en het visualiseren daarvan. Mogelijke gastcolleges en
   samenwerking met Medische Informatiekunde. De volgende werkvormen komen aan
   bod: werkcollege, discussie, practicum, zelfstudie, werkstuk.
}{
   \item -
}
}
\Ragged{\chapter{Toetsing aan IS2002 -- Model Curriculum}
\label{IS2002}

De ACM heeft, in samenwerking met de AIS (Association for Information Systems)
en de AITP (Association of Information Technology Professionals) een model
curriculum opgesteld voor bacheloropleidingen op het gebied van
informatiesystemen~\cite{IS2002} (vanaf nu verwijzen we hiernaar als IS2002). 

Voor een opleiding informatiekunde is het uiteraard relevant op zich te toetsen
aan een dergelijk model curriculum. De recente visitatie
Informatiekunde~\cite{Visitatie2002}, spreekt zich ook uit voor het gebruik van
modelcurricula als toetssteen voor informatiekunde opleidingen.

Twee belangrijke observaties die vooraf gemaakt moet worden zijn:
\begin{enumerate}
   \item De huidige versie van het model curriculum is vooral ge-ent is
   studenten met een vooropleiding volgens Noord-Amerikaans model. Deze wijken,
   zoals bekend, nogal veel af van de situatie zoals we deze in de meeste EU
   landen, en in het bijzonder in Nederland, kennen. Dit betekent dat het
   letterlijk overnemen van de inhoud van het modelcurriculum naar onze mening
   niet aan te raden is.

   \item Daarnaast is het natuurlijk zo dat elke Universiteit zich toch zal
   wensen te profileren op basis van haar expertise binnen de Informatiekunde.
   In ons geval is dat de focus op informatiearchitectuur, een exacte
   wetenschappelijke ori\"entering op het vakgebied, en de natuurlijke binding
   met de informatica opleiding. 
   
   Een vergelijking met een standaard curriculum is dan weliswaar relevant,
   maar afwijkingen zullen onvermijdelijk zijn.
\end{enumerate}
Echter, ook met deze observaties blijft het relevant om een vergelijking te maken
tussen de informatiekunde opleiding van het \NIII\ en het IS2002 model curriculum.

\section{Uitgangspunten}
In IS2002 worden een aantal ``guiding assumptions about the information
systems profession'' gemaakt:
\begin{quote}
   In conceptualizing the role of information systems in the future and the
   requirements for IS curricula, several elements remain important and
   characteristic of the discipline. These characteristics evolve around four
   major areas of the IS profession and therefore must be integrated into any
   IS curriculum:
   \begin{enumerate}
      \item IS professionals must have a broad business and real world perspective. 
      Students must therefore understand that:
      \begin{itemize}
         \item IS are enablers of successful performance in organizations
         \item IS span and integrate all organizational levels and business functions
	 \item IS are increasingly of strategic significance because of the
	 scope of the organizational systems involved and the role systems play
	 in enabling organizational strategy
      \end{itemize}

      \item IS professionals must have strong analytical and critical thinking skills. 
      Students must therefore:
      \begin{itemize}
         \item Be problem solvers and critical thinkers
         \item Use systems concepts for understanding and framing problems
         \item Be capable of applying both traditional and new concepts and skills
         \item Understand that a system consists of people, procedures, hardware, 
	 software, and data
      \end{itemize}

      \item IS professionals must exhibit strong ethical principles and have
      good interpersonal communication and team skills. Students must
      understand that:
      \begin{itemize}
         \item IS require the application of professional codes of conduct
         \item IS require collaboration as well as successful individual effort
         \item IS design and management demand excellent communication skills 
	 (oral, written, and listening)
	 \item IS require persistence, curiosity, creativity, risk taking, and
	 a tolerance of these abilities in others
      \end{itemize}

      \item IS professionals must design and implement information technology
      solutions that enhance organizational performance. Students must
      therefore:
      \begin{itemize}
	 \item Possess skills in understanding and modeling organizational
	 processes and data, defining and implementing technical and process
	 solutions, managing projects, and integrating systems
	 \item Be fluent in techniques for acquiring, converting, transmitting,
	 and storing data and information
	 \item Focus on the application of information technology in helping
	 individuals, groups, and organizations achieve their goals
      \end{itemize}
   \end{enumerate}
\end{quote}
Naar onze mening liggen deze uitgangspunten volledig in lijn met de doelstellingen
en eindtermen van onze opleiding (hoofdstuk~\ref{h:wat-doelstellingen}), zoals deze
volgen uit onze visie op het vakgebied~\cite{Visie2003}.

\section{Definitie van het vakgebied}
In IS2002 staat de volgende ``scope of information systems'':
\begin{quote}
  Information Systems as a field of academic study encompasses the concepts,
  principles, and processes for two broad areas of activity within
  organizations: (1) acquisition, deployment, and management of information
  technology resources and services (the information systems function) and (2)
  development, operation, and evolution of infrastructure and systems for use
  in organizational processes (system development, system operation, and system
  maintenance). The systems that deliver information and communications
  services in an organization combine both technical components and human
  operators and users. They capture, store, process, and communicate data,
  information, and knowledge.

  The information systems function in an organization has a broad
  responsibility to plan, develop or acquire, implement, and manage an
  infrastructure of information technology (computers and communications), data
  (both internal and external), and enterprise-wide information processing
  systems. It has the responsibility to track new information technology and
  assist in incorporating it into the organization's strategy, planning, and
  practices. The function also supports departmental and individual information
  technology systems. The technology employed may range from large centralized
  to mobile distributed systems. The development and management of the
  information technology infrastructure and processing systems may involve
  organizational employees, consultants, and outsourcing services.

  The activity of developing or acquiring information technology applications
  for organizational and inter-organizational processes involves projects that
  define creative and productive use of information technology for transaction
  processing, data acquisition, communication, coordination, analysis, and
  decision support. Design, development or acquisition, and implementation
  techniques, technology, and methodologies are employed. Processes for
  creating and implementing information systems in organizations incorporate
  concepts of business process design, innovation, quality, human-machine
  systems, human-machine interfaces, e-business design, sociotechnical systems,
  and change management.

  Information systems professionals work with information technology and must
  have sound technical knowledge of computers, communications, and software.
  Since they operate within organizations and with organizational systems, they
  must also understand organizations and the functions within organizations
  (accounting, finance, marketing, operations, human resources, and so forth).
  They must understand concepts and processes for achieving organizational
  goals with information technology. The academic content of an information
  systems degree program therefore includes information technology, information
  systems management, information systems development and implementation,
  organizational functions, and concepts and processes of organizational
  management.
\end{quote}
Deze definitie is, naar onze mening, in lijn met de definitie van het vakgebied
zoals wij deze uiteen hebben gezet in~\cite{Visie2003}, welke we als basis voor
het curriculum gebruikt hebben.

\section{Overzicht van vaardigheden}
In~\cite{IS2002} staat aan tabel met daarin de belangrijkste vaardigheden benoemd
waaraam een `IS Program Graduate'' moet voldoen. Hieronder staat deze tabel 
weergegeven.

\begin{center} \footnotesize \begin{tabular}{|l|l|l|}
   \hline
   \multicolumn{3}{|c|}{\bf Analytical and Critical Thinking}\\
   \hline
   Organisational Problem Solving & Ethics and Professionalism & Creativity\\
   \hline
   \Stack{
         Problem solving models,\\
         ~~techniques, and\\
         ~~approaches\\
         Personal decision making\\
         Critical thinking\\
         Methods to collect, summarize,\\
         ~~and interpret data\\
         Statistical and mathematical\\
         methods
   } & \Stack{
         Codes of conduct\\
         Ethical theory\\
         Leadership\\
         Legal and regulatory standards\\
         Professionalism - self directed\\
         ~~leadership, time management\\
         Professionalism - commitment\\
	 ~~to and completion of work
   } & \Stack{
         Creativity concepts\\
         Creativity techniques\\
         The systems approach
   } \\
   \hline
   \multicolumn{3}{|c|}{\bf Business Fundamentals}\\
   \hline
   Business Models & Functional Business Areas & Evaluation of Business Performance\\
   \hline
   \Stack{
         Contemporary and emerging\\
         ~~business models\\
         Organizational theory,\\
         ~~structure, and functions\\
         System concepts and theories\\
   } & \Stack{
         Accounting\\
         Finance\\
         Marketing\\
         Human Resources\\
         Logistics and Manufacturing\\
   } & \Stack{
         Benchmarking\\
         Value chain and value network\\
	 ~~analysis\\
         Quality, effectiveness, and
	 ~~efficiency\\
         Valuation of organizations\\
         Evaluation of investment\\
	 ~~performance
   }\\
   \hline
   \multicolumn{3}{|c|}{\bf Interpersonal, Communication and Team Skills}\\
   \hline
   Interpersonal & Team Work and Leadership & Communication \\
   \hline
   \Stack{
         Listening\\
         Encouraging
         Motivating\\
         Operating in a global, culturally\\
         ~~diverse environment\\
   } & \Stack{
         Building a team\\
         Trusting and empowering\\
         Encouraging\\
         Developing and communicating a\\
         vision/mission\\
         Setting and tracking team goals\\
         Negotiating and facilitating\\
         Team decision making\\
         Operating in a virtual team\\
         environment\\
         Being an effective leader\\
   } & \Stack{
         Listening, observing, interviewing,\\
	 and documenting\\
         Abstraction and precise writing\\
         Developing multimedia content\\
         Writing memos, reports, and\\
         documentation\\
         Giving effective presentations\\
   } \\
   \hline
   \multicolumn{3}{|c|}{\bf Technology}\\
   \hline
   \multicolumn{3}{|c|}{
    \tabcolsep=0pt 
     \begin{tabular}{l|l|l|l}
      ~Application ~&~ Internet Systems ~&~ Database Design and ~&~ System Infrastructure\\
      ~Development ~&~ Architecture and ~&~ Administration      ~&~ and Integration\\
      ~            ~&~ Development      ~&~                     ~&~ \\
      \hline
      ~\Stack{
         Programmingprinciples,\\
	 ~~objects, algorithms,\\
	 ~~modules, testing\\
         Application development --\\
         ~~requirements, specs,\\
         ~~development\\
         Algorithmic design,\\
	 ~~data, object, and\\
	 ~~file structures\\
         Client-server software\\
         ~~development
      } ~&~ \Stack{
         Web page development\\
         Web architecture design and\\
         ~~development\\
         Design and development of\\
         ~~multi-tiered architectures
      } ~&~ \Stack{
         Modeling and design,\\
         ~~construction, schema tools,\\
         ~~and DB Systems\\
         Triggers, stored procedures,\\
         ~~design and development of\\
         ~~audit controls\\
         Administration: security,\\
         ~~safety, backup, repairs,\\
         ~~and replicating
      } ~&~ \Stack{
         Computer systems, hardware\\
         Networking (LAN/WAN)\\
         ~~and telecommunications\\
         LAN/WAN design and\\
	 ~~management\\
         Systems software\\
         Operating systems\\
	 ~~management\\
         Systems configuration,\\
	 ~~operation, and\\
	 ~~administration
      }
   \end{tabular}}\\
   \hline
   \multicolumn{3}{|c|}{\bf Information systems = Technology-enabled business development}\\
   \hline
   \multicolumn{3}{|c|}{Systems Analysis and Design, Business Process Design, 
                        Systems Implementation, IS Project Management}\\
   \hline
   \Stack{
         Strategic utilization of information\\
         ~~technology and systems\\
         IS planning\\
         IT and organizational systems\\
   } & \Stack{
         Systems analysis\\
         Logical and physical design\\
         Design execution\\
         Testing\\
   } & \Stack{
         Deployment\\
         Maintenance\\
         Use of IT\\
         Customer service\\
   }\\
   \hline
\end{tabular}\end{center}

Merk op dat het hierbij gaat om vaardigheden na afloop van de bachelorfase!
Wanneer we deze lijst van vaardigheden vergelijken met de vaardigheden uit
hoofdstuk~\ref{h:wat-doelstellingen}, dan zijn er vooral verschillen te zien op
het vlak van:
\begin{enumerate}
   \item Vaardigheden met betrekking tot ``Business fundamentals''.
   
   Wel in IS2002, niet prominent in Curriculum 2003.

   \item Vaardigheden met betrekking tot ``Professionalism''.

   Wel in IS2002, niet in Curriculum 2003.
   
   \item Vaardigheden met betrekking tot formele en systeemtheoretische 
   grondslagen.

   Niet in IS2002, wel in Curriculum 2003.

   \item Vaardigheden met betrekking tot onderzoek en reflectie.

   Niet in IS2002, wel in Curriculum 2003.
\end{enumerate}
Deze verschillen zijn als volgt te motiveren:
\begin{itemize}
   \item Informatiekunde aan het \NIII\ heeft, om voor de hand liggende redenen,
   een exact wetenschappelijke invulling gekregen. Dit betekent dat er relatief gezien
   meer aandacht is voor de formele en systeemtheoretische grondslagen, ten koste
   van de bedrijfsmatige fundamenten.

   Daarnaast is het in de Nederlandse context zeker zo dat een deel van de ``Business
   Fundamentals'' zoals accounting en finance al op de middelbare school aan bod zijn
   gekomen.

   \item Het lijkt erop alsof het IS2002 curriculum zich meer profileerd als 
   beroepsopleiding, met meer aandacht voor ``Professionalism'', terwijl de informatiekunde
   opleiding van het \NIII\ de meer typische academische vaardigheden hoger in het
   vaandel heeft. Voor een Universitaire opleiding informatiekunde lijkt ons dit laatste
   een verdedigbaar standpunt.
\end{itemize}

}
\chapter{Toetsing aan MSIS2000 -- Model Curriculum}
\label{MSIS2000}

De ACM heeft, in samenwerking met de AIS (Association for Information Systems)
een model curriculum opgesteld voor masteropleidingen op het gebied van
informatiesystemen~\cite{MSIS2000} (vanaf nu verwijzen we hiernaar als
MSIS2000). 

Voor een opleiding informatiekunde is het uiteraard relevant op zich te toetsen
aan een dergelijk model curriculum. De recente visitatie
Informatiekunde~\cite{Visitatie2002}, spreekt zich ook uit voor het gebruik van
dit modelcurriculum als toetssteen voor informatiekunde opleidingen.

Twee belangrijke observaties die vooraf gemaakt moet worden zijn:
\begin{enumerate}
   \item De huidige versie van het model curriculum is vooral ge-ent is
   studenten met een vooropleiding volgens Noord-Amerikaans model. Deze wijken,
   zoals bekend, nogal veel af van de situatie zoals we deze in de meeste EU
   landen, en in het bijzonder in Nederland, kennen. Dit betekent dat het
   letterlijk overnemen van de inhoud van het modelcurriculum naar onze mening
   niet aan te raden is. Zoals men zelf stelt:
   \begin{quote}
      MSIS 2000 is based on the educational system and degree structures common
      to the United States and Canada. This limits is use outside these
      systems, but the report still has relevance for teh reasoning and design
      process for curriculum development in other environments.
   \end{quote}

   \item Daarnaast is het natuurlijk zo dat elke Universiteit zich toch zal
   wensen te profileren op basis van haar expertise binnen de Informatiekunde.
   In ons geval is dat de focus op informatiearchitectuur, een exacte
   wetenschappelijke ori\"entering op het vakgebied, en de natuurlijke binding
   met de informatica opleiding. 
   
   Een vergelijking met een curriculum ``information management'' is dan 
   weliswaar relevant, maar afwijkingen zullen onvermijdelijk zijn.
\end{enumerate}
Echter, ook met deze observaties blijft het relevant om een vergelijking te
maken tussen de informatiekunde opleiding van het \NIII\ en het MSIS2000 model
curriculum.

\section{Componenten}
Het MSIS2000 model bouwt op vier basiscomponenten:
\begin{description}
   \item[Foundations --] At the foundation level, the curriculum is designed to
   accommodate students from a wide variety of backgrounds. In particular, the
   model specifies the business and information systems skills required as
   prerequisite to the rest of the curriculum.
   \item[Core --] The next level, or core, is a set of primary courses. All
   graduates require this common core. Some of the core courses are similar in
   name to those in the 1982 Curriculum, but the contents are a major revision
   reflecting the changes in the Information Systems field. The core courses
   are:
   \begin{itemize}
      \item Data management
      \item Analysis, modeling, and design
      \item Data communications and networking
      \item Project and change management
      \item IS policy and strategy
   \end{itemize}
   \item[Integration --] A major innovation in this curriculum is in the
   integration component required after the core. This component addresses the
   increasing need to integrate a broad range of technologies and offers the
   students the opportunity to synthesize the ideas presented earlier and to
   help students implement comprehensive systems across an organization.
   \item[Career Tracks --] Another innovation is that the program architecture
   is flexible to accommodate individual institutional requirements for an MS
   degree. This flexibility occurs at both the entry level with the foundation
   courses that can be tailored to meet individual needs and at the highest
   level where institutions and students may select specific career tracks that
   are representative of current organizational needs.
\end{description}
In MSIS2002 wordt het ``foundations'' deel opgesplitst naar ``business'' en
``information systems''. Het foundations deel wordt verondersteld voorkennis
vanuit de bachelorfase te zijn. Men verwijst hierbij in MSIS2000 naar een
voorloper van het modelcurriculum voor de bachelorfase zoals besproken zal
worden in appendix~\ref{IS2002}.

Voor de componenten: IS Core, Integration en Career Tracks bespreken we hieronder
de relatie met het Curriculum 2003.

\section{IS Core}
Elementen uit vakken zoals ``Data management'', ``Analysis, modelling and design'', ``Data
communications and networking'' komen in de master weliswaar aan bod in vakken zoals:
\begin{quote}
   Informatiearchitectuur, Systeemtheorie: Structuur \& Co\"ordinatie, Security Protocols.
\end{quote}
Echter, vakken op dit gebied komen juist in de bachelor naar voren. Denk hierbij aan:
\begin{quote}
   Integratie van Software Systemen, ICT Infrastructuur, Opslaan \& Terugvinden en 
   Informatiesystemen
\end{quote}
Onze verwachting is dat in Europa de bachelorfase met name gericht zal zijn op de
generieke kennis voor Informatiekundigen, terwijl de masterfase zich juist kan richten
op de ``locale specialiteiten''. Vandaar dat in de huidige master een duidelijke plek
is ingeruimd voor:
\begin{quote}
   Informatiearchitectuur, Kwaliteit van Informatiesystemen en Security Protocols
\end{quote}
welke alle drie voortvloeien uit core onderzoeksactiviteiten van het \NIII.

De vakken ``Project and change management'' en ``IS Policy and strategy'' worden
afgedekt door:
\begin{quote}
   Systeemontwikkeling: Sturen en Informatiearchitectuur (deels)
\end{quote}

\section{Integration}
Volgens MSIS2000 kan integratie vanuit drie perspectieven benaderd worden:
\begin{itemize}
   \item Integrating the Enterprise
   \begin{itemize}
      \item provide an integrated view of the firm and its relations with
      suppliers and customers 
      \item demonstrate an integrated set of business processes and functional
      applications that meet business needs
   \end{itemize}
   \item Integrating the IS Function
   \begin{itemize}
      \item design effective/efficient IS organizational processes
      \item assess the impact of emerging technologies
      \item define human resource needs and management methods
      \item IS governance alternatives
      \item define the role of the CIO
      \item apply methods to measure and demonstrate the value of IS
   \end{itemize}
   \item Integrating IS Technologies
   \begin{itemize}
      \item evaluate and select from architectural and platform choices,
      priorities, and policies 
      \item assessment of the impact of emerging technologies
      \item evaluate the role of standards
      \item evaluate effect of vendor strategies
   \end{itemize}
\end{itemize}
waarbij aangegeven wordt dat een keuze nodig is. Met andere woorden, men
hoeft niet pers\'e aan alle drie perspectieven aandacht te besteden.
In het master vak: ``Informatiearchitectuur'' wordt aandacht besteed aan
de ``Integrating IS Technologies'' en ``Integrating the IS Function''.
Daarnaast wordt overwogen een keuzevak ``Bestendiging van Informatiesystemen''
in te voeren wat juist aandacht besteed aan ``Integrating the IS Function''.
Studenten zouden dit laatste vak in plaats van Visualiseren \& Communiceren
kunnen volgen.

\section{Career tracks}
De \FNWI\ maakt voor alle 5-jarige $\beta$-studies in principe onderscheid tussen drie,
beroepsgeorienteerde uitstroomprofielen:
\begin{itemize}
  \item Onderzoek.
  \item Communicatie \& Educatie.
  \item Management \& Toepassing.
\end{itemize}
In de Informatiekunde bachelor is er bewust niet gekozen voor enige voorsortering op
deze uitstroomprofielen. In plaats daarvan is de opleiding gericht op vaardigheden die
nodig zijn voor alle drie profielen.

In de Informatiekunde master is er wel mogelijkheid tot differentiatie. Hierbij
dient Informatiekunde uiteraard de indeling van de \FNWI\ te volgen. De
standaardinvulling van de masterfase is gericht op een student die
informatiearchitect of informatiemanager wil worden. In het vak
``Systeemontwikkeling; Sturen'' kunnen de studenten zich in een praktijk
situatie de rol van informatiearchitect of informatiemanager bekwamen, terwijl
ze in het theoriedeel van dat vak nader geschoold worden in inhoudelijk en/of
procesmatig (be-)sturen.

Studenten die zich sterker willen profileren in de richting van het onderzoek
kunnen het vak ``Systeemontwikkeling; Sturen'' vervangen door het vak
``Onderzoekslaboratorium''.

\chapter{Visitatie Informatiekunde 2002}
\label{Visitatie2002}

In 2002 heeft er een visitatie van drie informatiekunde opleidingen
plaatsgevonden~\cite{Visitatie2002}:
\begin{itemize}
   \item Vrije Universiteit: Informatiekunde.
   \item Universiteit Twente: Bedrijfs Informatie Technologie.
   \item Universiteit van Tilburg: Informatiekunde.
\end{itemize}
Voor de Nijmeegse Informatiekunde opleiding was het nog te vroeg
om deel te nemen aan deze visitatie. Tegelijkertijd is het zo dat
er in de visitatie een aantal aandachtspunten zijn genoemd waar
we het Curriculum 2003 zinnig mee kunnen confronteren.

\section{Internationaal kader}
De visitatie van Informatiekunde opleidingen verwijst expliciet naar
de rol van het MSIS 2000 als modelcurriculum
waar Informatiekunde opleidingen zich aan zouden moeten spiegelen.
In appendix~\ref{MSIS2000} zijn we reeds deze confrontatie aangegaan.
Het bachelor-deel van dit modelcurriculum heeft in de vorm van
IS2002 een update ondergaan. Zie voor een bespreking van de relatie
tussen Curriculum 2003 en IS2002 appendex~\ref{IS2002}.

\section{Specifieke opleidingseisen}
In het visitatierapport wordt melding gemaakt van de volgende specifieke
opleidingseisen op basis waarvan de verschillende informatiekunde opleidingen
zijn getoetst:
\begin{enumerate}
   \item Inleidingen in de belangrijkste deelgebieden van de Informatiekunde,
   en kennismaking met de manier van denken in systeemtheoretische concepten.
   \item Steunvakken die het mogelijk maken een abstracte en formele denktrant
   aan te leren.
   \item Integrerende vakken waarin meedere disciplines samenkomen.
   \item Naast globale kennis van de genoemde deelgebieden, specialisatie in
   ten minste \'e\'en deelgebied.
   \item Kennismaking met de (beroeps)praktijk, bijvooreeld in de vorm van
   toegepaste vakken en/of een stage.
   \item Het verrichten van zowel literatuuronderzoek als eigen onderzoek.
\end{enumerate}
Wanneer we Curriculum 2003 bezien in het licht van de bovenstaande
eisen, dan kunnen we naar onze mening, respectievelijk, stellen dat:
\begin{enumerate}
   \item Curriculum 2003 is opgesteld op basis van een uitgebreide visie
   op het vakgebied~\cite{Visie2003}, en dat de verschillende deelgebieden
   van de informatiekunde daadwerkelijk aan bod komen in de opleiding.
   In het aspect `Grondslagen' wordt zo'n 6 ECTS aandacht besteed aan
   systeemtheoretische grondslagen van het vakgebied.

   \item In het informatiekunde curriculum zijn de vakken ``Beweren \&
   Bewijzen'' en ``Formeel Denken'' expliciet opgenomen om deze rol te
   vervullen.

   \item De vakken R\&D 1 en R\&D 2 zijn specifiek voor dit doel in het
   leven geroepen. Daarnaast bieden de systeemontwikkelingsvakken ook
   nog een extra kans voor studenten om de door hun opgedane vaardigheden
   verder te integreren in een praktische setting.

   \item Informatiekunde Nijmegen heeft vooralsnog \'e\'en inhoudelijke
   specialisatierichting: Informatiearchitectuur. Wellicht dat daar in de
   nabije toekomst nog een specialisatierichting omtrent kwaliteit van
   informatiesystemen bij komt. Daarnaast hebben de studenten de mogelijkheid
   zich te specialiseren in \'e\'en of twee verbredingsgebieden van de
   Informatiekunde.

   \item Reeds in het vak ``Orientatiecollege Toepassingen'' krijgen de
   studenten in het eerste jaar al te maken met verbredingsgebieden van
   de Informatiekunde, en dit zowel vanuit theoretisch als \emph{praktisch}
   perspectief. Daarnaast worden docenten zoveel mogelijk gestimuleerd om
   waar relevant een gastspreker uit de praktijk uit te nodigen als
   aanvulling op de theorie.

   \item De R\&D 1 en R\&D 2 vakken, samen met de bachelor- en master-scriptie,
   bieden genoeg mogelijkheden om zowel literatuuronderzoeken als eigen onderzoek
   uit te voeren.
\end{enumerate}

\section{Specifieke aandachtspunten}
Naast de opleidingseisen heeft de visitatiecommissie ook nog een aantal specifieke
aandachtspunten benoemd:
\begin{itemize}
   \item De relatie tussen het basisprogramma en de afstudeermogelijkheden.
   \item De integratie van de verworven kennis.
   \item De mate van specialisatie van de opleiding.
   \item De verhouding tussen theorie en praktijk in de opleiding, en de
   professionele vorming.
   \item De verwevenheid van onderwijs en onderzoek.
\end{itemize}
Wanneer we Curriculum 2003 bezien in het licht van de bovenstaande
eisen, dan kunnen we naar onze mening, respectievelijk, stellen dat:
\begin{itemize}
   \item Omdat het Curriculum is ontworpen vanuit een overall visie op het vakgebied,
   is er een duidelijke relatie tussen de inhoud en structuur van het basisprogramma en
   de afstudeermogelijkheden. Zowel qua inhoudelijke specialisaties als qua keuzen
   met betrekking tot een verbredingsgebied. Merk op: een eis van de bachelor-scriptie is
   ook dat deze inhoudelijk aansluit op \'e\'en van de twee gekozen verbredingsgebieden.
   \item Zoals eerder aangegeven, zijn de vakken R\&D 1 en R\&D 2 specifiek voor dit doel in het
   leven geroepen. Daarnaast bieden de systeemontwikkelingsvakken ook
   nog een extra kans voor studenten om de door hun opgedane vaardigheden
   verder te integreren in een praktische setting.
   \item Er een balans is tussen theoretische aspecten (grondslagen), praktisch
   gerichte vakken, en de verdere professionele ontwikkeling (systeemontwikkeling).
   \item Diverse vakken zijn direct gekoppeld aan lopende onderzoeksprogramma's binnen het \NIII.
   Denk aan: kwaliteit van software, security, information retrieval, computational intelligence
   en informatiearchitectuur.
\end{itemize}

\bibliography{refs}
\bibliographystyle{plain}
\end{document}